%% file: STAR_hJet_PRC_Master.tex
\documentclass[10pt,aps,prc,twocolumn,floatfix,showpacs,groupedaddress,superscriptaddress]{revtex4-1}
\usepackage{graphicx}
\usepackage{array}
\usepackage{color}
\usepackage{lineno}
\usepackage{placeins}
\usepackage{cleveref}
\usepackage{epstopdf}
\usepackage{lipsum}

\usepackage{hyperref}

\usepackage[percent]{overpic}
\usepackage{datetime}
\usepackage{multirow}
\usdate

\newcommand{\sNN}{\mbox{$\sqrt{s_{\rm{NN}}}$}}

\input{Include}

\begin{document}


\title{Measurements of jet quenching with semi-inclusive hadron+jet distributions in \AuAu\  collisions at \sqrtsNN\ = 200 GeV}

\input{Author_v2}

\begin{abstract}
The STAR Collaboration reports the measurement of semi-inclusive distributions 
of charged-particle jets recoiling from a high transverse momentum hadron trigger, 
in central and peripheral \AuAu\ 
collisions at \sNN\ = 200 GeV.  Charged jets are 
reconstructed with the \antikT\ algorithm for jet radii \rr\ between 0.2 and 
0.5 and with low infrared cutoff of track 
constituents ($\pT>0.2$ \gev). A novel mixed-event technique is used to 
correct the large uncorrelated background present in heavy ion 
collisions. Corrected recoil jet 
distributions are reported at mid-rapidity, for charged-jet transverse momentum $\pTjetch<30$ \gev. Comparison is made to similar measurements for \PbPb\ collisions at 
\sqrts\ = 2.76 TeV, to 
calculations for \pp\ collisions at \sqrts\ = 200 GeV based on the PYTHIA Monte 
Carlo generator and on a Next-to-Leading Order perturbative QCD approach, and to theoretical calculations incorporating jet quenching. The recoil jet 
yield is suppressed in central relative to peripheral collisions, with 
the magnitude of the suppression  corresponding to medium-induced charged energy 
transport out of the jet cone of 
$2.8\pm0.2\mathrm{(stat)}\pm1.5\mathrm{(sys)}$ \gev, for $10<\pTjetch<20$ \gev\ and \rr\ = 0.5. 
No medium-induced change in jet shape is observed for $\rr<0.5$.
The azimuthal distribution of low-\pTjetch\ recoil jets may be enhanced at large azimuthal 
angles to the trigger axis, due to scattering off quasi-particles in the hot QCD medium. Measurement of this distribution gives a 90\% statistical confidence upper limit to the yield enhancement at large deflection angles in central \AuAu\ collisions of $50\pm30\mathrm{(sys)\%}$ of the large-angle yield in \pp\ collisions predicted by PYTHIA.


\end{abstract}




\maketitle




\input{Intro}
\input{Experiment_and_Dataset}

\input{Observables}

\input{RawDistributions}

\input{Corrections}

\input{SystematicUncertainties}

\input{NLO}

\input{Results}
\input{Summary}

\input{Acknowledgements}

\bibliographystyle{h-physrev5}

\bibliography{references}

\end{document}

%% file: Include.tex
\newcommand{\CommentBlock}[1]{}

\newcommand{\AuAu}{Au+Au}
\newcommand{\PbPb}{Pb+Pb}

\newcommand{\pp}{p+p}

\newcommand{\sqrtsNN}{\ensuremath{\sqrt{s_{\mathrm {NN}}}}}
\newcommand{\sqrts}{\ensuremath{\sqrt{s}}}
\newcommand{\rr}{\ensuremath{R}}

\newcommand{\gev}{\ensuremath{\mathrm{GeV/}c}}

\newcommand{\kT}{\ensuremath{k_\mathrm{T}}}
\newcommand{\antikT}{anti-\ensuremath{k_\mathrm{T}}}

\newcommand{\pT}{\ensuremath{p_\mathrm{T}}}

\newcommand{\pTjet}{\ensuremath{p_{\mathrm{T,jet}}}}
\newcommand{\pTjetch}{\ensuremath{p_\mathrm{T,jet}^\mathrm{ch}}}
\newcommand{\pTtrig}{\ensuremath{p_{\mathrm{T,trig}}}}

\newcommand{\pTjetchlow}{\ensuremath{p_\mathrm{T,jet;low}^\mathrm{ch}}}
\newcommand{\pTjetchhigh}{\ensuremath{p_\mathrm{T,jet;high}^\mathrm{ch}}}

\newcommand{\pTjetpart}{\ensuremath{p_\mathrm{T,jet}^\mathrm{part}}}
\newcommand{\pTjetdet}{\ensuremath{p_\mathrm{T,jet}^\mathrm{det}}}

\newcommand{\pTthresh}{\ensuremath{p_{\mathrm{T,thresh}}}}

\newcommand{\pizero}{\ensuremath{\pi^0}}

\newcommand{\dNjetdpTdphinovar}{\ensuremath{\frac{\rm{d}^{3}N^\mathrm{AA}_{jet}}{\mathrm{d}\pTjetch\mathrm{d}\dphinovar\mathrm{d}\etajet}}}

\newcommand{\etajet}{\ensuremath{\eta_\mathrm{jet}}}

\newcommand{\Yjet}{\ensuremath{Y\left(\pTjetch\right)}}
\newcommand{\Phijet}{\ensuremath{\Phi\left(\dphinovar\right)}}

\newcommand{\dpT}{\ensuremath{\delta{\pT}}}

\newcommand{\qhat}{\ensuremath{\hat{q}}}
\newcommand{\vtwo}{\ensuremath{v_2}}
\newcommand{\vthree}{\ensuremath{v_3}}

\newcommand{\phiEP}{\ensuremath{\phi_{EP}}}

\newcommand{\Ntrig}{\ensuremath{\mathrm{N}^\mathrm{AA}_{\rm{trig}}}}

\newcommand{\RAA}{\ensuremath{R_\mathrm{AA}}}

\newcommand{\ICP}{\ensuremath{I_\mathrm{CP}}}

\newcommand{\Drecoil}{\ensuremath{\Delta_\mathrm{recoil}}}

\newcommand{\pTcorr}{\ensuremath{p_\mathrm{T,jet}^\mathrm{corr,ch}}}

\newcommand{\pTraw}{\ensuremath{p_\mathrm{T,jet}^\mathrm{raw,ch}}}
\newcommand{\pTrawi}{\ensuremath{p_\mathrm{T,jet}^\mathrm{raw,i}}}

\newcommand{\pTreco}{\ensuremath{p_\mathrm{T,jet}^\mathrm{reco,ch}}}
\newcommand{\pTdetch}{\ensuremath{p_\mathrm{T,jet}^\mathrm{det,ch}}}
\newcommand{\pTpartch}{\ensuremath{p_\mathrm{T,jet}^\mathrm{part,ch}}}

\newcommand{\Ajet}{\ensuremath{A_\mathrm{jet}}}

\newcommand{\pTpart}{\ensuremath{p_\mathrm{T,jet}^\mathrm{part}}}

\newcommand{\pTrecoi}{\ensuremath{p_\mathrm{T,jet}^\mathrm{reco,i}}}

\newcommand{\Ajeti}{\ensuremath{A_\mathrm{jet}^\mathrm{i}}}

\newcommand{\rhoAi}{\ensuremath{\rho\cdot{A_\mathrm{jet}^\mathrm{i}}}}

\newcommand{\dphi}{\ensuremath{\Delta\varphi}}
\newcommand{\dphinovar}{\ensuremath{\Delta\phi}}

\newcommand{\Rdet}{\ensuremath{R_{\mathrm{det}}}}
\newcommand{\Rbkg}{\ensuremath{R_{\mathrm{bkg}}}}

\newcommand{\pTrec}{\ensuremath{p_{\mathrm{T,jet}}^{\mathrm{det}}}}
\newcommand{\pTgen}{\ensuremath{p_{\mathrm{T,jet}}^{\mathrm{part}}}}

\newcommand{\zvtx}{\ensuremath{z_\mathrm{vtx}}}

\newcommand{\AAtohjet}{\mathrm{AA}\rightarrow\rm{h}+{jet}+X}
\newcommand{\AAtoh}{\mathrm{AA}\rightarrow\rm{h}+X}

\newcommand{\fME}{\ensuremath{f^\mathrm{ME}}}

\newcommand{\pTembed}{\ensuremath{p_\mathrm{T}^\mathrm{embed}}}

%% file: Author_v2.tex



\affiliation{AGH University of Science and Technology, FPACS, Cracow 30-059, Poland}
\affiliation{Argonne National Laboratory, Argonne, Illinois 60439}
\affiliation{Brookhaven National Laboratory, Upton, New York 11973}
\affiliation{University of California, Berkeley, California 94720}
\affiliation{University of California, Davis, California 95616}
\affiliation{University of California, Los Angeles, California 90095}
\affiliation{Central China Normal University, Wuhan, Hubei 430079}
\affiliation{University of Illinois at Chicago, Chicago, Illinois 60607}
\affiliation{Creighton University, Omaha, Nebraska 68178}
\affiliation{Czech Technical University in Prague, FNSPE, Prague, 115 19, Czech Republic}
\affiliation{Nuclear Physics Institute AS CR, 250 68 Prague, Czech Republic}
\affiliation{Frankfurt Institute for Advanced Studies FIAS, Frankfurt 60438, Germany}
\affiliation{Institute of Physics, Bhubaneswar 751005, India}
\affiliation{Indiana University, Bloomington, Indiana 47408}
\affiliation{Alikhanov Institute for Theoretical and Experimental Physics, Moscow 117218, Russia}
\affiliation{University of Jammu, Jammu 180001, India}
\affiliation{Joint Institute for Nuclear Research, Dubna, 141 980, Russia}
\affiliation{Kent State University, Kent, Ohio 44242}
\affiliation{University of Kentucky, Lexington, Kentucky, 40506-0055}
\affiliation{Lamar University, Physics Department, Beaumont, Texas 77710}
\affiliation{Institute of Modern Physics, Chinese Academy of Sciences, Lanzhou, Gansu 730000}
\affiliation{Lawrence Berkeley National Laboratory, Berkeley, California 94720}
\affiliation{Lehigh University, Bethlehem, PA, 18015}
\affiliation{Max-Planck-Institut fur Physik, Munich 80805, Germany}
\affiliation{Michigan State University, East Lansing, Michigan 48824}
\affiliation{National Research Nuclear University MEPhI, Moscow 115409, Russia}
\affiliation{National Institute of Science Education and Research, HBNI, Jatni 752050, India}
\affiliation{National Cheng Kung University, Tainan 70101 }
\affiliation{Ohio State University, Columbus, Ohio 43210}
\affiliation{Institute of Nuclear Physics PAN, Cracow 31-342, Poland}
\affiliation{Panjab University, Chandigarh 160014, India}
\affiliation{Pennsylvania State University, University Park, Pennsylvania 16802}
\affiliation{Institute of High Energy Physics, Protvino 142281, Russia}
\affiliation{Purdue University, West Lafayette, Indiana 47907}
\affiliation{Pusan National University, Pusan 46241, Korea}
\affiliation{Rice University, Houston, Texas 77251}
\affiliation{University of Science and Technology of China, Hefei, Anhui 230026}
\affiliation{Shandong University, Jinan, Shandong 250100}
\affiliation{Shanghai Institute of Applied Physics, Chinese Academy of Sciences, Shanghai 201800}
\affiliation{State University Of New York, Stony Brook, NY 11794}
\affiliation{Temple University, Philadelphia, Pennsylvania 19122}
\affiliation{Texas A\&M University, College Station, Texas 77843}
\affiliation{University of Texas, Austin, Texas 78712}
\affiliation{University of Houston, Houston, Texas 77204}
\affiliation{Tsinghua University, Beijing 100084}
\affiliation{University of Tsukuba, Tsukuba, Ibaraki, Japan,}
\affiliation{Southern Connecticut State University, New Haven, CT, 06515}
\affiliation{United States Naval Academy, Annapolis, Maryland, 21402}
\affiliation{Valparaiso University, Valparaiso, Indiana 46383}
\affiliation{Variable Energy Cyclotron Centre, Kolkata 700064, India}
\affiliation{Warsaw University of Technology, Warsaw 00-661, Poland}
\affiliation{Wayne State University, Detroit, Michigan 48201}
\affiliation{World Laboratory for Cosmology and Particle Physics (WLCAPP), Cairo 11571, Egypt}
\affiliation{Yale University, New Haven, Connecticut 06520}

\author{L.~Adamczyk}\affiliation{AGH University of Science and Technology, FPACS, Cracow 30-059, Poland}
\author{J.~K.~Adkins}\affiliation{University of Kentucky, Lexington, Kentucky, 40506-0055}
\author{G.~Agakishiev}\affiliation{Joint Institute for Nuclear Research, Dubna, 141 980, Russia}
\author{M.~M.~Aggarwal}\affiliation{Panjab University, Chandigarh 160014, India}
\author{Z.~Ahammed}\affiliation{Variable Energy Cyclotron Centre, Kolkata 700064, India}
\author{N.~N.~Ajitanand}\affiliation{State University Of New York, Stony Brook, NY 11794}
\author{I.~Alekseev}\affiliation{Alikhanov Institute for Theoretical and Experimental Physics, Moscow 117218, Russia}\affiliation{National Research Nuclear University MEPhI, Moscow 115409, Russia}
\author{D.~M.~Anderson}\affiliation{Texas A\&M University, College Station, Texas 77843}
\author{R.~Aoyama}\affiliation{University of Tsukuba, Tsukuba, Ibaraki, Japan,}
\author{A.~Aparin}\affiliation{Joint Institute for Nuclear Research, Dubna, 141 980, Russia}
% ...
\collaboration{STAR Collaboration}\noaffiliation

%% file: Intro.tex
\section{Introduction}
\label{sect:intro}

The interaction of energetic jets with hot QCD matter provides unique
probes of the Quark-Gluon Plasma (QGP) generated in high-energy
collisions of heavy nuclei (``jet quenching'', 
~\cite{Majumder:2010qh} and references therein). Jet quenching was first observed experimentally as the suppression of inclusive hadron production and hadron correlations
at high transverse momentum (high \pT) 
~\cite{Adare:2010ry,Adler:2002xw,Adler:2002tq,
  Adams:2003kv,Adams:2006yt,Adamczyk:2013jei,Adcox:2001jp,Adare:2012wg,Adare:2012qi,
  Abelev:2012hxa, CMS:2012aa, Aamodt:2011vg, Chatrchyan:2012wg}. Jet quenching is calculable theoretically, using approaches based on perturbative QCD and on strong coupling. Comparison of theoretical
calculations with measurements of inclusive hadron suppression at the Relativistic Heavy Ion Collider (RHIC) 
and Large Hadron Collider (LHC) has been used to constrain the jet transport parameter \qhat\ in the 
QGP~\cite{Burke:2013yra}.

Measurements based on high-\pT\ hadrons, which are leading fragments
of jets, bias towards jets that have lost relatively little energy in the
medium~\cite{Baier:2002tc}. These are therefore {\it disappearance} measurements, in which the contribution of jets that interact most strongly in the medium is suppressed. Such measurements have limited 
sensitivity to the detailed dynamics of parton shower
modification and the response of the medium to the passage of the
jet. Comprehensive exploration of jet quenching therefore requires measurements of reconstructed jets and their correlations. Jet measurements in the high-multiplicity environment of heavy ion
collisions are challenging, however, because of the large and dynamically fluctuating backgrounds in such events.

Heavy ion jet measurements at the LHC have reported medium-induced suppression 
in inclusive jet production~\cite{Abelev:2013kqa,Aad:2014bxa,Adam:2015ewa,Khachatryan:2016jfl}, as 
well as modification of di-jet and $\gamma$-jet correlations~\cite{Aad:2010bu,Chatrchyan:2012nia,Chatrchyan:2012gt}. These measurements 
suppress the contribution of uncorrelated background to the jet signal by 
rejecting reconstructed jets on a jet-by-jet basis based on measured jet 
\pT\ adjusted by an estimate of the uncorrelated background contribution, which may 
induce bias in the accepted jet population. 
The ALICE Collaboration at the LHC has measured
jet quenching in central \PbPb\ collisions at \sqrtsNN\ = 2.76 TeV with a 
different approach to the suppression of uncorrelated background, using the 
semi-inclusive distribution of reconstructed jets recoiling from a 
high-\pT\ trigger hadron~\cite{Adam:2015doa}. In the ALICE approach, 
correction for large uncorrelated jet background is carried out at the 
level of ensemble-averaged distributions, without discrimination on a jet-by-jet basis of correlated jet signal from uncorrelated background jets. This background
suppression procedure, which does not impose bias on the reported jet 
population, enables heavy ion jet measurements over a broad kinematic range, 
including large jet radius \rr\ and low \pTjet.

This manuscript reports new measurements of jet quenching in central and
peripheral \AuAu\ collisions at \sqrtsNN\ = 200 GeV, by the STAR
Collaboration at RHIC. These
measurements are also based on the semi-inclusive distribution of
reconstructed charged-particle jets recoiling from a high-\pT\ trigger
hadron. We apply a novel mixed-event technique for correcting
uncorrelated jet background, and compare it to the
approach used in the ALICE measurement~\cite{Adam:2015doa}. 
Distributions of charged particle recoil jets with $\pTjetch<30$ \gev\ 
and jet resolution parameters (or jet radius) \rr\ = 0.2, 0.3, 0.4 and 0.5 are reported as a function of \pTjetch\ and \dphinovar, the 
azimuthal angle of the jet centroid relative to that of the trigger axis.

These measurements probe medium-induced modification of jet production and 
internal jet structure in several ways. Suppression of jet yield in central 
compared to the yield in peripheral collisions with the same jet cone radius \rr\ 
measures the energy transported to angles larger than \rr.
Comparison of recoil jet
yield at different \rr\ measures medium-induced modification of
jet shape (intra-jet 
broadening~\cite{Wang:2013cia,Kurkela:2014tla,Casalderrey-Solana:2016jvj}). 
The distribution of \dphinovar\ measures medium-induced acoplanarity (inter-jet
broadening). Yield enhancement in the tail of the \dphinovar\ distribution could 
indicate medium-induced Moli{\`e}re scattering off quasi-particles in the hot QCD 
medium~\cite{D'Eramo:2012jh,Wang:2013cia}. The acoplanarity distribution of low energy 
jets is sensitive to $\langle\qhat\cdot{L}\rangle$, where \qhat\ is the jet transport parameter and 
$L$ is the in-medium path length~\cite{Chen:2016vem}.  

We compare these results with 
those from ALICE~\cite{Adam:2015doa}, providing a direct comparison of jet quenching 
measured by reconstructed jets at RHIC and the LHC.
Comparison of these measurements to distributions from \pp\ collisions 
at \sqrts\ = 200 GeV can be used to identify nuclear effects that 
are present in peripheral \AuAu\ collisions. However, due to 
the lack of a measured reference distribution for \pp\ collisions
at present, we compare the \AuAu\ measurements to expectations for \pp\ 
collisions at 
\sqrts\ = 200 GeV from the PYTHIA Monte Carlo event generator, tune A~\cite{Sjostrand:2006za}, and from a perturbative QCD calculation at 
Next-to-Leading Order (NLO)~\cite{deFlorian:2009fw}.

The paper is organized as follows: Sect. II, experiment, dataset, and offline 
analysis; Sect. III, jet reconstruction; Sect. IV, semi-inclusive hadron+jet 
distributions; Sect. V, uncorrelated background and event mixing; Sect. VI, raw 
distributions; Sect. VII, corrections; Sect. VIII, systematic uncertainties; 
Sect. IX, closure test; Sect. X, perturbative QCD calculation; Sect. XI, 
results; and Sect. XII, summary.

%% file: Experiment_and_Dataset.tex
\section{Experiment, Dataset, and Offline Analysis}
\label{sect:ExpDatasetOffline}

STAR is a large, multi-purpose experiment at RHIC, consisting of a solenoidal magnet 
and detectors for triggering, tracking, particle identification, calorimetry, 
and event categorization~\cite{Ackermann:2002ad}.

The data used in this analysis were recorded during the 2011 RHIC run with 
\AuAu\ collisions at \sqrtsNN\ = 200 GeV. Events were accepted online with a 
minimum-bias trigger requiring the coincidence of signals from the Zero 
Degree Calorimeters (ZDC) and the Vertex Position Detectors (VPD)~\cite{Llope:2003ti}. 
The trigger included the requirement that the $z$-position of the primary vertex of the event (\zvtx) was within 
$\pm$ 30 cm of the nominal center of the STAR detector.

Offline analysis was carried out using charged tracks measured by the STAR Time 
Projection Chamber (TPC)~\cite{Anderson:2003ur}. The TPC has inner radius of 50 cm 
and outer radius of 200 cm, with acceptance $|\eta|<1.0$ over the full azimuth. The TPC registers a maximum of 45 independent points for a charged track. The 
primary vertex is defined using global tracks, based on fitting of TPC clusters.
The vertex position 
resolution in the beam direction is 
$\delta{z_{vtx}}=350 \mu{m}$ for the highest multiplicity events in the 
analysis, which contain around 1000 primary tracks. 

The analysis utilizes primary tracks, which are global tracks whose
distance of closest approach to the primary 
vertex in the transverse plane ($\mathrm{DCA}_{xy}$) is less than 1 cm.  
The primary track momentum is determined by a fit that includes the primary 
vertex. Primary tracks 
with $\pT>0.2$ \gev\ are accepted for further analysis. 

The primary charged track transverse momentum resolution is 
$\sigma_{p_T}/\pT=0.01\times\pT$ [\gev]. The STAR tracking system momentum 
resolution at high \pT\ has been verified by matching tracks to a shower in the 
Barrel Electromagnetic Calorimeter (BEMC) for electrons from W-decay in \pp\ 
collisions~\cite{Aggarwal:2010vc}. Tracks with primary \pT\ larger than 30 \gev\ are 
excluded from the analysis. The probability for an event to have both a 
track with $9<\pT<30$ \gev\ and a track with $\pT>30$ \gev\ is negligible.

Tracking efficiency is determined by embedding simulated tracks into real \AuAu\ 
events. Primary track efficiency for charged pions is 48\% at \pT\ = 0.2 \gev, 67\% 
at \pT\ = 0.4 \gev, and 73\% at \pT\ = 20 \gev\ for central \AuAu\ collisions; and 
66\% at \pT\ = 0.2 
\gev, 86\% at \pT\ = 0.4 \gev, and 89\% at \pT\ = 20 \gev\ for peripheral \AuAu\ 
collisions. At high transverse momentum the tracking efficiency of charged pions, kaons, 
and protons is similar, while the efficiency of protons and kaons is 
significantly lower than that of pions 
for $\pT<0.5$ \gev. 

Pile-up events, due to high instantaneous 
luminosity, are excluded offline by requiring at 
least two tracks from the 
primary vertex to be matched to cells of the Time-of-Flight (TOF) detector, which is a fast detector that can identify out-of-time tracks.
Quality assurance is carried out on a run-wise basis, with a run corresponding 
to several hours of 
online data-taking. A run was rejected if its deviation from global mean 
values exceeded $5 \sigma$ 
for mean transverse momenta $\langle p_{T} \rangle$  or $2 \sigma$ for 
multiplicity $\langle M \rangle$, measured using uncorrected charged track 
distributions in $|\eta|<0.5$; or $2.5 \sigma$ for the 
interaction rate measured in the forward scintillator Beam-Beam Counters, 
$\langle BBCx \rangle$.

\begin{figure}[htbp] 
\includegraphics[width=0.5\textwidth]{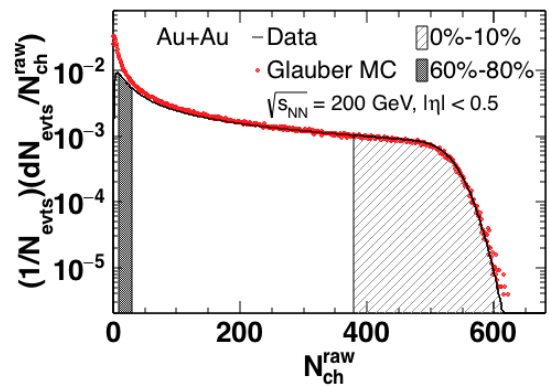} 
\caption{(Color online) Centrality selection for \AuAu\ collisions at \sqrtsNN\ 
= 200 GeV: distribution of uncorrected charged track multiplicity in 
$|\eta|<0.5$ (black histogram), with comparison to the result of a Glauber 
model~\cite{Miller:2007ri} calculation (red points). The shaded regions show the windows for 0\%-10\% 
(central) and 60\%-80\% (peripheral) \AuAu\ collisions.} 
\label{fig:RefMult}
\end{figure}

Figure~\ref{fig:RefMult} shows the distribution of uncorrected multiplicity of 
charged particle tracks within $|\eta|$ $<$ 0.5. Events are classified offline 
using percentile intervals of this distribution, with the 0\%--10\% 
(``central'') and 60\%--80\% (``peripheral'') intervals shown in the figure. 
The figure also shows the charged particle 
multiplicity distribution from a Monte Carlo Glauber calculation~\cite{Miller:2007ri}. Comparison of the distributions from the Monte Carlo
calculation and data gives an online trigger efficiency of 100\% for 
central collisions and 70\% for peripheral collisions. 

After event selection 
cuts, the data set consists of 56.5 M central (0\%--10\%) and 106.7 M peripheral 
(60\%--80\%) events. The effect of trigger inefficiency in peripheral collisions is accounted for by a multiplicity-dependent weighting of events.

Simulated events are generated using PYTHIA 6.416 tune A~\cite{Sjostrand:2006za} 
folded with a detector response based on GEANT3~\cite{GEANT3}. Distributions 
calculated without incorporating detector response are denoted ``particle 
level'', while distributions that include detector response are denoted 
``detector level.'' 
Fast generation of detector-level events from particle-level PYTHIA
simulations is carried out by random rejection of charged
tracks to model tracking efficiency, and smearing of track \pT\ to model 
momentum 
resolution, with \pT-dependent efficiency and resolution.

Hybrid events for embedding studies are constructed by generating PYTHIA events 
for \pp\ collisions at \sqrts\ = 200 GeV, selecting events containing a high-\pT\ hadron in the trigger acceptance (Sect. \ref{sect:Observables}), and 
applying the ``fast generation" detector-level effects. Each simulated event is 
combined with a real \AuAu\ 
event at the track level from the central or peripheral population, without requiring a track 
in the trigger acceptance in the real event. Since embedding is carried out at the track level, tracks are specified in terms of $(\pT,\eta,\phi)$, with no need to specify a vertex position. The hybrid events are analyzed using the 
same procedure used for real 
data analysis.

We also compare these measurements to theoretical expectations for \pp\ 
collisions at \sqrts\ = 200 GeV based on an NLO pQCD calculation~\cite{deFlorian:2009fw} (Sect.~\ref{sect:pQCD}).

%% file: Observables.tex
\section{Jet reconstruction}
\label{sect:JetReco}

The analysis utilizes charged jets, which are composed of charged tracks. Jet reconstruction is carried out with the \kT~\cite{Cacciari:2011ma} 
and \antikT~\cite{FastJetAntikt} algorithms applied to all accepted charged
tracks using the E-recombination scheme~\cite{Cacciari:2011ma}. Jet distributions are corrected to the charged particle level for the effects of uncorrelated background and instrumental response. 

Jet area is determined using the Fastjet area algorithm~\cite{FastJetArea} with 
ghost particle area of 0.01. Ghost particles are randomly generated particles 
with negligible \pT\ that are distributed uniformly in the acceptance with known 
density, and are clustered during jet reconstruction together with real tracks. 
The 
number of ghost particles in a jet thereby provides an infrared and 
collinear-safe (IRC-safe) measurement of jet area, for jets of arbitrary 
shape~\cite{FastJetArea}.

We utilize the following notation to distinguish \pT\ of various types of jet in 
the analysis: \pTraw\ is \pT\ of jets generated by the jet reconstruction 
algorithm; \pTreco\ is \pTraw\ adjusted by an estimate of the uncorrelated 
background contribution; and \pTjetch\ is \pT\ of jets after full 
correction for the effects of instrumental response and background fluctuations. 
For the simulation of \pp\ collisions, \pTgen\ is the reconstructed jet 
energy at the particle-level and \pTrec\ is at
the detector-level, with no correction for uncorrelated background considered; 
i.e. these are equivalent to \pTraw\ at the two levels of simulation.

Discrimination of correlated jet signal from uncorrelated background in 
this analysis is carried out at the level of ensemble-averaged 
distributions. Specifically, we do not discriminate the individual objects 
generated by the jet reconstruction algorithm based on features that may indicate
contribution from high-Q$^2$ partonic scattering processes. 
We therefore refer to all such objects as ``jet candidates", 
rather than simply as ``jets", to denote that a 
significant fraction of such objects are purely combinatoric in origin; i.e. 
without a component arising from a high-Q$^2$ scattering process, in contrast to 
what is 
conventionally meant by the term ``jet" in QCD.

Jet reconstruction is carried out multiple times for each event. The first jet 
reconstruction pass uses the 
\kT\ algorithm with \rr\ = 0.3 to estimate
the background transverse energy density $\rho$ in the event~\cite{FastJetPileup},

\begin{equation}
\rho=\mathrm{median}\left\{ \frac{\pTrawi}{\Ajeti} \right\},
\label{eq:rho}
\end{equation}

\noindent
where i labels the jet candidates in the event, and \pTrawi\ and
\Ajeti\ are the transverse momentum and area of jet candidate i. The median is 
calculated by excluding the two hardest jets in the event for peripheral \AuAu\ 
collisions, and the three hardest jets for central \AuAu\ collisions.

\begin{figure}[htbp]
\includegraphics[width=0.5\textwidth]{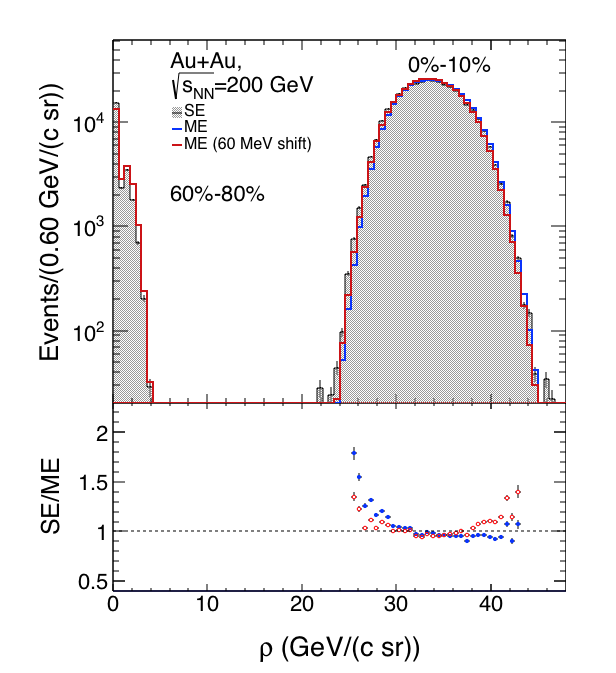}
\caption{(Color online) Upper panel: distribution of $\rho$ 
for central and peripheral \AuAu\ collisions (SE), 
and for mixed events (ME, see Sect.~\ref{sect:ME}). Lower panel: ratio of distributions SE/ME for central \AuAu\ collisions. Blue points are ME distribution used in analysis; red points are same distribution shifted by 60 MeV/($c$ sr). See discussion in Sect.~\ref{sect:ME}.}
\label{fig:rho}
\end{figure} 

Figure \ref{fig:rho} shows the distribution of $\rho$ for central and peripheral 
\AuAu\ collisions. Distributions are shown for STAR data (SE) and for 
mixed events (ME, see Sect.~\ref{sect:ME}). The term SE refers to "same events", 
in contrast to mixed events. The value of $\rho$ varies event-to-event due to 
variation in gross event features within each centrality class, in particular 
multiplicity and transverse energy. There are peripheral \AuAu\ events with 
$\rho=0$, which can occur for low multiplicity events since $\rho$ is calculated 
as the median of the jet energy density distribution. 

Successive jet reconstruction passes are then carried out using the 
\antikT\ algorithm, with \rr\ = 0.2, 0.3, 0.4, and 0.5. For each jet candidate 
generated in these passes, the 
value of \pTrawi\  is adjusted by the estimated background
energy density scaled by jet area~\cite{FastJetPileup},

\begin{equation}
\pTrecoi=\pTrawi - \rhoAi.
\label{eq:pTraw}
\end{equation}

The jet candidate acceptance is $|\eta_{\rm{jet}}|<(1.0-\rr)$, where
$\eta_{\mathrm{jet}}$ is the pseudo-rapidity of the jet
centroid. A jet area cut suppresses jets comprising uncorrelated background, 
while preserving high efficiency for jet candidates containing a true jet. Jet candidates are rejected if $\Ajeti<0.05$ for \rr\ = 0.2; $\Ajeti<0.20$
for \rr\ = 0.3; $\Ajeti<0.35$ for \rr\ = 0.4; and $\Ajeti<0.65$ for \rr\ =
0.5. The jet area cut is discussed further in Sect.~\ref{sect:ME}.

\begin{figure}[htbp]
\includegraphics[width=0.5\textwidth]{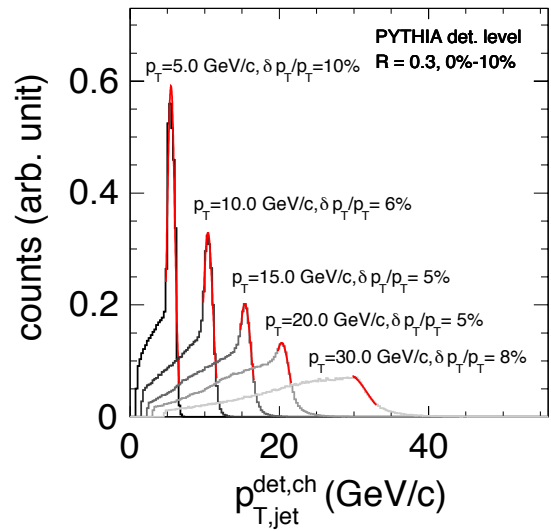}
\caption{(Color online) Distribution of jets with \rr=0.3 in \pp\ collisions at \sqrts=200 GeV, generated by PYTHIA: \pTrec\ (detector level) for fixed values of \pTgen\ (particle level). Detector-level effects are for the environment of central \AuAu\ collisions. The red lines are Gaussian fits to the narrow peak, with relative width given as $\delta\pT/\pT$.}
\label{fig:JER}
\end{figure}

Figure~\ref{fig:JER} shows the distribution of jets simulated by PYTHIA for fixed values of particle-level \pTgen, as a function of detector-level \pTrec. The detector-level effects correspond to conditions in central \AuAu\ collisions. These distributions represent the instrumental response to charged jets, and are non-Gaussian. Correction for these instrumental effects is carried out by an unfolding procedure~\cite{Cowan:2002in,Hocker:1995kb} utilizing an
instrumental response matrix. It is nevertheless illustrative to quantify the 
main features of the instrumental response. For charged jets in the range 
$5<\pTreco<30$ \gev, jet energy resolution (JER) due to instrumental effects has a peak with 
$\sigma=5-10\%$ and tail to low jet energy, . The complete JER distribution has 
RMS = 25\%, with negligible dependence of the JER on \rr. The  jet energy 
scale (JES) uncertainty due to instrumental effects, which arises predominantly 
from uncertainty in tracking efficiency, is 5\%, likewise with negligible 
\rr-dependence. 
          
\begin{figure*}[htbp]
\includegraphics[width=1.0\textwidth]{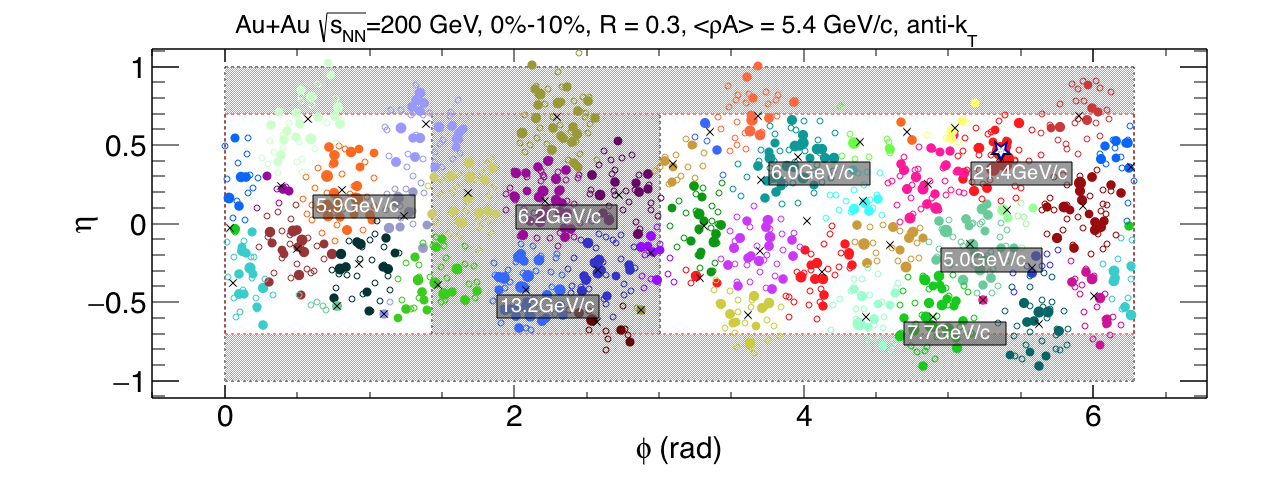}
\caption{(Color online) Event display showing the distribution of 
charged tracks 
and jets (\antikT,\rr=0.3) in one \AuAu\ collision from the central event 
population, as 
a function of $\eta$ and 
$\phi$. 
Filled circles show charged tracks, open circles show ghost 
particles, and the centroid of each accepted jet is indicated by ``x". 
Charged tracks 
and ghost particles clustered into each reconstructed jet have the same 
color. The shaded text boxes give \pTraw\ for all jet candidates with 
$\pTraw>5$ \gev. The outer dashed rectangle is the tracking acceptance, while the red shaded area is the region of the tracking acceptance that is excluded by the \rr-dependent jet fiducial cut. The trigger particle is indicated by the star, while the blue 
shaded area is the recoil jet acceptance. The trigger particle in this event is associated 
with the jet candidate with largest \pTraw.}
\label{fig:Eta_phi}
\end{figure*}

There is no absolute definition of
uncorrelated background energy density in an event. The definition of 
$\rho$ outlined above is not unique; different choices 
of reconstruction algorithm, jet radius \rr, and number of excluded jets, 
provide equally valid background estimates. As discussed below, the jet-wise 
adjustment in 
Eq.~\ref{eq:pTraw} is the first step in a multi-step process in which 
full correction 
for uncorrelated background utilizes an instrumental response matrix 
incorporating the same choice of $\rho$. 
Since no jet candidates are excluded based on their value 
of \pTrecoi\ in this 
analysis, the final corrected spectrum 
is independent of the specific choices made in the definition of 
$\rho$. The above choices for $\rho$ are made for technical reasons, to 
ensure numerical stability of the unfolding procedures.

\section{Semi-inclusive hadron+jet distributions}
\label{sect:Observables}

\subsection{Specification of observables}

The analysis is based on the semi-inclusive distribution
of charged jets recoiling from a high-\pT\ trigger hadron (``h+jet'')~\cite{deFlorian:2009fw,deBarros:2012ws,Adam:2015doa}. The trigger hadron is a 
charged particle with \pTtrig\ within a specified interval. The interval for the 
primary analysis is $9<\pTtrig<30$ \gev, while lower \pTtrig\ is used for 
systematic studies. 

The trigger hadron is selected inclusively: if there 
is a charged hadron observed within the \pTtrig\ interval the event is accepted, 
otherwise it is rejected. The
probability per central \AuAu\ collision to find a hadron within the interval 
$9<\pTtrig<30$ \gev\
is about 0.1\%, while the probability to observe multiple trigger 
hadron candidates is negligible. The resulting \pT-distribution\ of trigger 
hadrons is therefore the same as that of the inclusive charged hadron 
distribution.
The trigger hadron is not 
necessarily the highest-\pT\ hadron in the event, because neutral 
hadrons are not considered in the analysis. 

Figure \ref{fig:Eta_phi} is an event display for an \AuAu\ collision in 
the central event population, showing charged tracks, ghost 
particles, and reconstructed jet candidates. The acceptance 
is densely populated with tracks, and all tracks shown are associated 
with an accepted jet candidate. Voids in the track distribution occur near the 
edges of the jet fiducial acceptance, where the region occupied by a 
jet candidate lies partially within the tracking acceptance but its centroid 
lies outside the jet acceptance. The most energetic jet in this 
event happens to contain the trigger hadron, but that is not required. The 
recoil acceptance contains two jets with $\pTreco>5$ \gev.

The measured observable is the number of recoil jets observed in a phase space bin, normalized by the number of trigger hadrons. Because the trigger hadron is chosen inclusively, the resulting 
distribution is semi-inclusive and is equivalent to the ratio 
of production cross sections,

\begin{widetext}
\begin{equation}
\frac{1}{\Ntrig}\cdot\dNjetdpTdphinovar\Bigg\vert_{\pTtrig}
= \left(
\frac{1}{\sigma^{\AAtoh}} \cdot
\frac{\rm{d}^3\sigma^{\AAtohjet}}{\mathrm{d}\pTjetch\mathrm{d}\dphinovar\mathrm{d}\etajet}\right) 
\Bigg\vert_{\pTtrig},
\label{eq:hJetDefinition}
\end{equation}
\end{widetext}

\noindent
where AA denotes \pp\ or \AuAu\ collisions; \Ntrig\ is the number of 
trigger hadrons; $\sigma^{\AAtoh}$ is the cross 
section to generate a hadron within the \pTtrig\ interval;
$\rm{d}^3\sigma^{\AAtohjet}/\rm{d}\pTjetch\mathrm{d}\dphinovar\mathrm{d}\etajet$ is the differential cross section for coincidence production of a trigger
hadron and recoil jet; \pTjetch\ and \etajet\ are the charged jet
transverse momentum and pseudo-rapidity; and \dphinovar\ is the azimuthal
separation between trigger hadron and recoil jet. 

We report two projections of Eq. \ref{eq:hJetDefinition}: the jet
yield integrated over a recoil region in azimuth relative to the
trigger hadron direction,

\begin{widetext}
\begin{equation}
\Yjet=\int_{3\pi/4}^{5\pi/4}\mathrm{d}\dphinovar  
\left[\frac{1}{\Ntrig}\cdot\dNjetdpTdphinovar\Bigg\vert_{\pTtrig>\pTthresh}\right];
\label{eq:hJetYield}
\end{equation}
\end{widetext}

\noindent
and the azimuthal distribution of recoil jets in an interval of \pTjetch,

\begin{widetext}
\begin{equation}
\Phijet=\int_{\pTjetchlow}^{\pTjetchhigh}\mathrm{d}\pTjetch  
\left[\frac{1}{\Ntrig}\cdot\dNjetdpTdphinovar\Bigg\vert_{\pTtrig>\pTthresh}\right].
\label{eq:hJetPhi}
\end{equation}
\end{widetext}

\subsection{Discussion of observables}

The semi-inclusive observable defined in Eq. 
\ref{eq:hJetDefinition} isolates a single high-$Q^2$ 
process in each event by the requirement of a high-\pT\ hadron, and then measures 
the distribution of correlated recoil jets. The main considerations for this 
choice 
of observable are as follows (see also~\cite{Adam:2015doa}).

The observable in Eq. 
\ref{eq:hJetDefinition} is equivalent to the 
ratio of inclusive cross sections, which we first discuss from a theoretical 
perspective.  Inclusive high-\pT\ 
hadron production in \pp\ collisions at \sqrts\ = 200 GeV is 
well-described by pQCD calculations at NLO~\cite{deFlorian:2005yj,d'Enterria:2013vba}, and the h+jet cross section in 
\pp\ collisions at \sqrts\ = 200 GeV has also
been calculated in pQCD at NLO~\cite{deFlorian:2009fw}. For \pp\ 
collisions at \sqrts\ = 200 GeV, the observable in Eq. \ref{eq:hJetDefinition} is 
therefore calculable in pQCD at NLO (Sect. 
\ref{sect:pQCD}). In \AuAu\ collisions at \sqrtsNN\ = 200 GeV, 
hadrons with $\pT>5$ \gev\ are expected to arise predominantly from jet 
fragmentation~\cite{Renk:2011gj}, and pQCD calculations incorporating 
medium-evolved fragmentation functions and other techniques are in good agreement with 
measurements of inclusive 
hadron suppression at high \pT~\cite{Armesto:2007dt,Chang:2014fba,Burke:2013yra}. 
Inclusive hadron production in \AuAu\ collisions is therefore well-understood in the 
trigger interval of this analysis, using perturbative approaches.

Any procedure to accept a subset of events from the Minimum Bias 
distribution 
imposes bias on the accepted event population. Event selection in this 
analysis is simple, requiring only the presence of a high-\pT\ charged hadron in 
the event, with no requirement that a jet satisfying certain criteria be found 
in the recoil acceptance. Specifically, no rejection of jet candidates is 
carried out based 
on \pTrecoi,  and 
discrimination of correlated from uncorrelated yield is carried 
out at the level of ensemble-averaged distributions. All jet 
candidates in the recoil acceptance therefore contribute to the recoil jet 
distribution, and no selection bias is imposed on the correlated recoil jet 
population by the 
procedure to discriminate correlated jet signal from background.

Trigger hadron selection is carried out inclusively, resulting in 
the same \pT-distribution as that of inclusive hadron production~\cite{Adams:2003kv,Adare:2012wg}. Although the same kinematic selection 
is used for central and peripheral \AuAu\
collisions ($9<\pTtrig<30$ \gev), the selected distribution of underlying hard processes 
may differ between collision centralities because of jet quenching 
effects on high-\pT\ hadron production, resulting in different trigger bias. However,  selection of high-\pT\ hadrons is expected from model studies to bias towards 
leading fragments of jets that have experienced little quenching, due to the 
interplay of jet energy loss, the shape of the jet production spectrum, and jet 
fragmentation~\cite{Baier:2002tc}, and thereby limiting the effects of quenching 
on the trigger bias.

Insight into the centrality dependence of the trigger bias can be obtained 
from measurements of inclusive high-\pT\ hadron production, whose yield is
strongly suppressed in central \AuAu\ collisions at \sqrtsNN\ = 200 GeV~\cite{Adare:2012wg,Adams:2003kv}. 
Yield suppression of \pizero\ production, measured by the ratio of the
inclusive  yield in \AuAu\ to that in \pp\ collisions (\RAA), has a rate of change with \pT\ in central \AuAu\ collisions 
(0\%-5\%) of $0.01\pm0.003\ (\gev)^{-1}$, over the 
range $7<\pT<20$ \gev~\cite{Adare:2012wg}.
Similar \pT-dependence is observed for peripheral collisions, though 
with larger uncertainty.
In other words, while inclusive hadron production is strongly 
suppressed in central relative to peripheral \AuAu\ collisions, the shape of the 
inclusive \pT-distribution is the same within uncertainties for the two 
centralities.
This supports the conjecture of high-\pT\ trigger hadrons being 
generated preferentially by non-interacting jets, 
thereby selecting a similar distribution of hard processes for peripheral and 
central 
collisions, though at a suppressed rate for central collisions.

Further exploration of the trigger bias 
in this measurement requires theoretical calculations that incorporate jet 
quenching. Since inclusive hadron \RAA\ is modeled accurately by such 
calculations (\cite{Burke:2013yra} and references therein), they will likewise 
model the trigger bias accurately by including effects of jet quenching on the 
generation of trigger hadrons.

\subsection{Interpretation of distributions}
\label{sect:Interp}

For jets in vacuum, a pQCD description is thought to be applicable for 
$\pTjetch\gtrsim{10}$ \gev, where jets are interpreted in terms of fragmentation of 
quarks and gluons. In this analysis, in contrast,  the terms ``jet" and ``jet 
candidate" refer generically to objects reconstructed by the \antikT\ algorithm 
with specified \rr, without regard to the interpretability of such objects in 
terms of quark or gluon fragmentation. The raw spectrum is measured as a 
function of \pTreco, with contribution to each bin in \pTreco\ from a broad 
range in \pTjetch\ due predominantly to large \pT-smearing by background 
fluctuations. No cuts are applied on \pTreco, in order not to bias the measured 
\pTjetch\ distributions. 

The corrected recoil jet distributions therefore contain entries for 
the entire range that is formally allowed, $\pTjetch>0$, and represent the 
distribution in \pTjetch\ of all jet-like objects that are correlated with the 
trigger. The per-trigger rate of such objects is finite for $\pTjetch\sim0$, 
since jet-like objects with $\rr>0$ subtend finite area, and a finite number of 
such objects fill the experimental acceptance.

The corrected recoil jet distributions are presented in 
Sect.~\ref{sect:Results} over their full measured range, $\pTjetch>0$. However, 
for interpretation of these distributions in terms of parton showers and their 
modification in-medium, we restrict consideration to $\pTjetch>10$ \gev, the 
range over which a perturbative description of jets is commonly thought to be 
applicable in vacuum.

%% file: RawDistributions.tex
\section{Uncorrelated background and event mixing}
\label{sect:ME}

Jet production in collisions of heavy nuclei occurs in a more complex 
environment than in \pp\ collisions, due to the high multiplicity of hadrons 
arising from copious soft interactions ($Q^2<$ few GeV$^2$) and the high rate 
of multiple, incoherently generated jets. Collective effects in the evolution of 
the system also shape the event structure. Hadrons from these various sources will contribute to the population within each phase-space region of 
dimension \rr\ that is characteristic of jet reconstruction. This 
renders jet measurements in 
nuclear collisions especially complex, necessitating precise definition of 
jet signal and uncorrelated background.

In this analysis, the raw jet yield distribution as a function of \pTreco\ 
requires correction for the large yield of 
background jets that are uncorrelated with the trigger hadron, and for the 
\pT-smearing of correlated jets by the background. The 
uncorrelated background jet yield is subtracted at the level of 
ensemble-averaged distributions using mixed events (ME), described below. 
Correction for \pT-smearing due to background 
fluctuations is carried out by the unfolding of ensemble-averaged distributions.

In the ME procedure, real events from the population without high-\pT\ trigger bias are assigned to exclusive classes, with each class 
corresponding to a narrow bin in $M$, the uncorrected charged particle
multiplicity; \zvtx, the $z$-position of reconstructed vertex; and \phiEP, the 
azimuthal orientation of the event plane (EP) in the
laboratory frame. The EP orientation is an approximation of the reaction plane orientation, defined by the collision impact parameter and the beam axis. Event plane reconstruction is described in~\cite{Adamczyk:2013gw}.

There are 8 bins in $M$, 20 bins in \zvtx, and 4
bins in \phiEP, corresponding to 640 distinct event mixing
classes. Within each multiplicity bin the distribution of track
multiplicity is sampled from the SE data set, to accurately reproduce the
multiplicity distribution of real events. This procedure accounts 
for the multiplicity bias in events containing a high-\pT\ trigger hadron, 
relative to the MB population. 

Each mixed event with $M$ tracks is 
generated by drawing one track from each of $M$ different events in a mixing class. For 
efficient construction of ME events, the event mixing algorithm draws from a buffer of 
about 1000 real events, with the algorithm terminating when any event in the 
buffer has had all its tracks used. All unused tracks remaining in the buffer 
are discarded, the event buffer is refilled, and the procedure is repeated.
Tracks are therefore used at most once in the mixing procedure.

The ME procedure generates an 
event population without multi-hadron correlations, but with the detailed 
features of real data in terms of non-uniformity in instrumental 
response and variation in detector acceptance due to the \zvtx\ distribution. 
Incorporation of such detector effects in the ME population is required  for accurate determination of the uncorrelated background 
distribution in the recoil jet population.

\begin{figure}[htbp]
\includegraphics[width=0.4\textwidth]{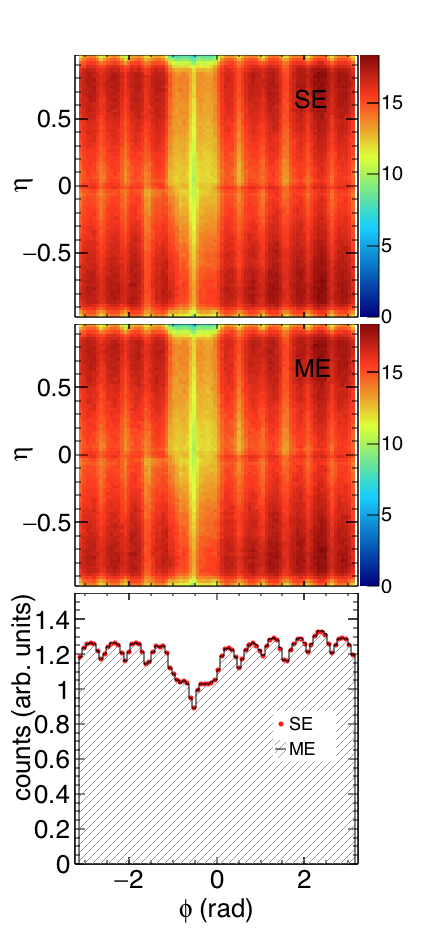}
\caption{(Color online) Distribution in ($\eta,\phi$) of charged particles from 
central \AuAu\ collisions, with $\pT<0.5$ \gev. Top panel: real or "same events" (SE); 
middle panel: mixed events (ME) for one event mixing class; lower panel: 
projection of SE and ME distributions onto $\phi$.}
\label{fig:SE_ME_EtaPhi}
\end{figure}

Figure \ref{fig:SE_ME_EtaPhi} shows the distribution of tracks with 
$\pT<0.5$ 
\gev\ for central \AuAu\ data, for SE events and for ME events from one mixing
class. The bottom panel shows the projection of the two distributions onto 
$\phi$. The periodic structure in the $\phi$ projection is due to reduced 
tracking efficiency near TPC sector boundaries, while the broad dip in the 
region $-1.0<\phi<0$ is due to reduced overall efficiency in two TPC sectors in 
this dataset. As noted above, only a subset of tracks from real events is used 
in the ME population. Nevertheless, the SE and ME projections agree 
in detail. Similar agreement is seen for all other ME mixing classes. This 
level of agreement is likewise stable throughout the data-taking period, with 
negligible time dependence.

The jet distribution due to uncorrelated background is determined by carrying 
out the same jet reconstruction procedure on the ME events as is used for the 
real data. However, no high-\pT\ trigger hadron is required for the ME analysis; 
rather, the trigger axis for ME events is chosen by selecting a random track, 
resulting in a similar azimuthal distribution to that in analysis of the SE population. 

No jet candidates are excluded in the calculation of $\rho$ for ME 
events, in contrast to the calculation of $\rho$ for SE events  
(Sect.~\ref{sect:JetReco}). This choice is motivated by fact that all 
multi-hadron 
correlations, including those due to jets, are suppressed in ME events. 
Figure~\ref{fig:rho} shows the distribution of $\rho$ in one event-mixing class, 
for both SE and ME events. The SE and ME $\rho$ distributions are in good 
agreement for both peripheral and central collisions, thereby 
validating the jet exclusion choices made for the various event 
populations. The fit of a Gaussian function to the central peak of the 
SE distribution gives $\sigma=3.7$ \gev.
Looking in detail at the tails of the distribution, the SE/ME ratio for central \AuAu\ collisions 
(lower panel, blue points) shows an excess in SE relative to ME of about 50\% in 
the left tail (smaller $\rho$), where the rate is a factor $\sim10^{3}$ smaller than at the peak of the distribution. This small relative change suggests that the ME $\rho$ 
distribution is slightly narrower than the SE $\rho$ distribution. In order to quantify this 
effect, the ME distribution is shifted towards smaller $\rho$ by 60 MeV/($c$ sr) (red 
points), where a similar increase in SE/ME ratio is now seen instead in the 
right tail at larger $\rho$. The width in the far tails of the ME $\rho$ 
distribution is therefore smaller than 
the SE width by less than 60 MeV/($c$ sr). We discuss this effect below, in the context of Fig.~\ref{fig:RawLowTrigpT}.


\begin{figure}[htbp]
\includegraphics[width=0.49\textwidth]{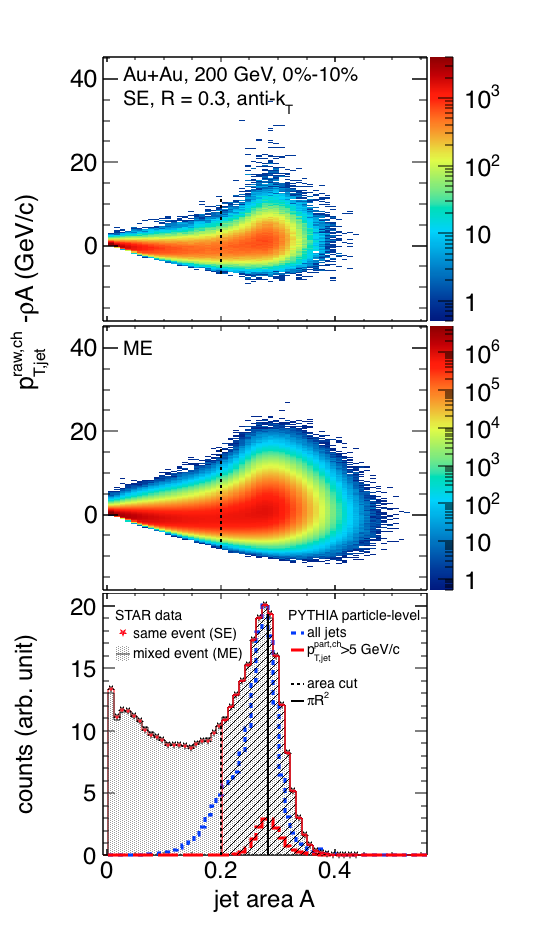}
\caption{(Color online) Distribution of SE and ME jet populations (\rr\ = 0.3) for 
one event-mixing class in central \AuAu\ collisions, as a function of \pTreco\ 
and \Ajet. Top panel: real events (SE); middle panel: mixed events (ME); bottom panel: 
projection of SE and ME distributions onto \Ajet. The lower panel also shows the 
recoil jet area distribution for \pp\ collisions at \sqrts\ = 200 GeV from 
PYTHIA-simulated events at the particle level with $\pTtrig>9$ \gev, for all 
recoil jets and for recoil $\pTjetpart>5$ \gev. The hatched region to the right 
of the dashed line is the accepted region for the \Ajet\ cut.}
\label{fig:SE_ME_pT_Area}
\end{figure}  

Figure \ref{fig:SE_ME_pT_Area} shows the distribution of jet candidates as a 
function of \pTreco\ and \Ajet\ for one 
event-mixing class,  for SE events (top panel); ME events 
(middle panel); and the projection of both
distributions onto \Ajet\ (bottom panel). The SE and ME distributions in Fig. 
\ref{fig:SE_ME_pT_Area} agree in detail, with a peak in \Ajet\ centered near 
$\pi\cdot\rr^2$. The bottom panel also shows 
the \Ajet\ distribution from a PYTHIA particle-level 
simulation of \pp\ collisions at \sqrts\ = 200 GeV, for all reconstructed jets and 
for jets with $\pTjetpart>5$ \gev\ recoiling from a trigger 
hadron with $\pT>9$ \gev. The area distribution for $\pTjetpart>5$ \gev\ 
coincides with the main peak, without the tail to smaller area. 

The detailed agreement of the \Ajet\ distributions for SE and ME events 
seen in Fig. \ref{fig:SE_ME_pT_Area}, lower panel, shows that the \Ajet\ 
distribution for high-multiplicity events is driven predominantly by geometric 
factors, specifically the experimental acceptance and \rr, together with 
response of the \antikT\ algorithm to the high-multiplicity environment. The 
correlated structure of true jets plays a less significant role. We note in 
addition that \Ajet\ for true jets reconstructed with the \antikT\ algorithm is 
insensitive to the presence of uncorrelated 
background~\cite{FastJetAntikt}. Reduction in the uncorrelated background jet 
yield can therefore be carried out by a cut on \Ajet, as indicated by the vertical 
dashed line. Based on the PYTHIA particle-level simulation, this cut suppresses 
about 15\% of the yield of correlated jets for $\pTjetpart<5$ \gev, with 
negligible suppression for $\pTjetpart>5$ \gev.

\section{Raw distributions}
\label{sect:pTcorrDistr}

\begin{figure*}[htbp] 
\includegraphics[width=0.45\textwidth]{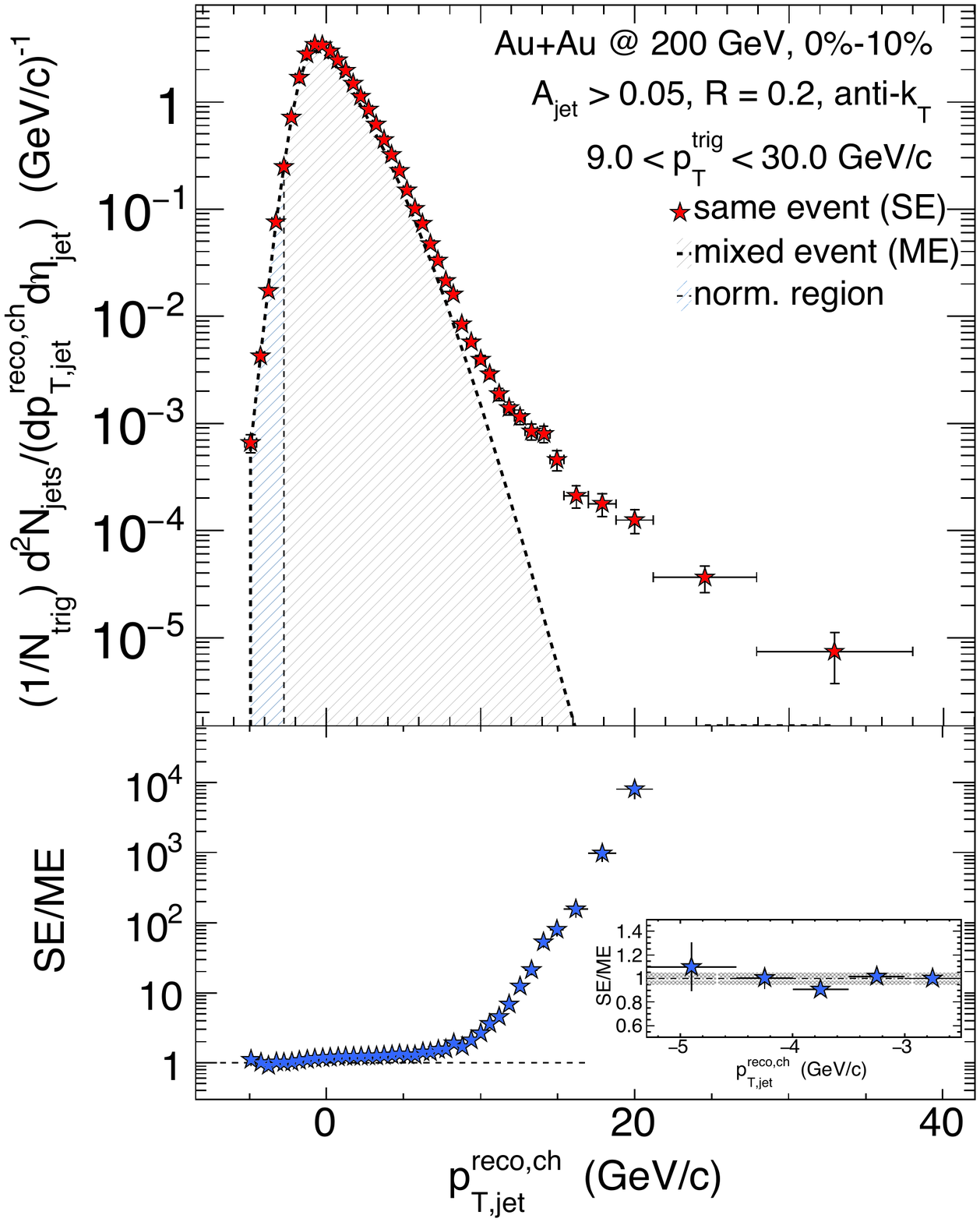} 
\includegraphics[width=0.45\textwidth]{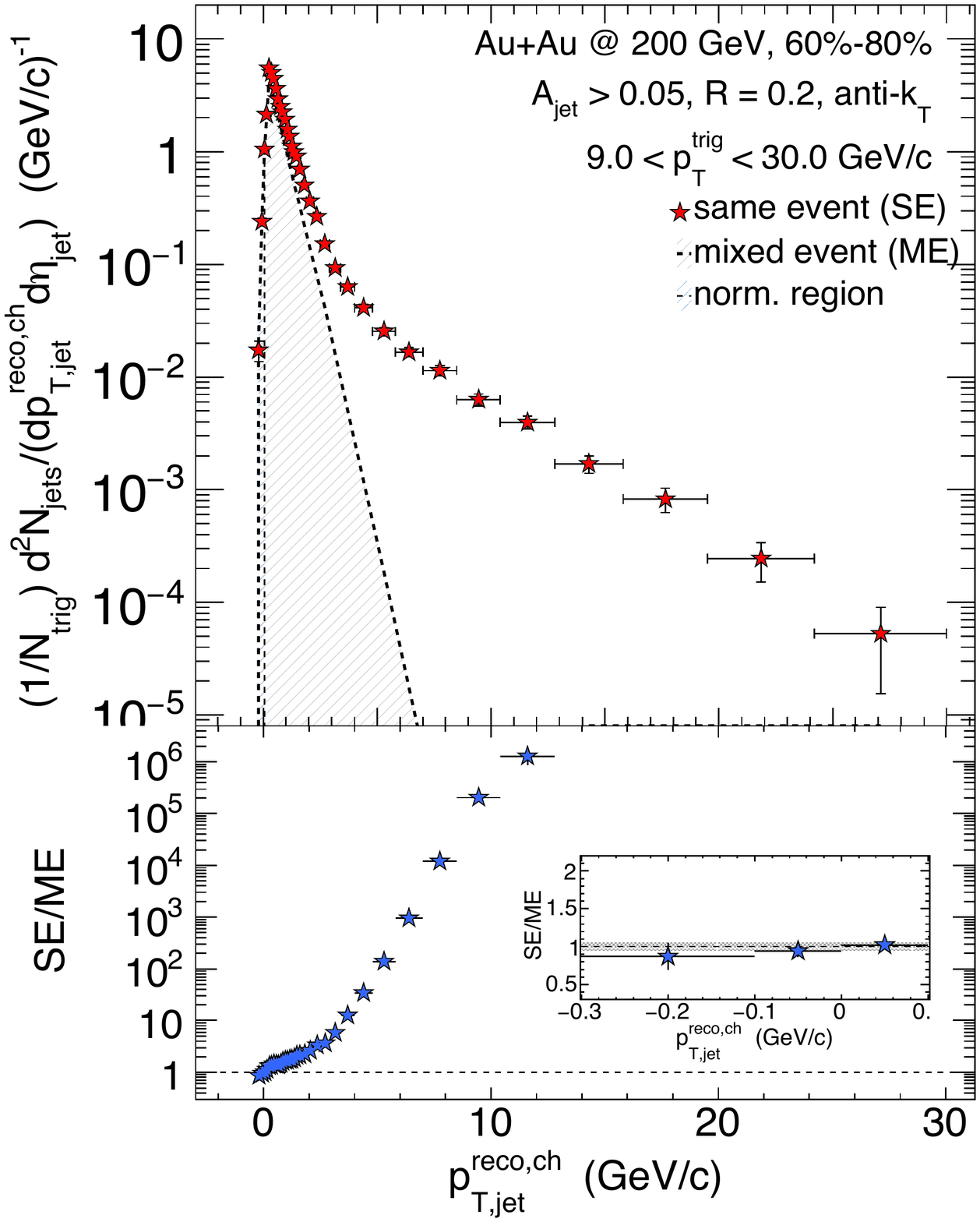}
\includegraphics[width=0.45\textwidth]{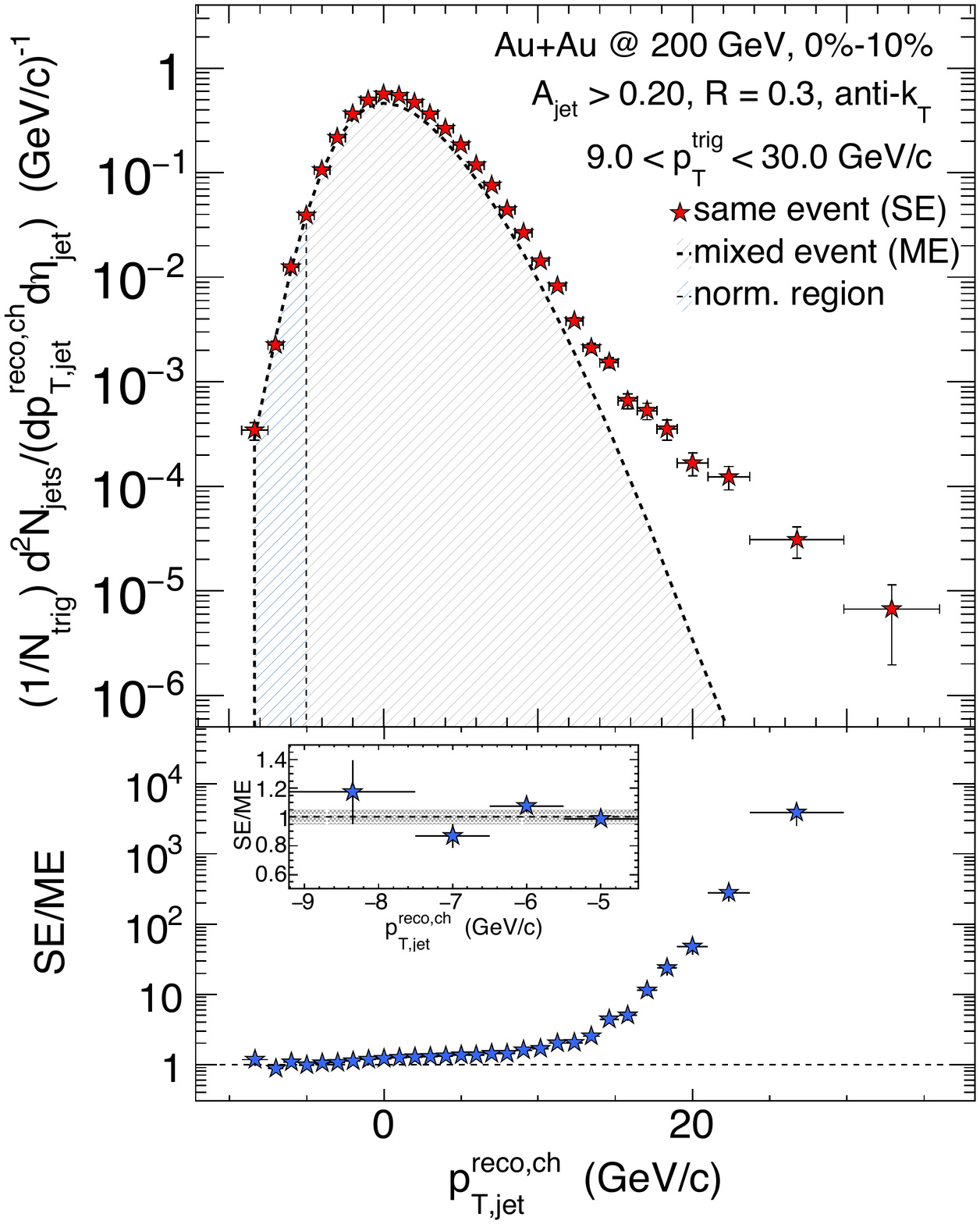}
\includegraphics[width=0.45\textwidth]{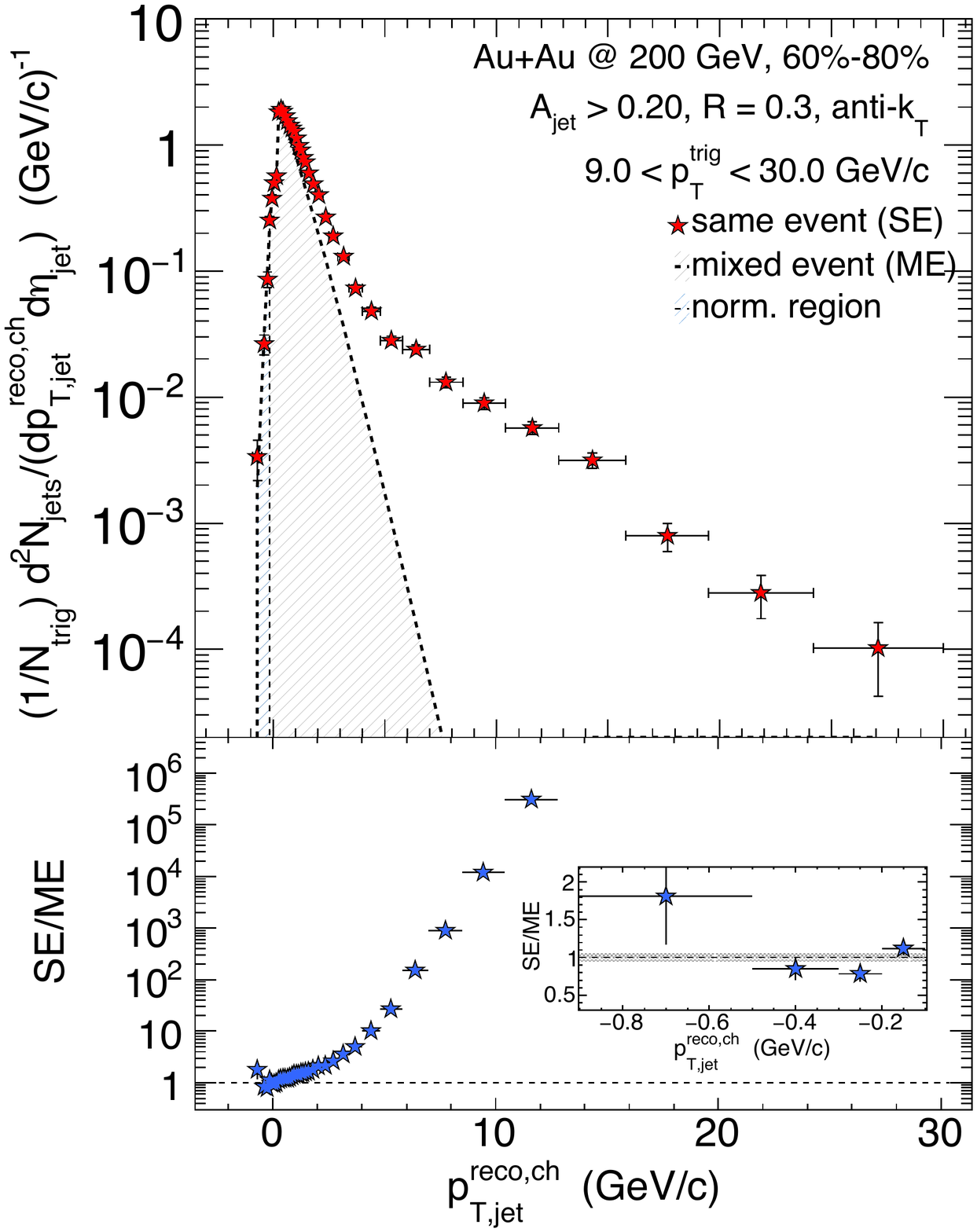}
\caption{(Color online) Distributions of \pTreco\ for \AuAu\ collisions at
  \sqrtsNN\ = 200 GeV. Left panels: central; right panels: peripheral. Upper 
panels: \rr\ = 0.2; lower panels: \rr\ = 0.3. The upper sub-panel shows the 
distributions for SE (red points)
   and ME (shaded region), with the blue shaded region indicating the
   range used for ME normalization. Error bars on SE distributions are
   statistical. The lower sub-panel shows the ratio of
   the SE and normalized ME distributions, while the insert shows the ratio in
   the normalization region. See text for details.}
\label{fig:RawData_2_3}
\end{figure*}

\begin{figure*}[htbp] 
\includegraphics[width=0.45\textwidth]{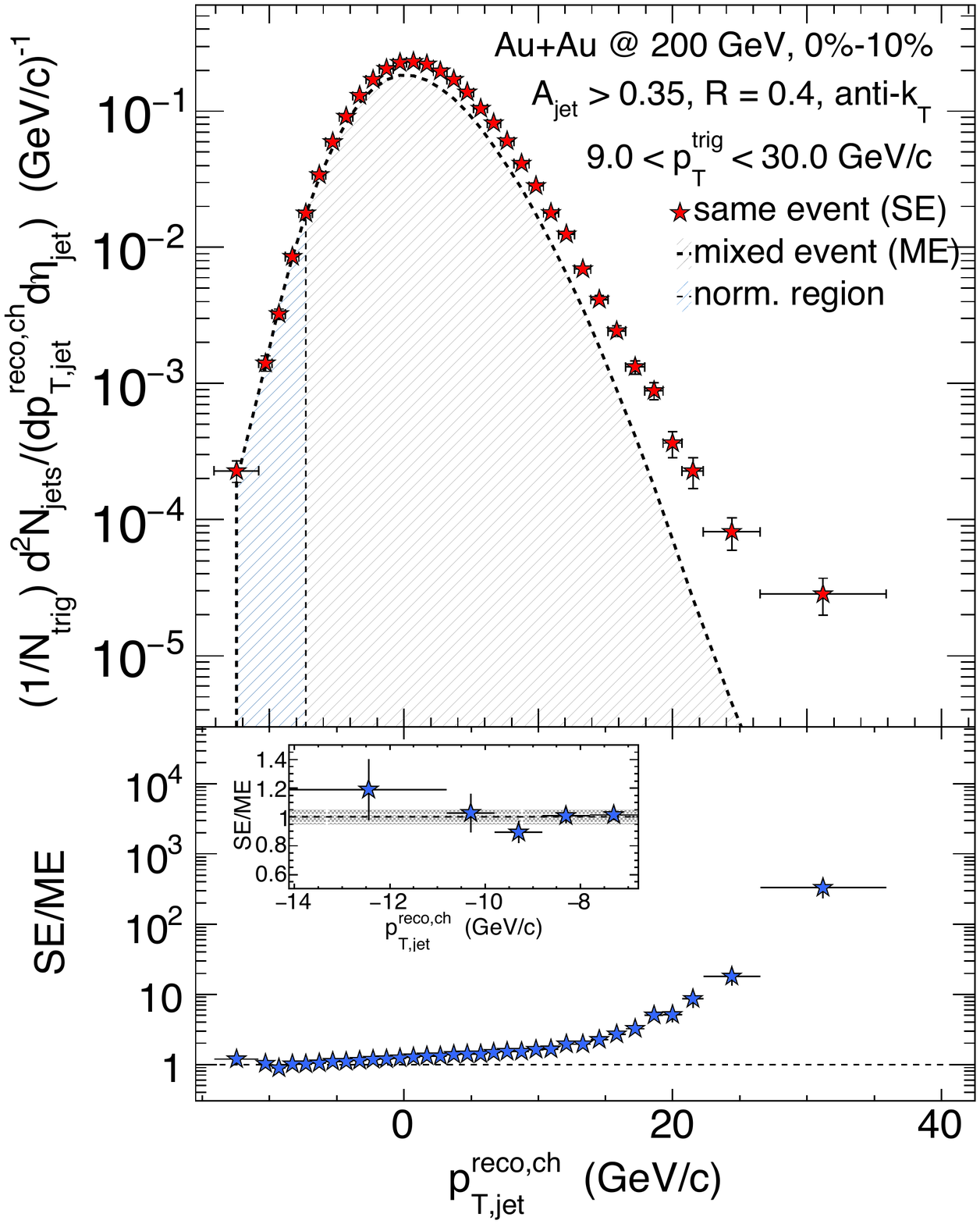}
\includegraphics[width=0.45\textwidth]{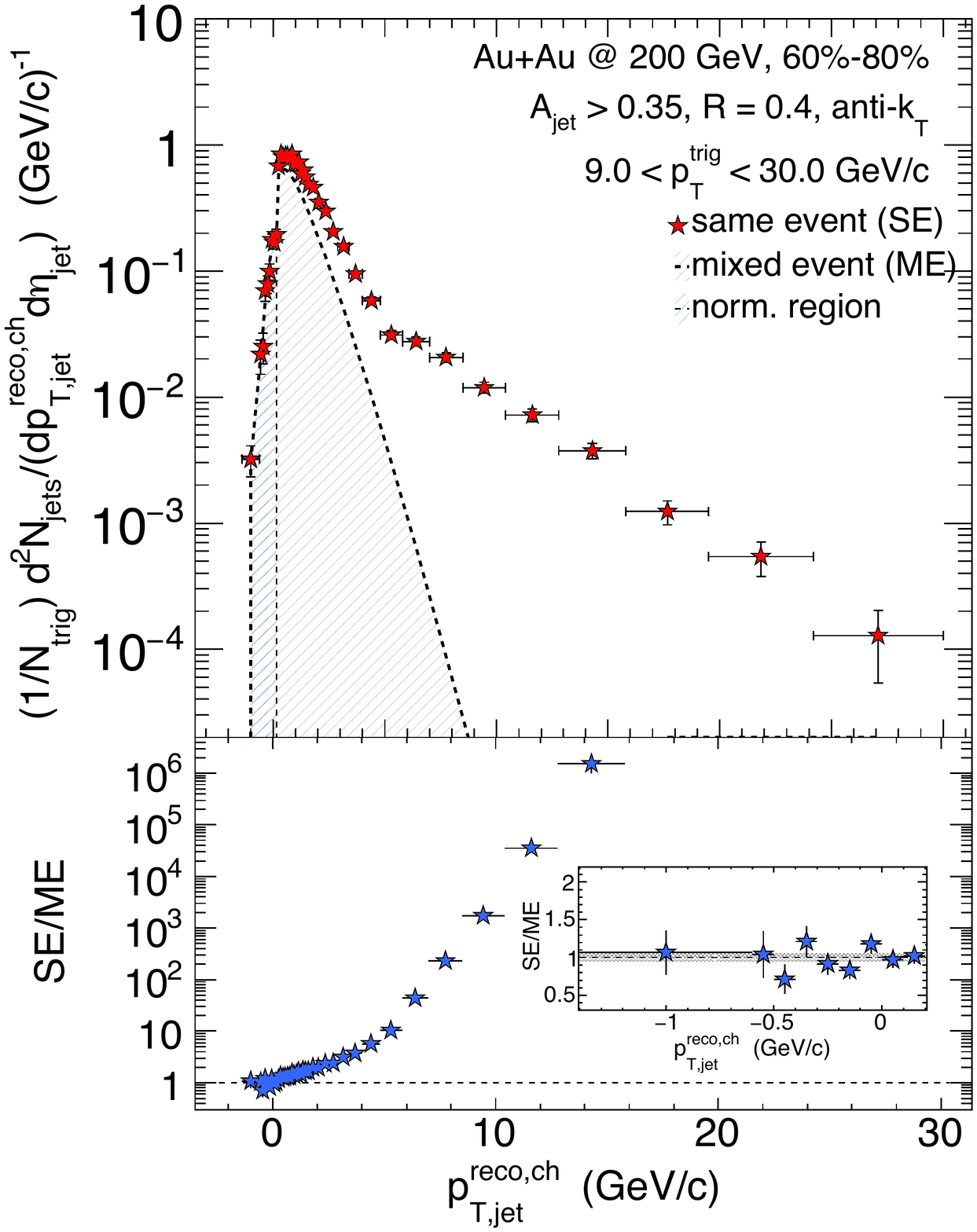}
\includegraphics[width=0.45\textwidth]{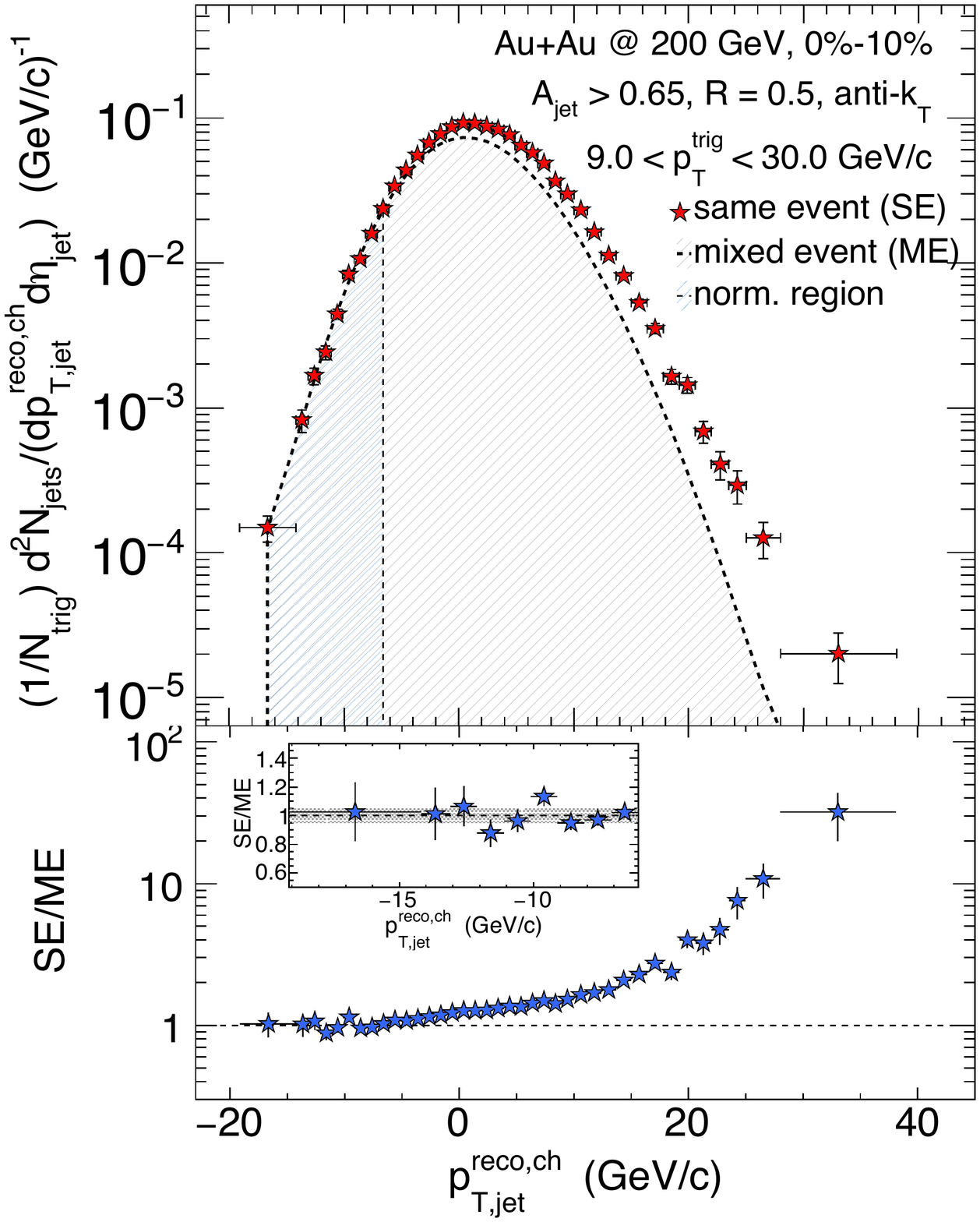}
\includegraphics[width=0.45\textwidth]{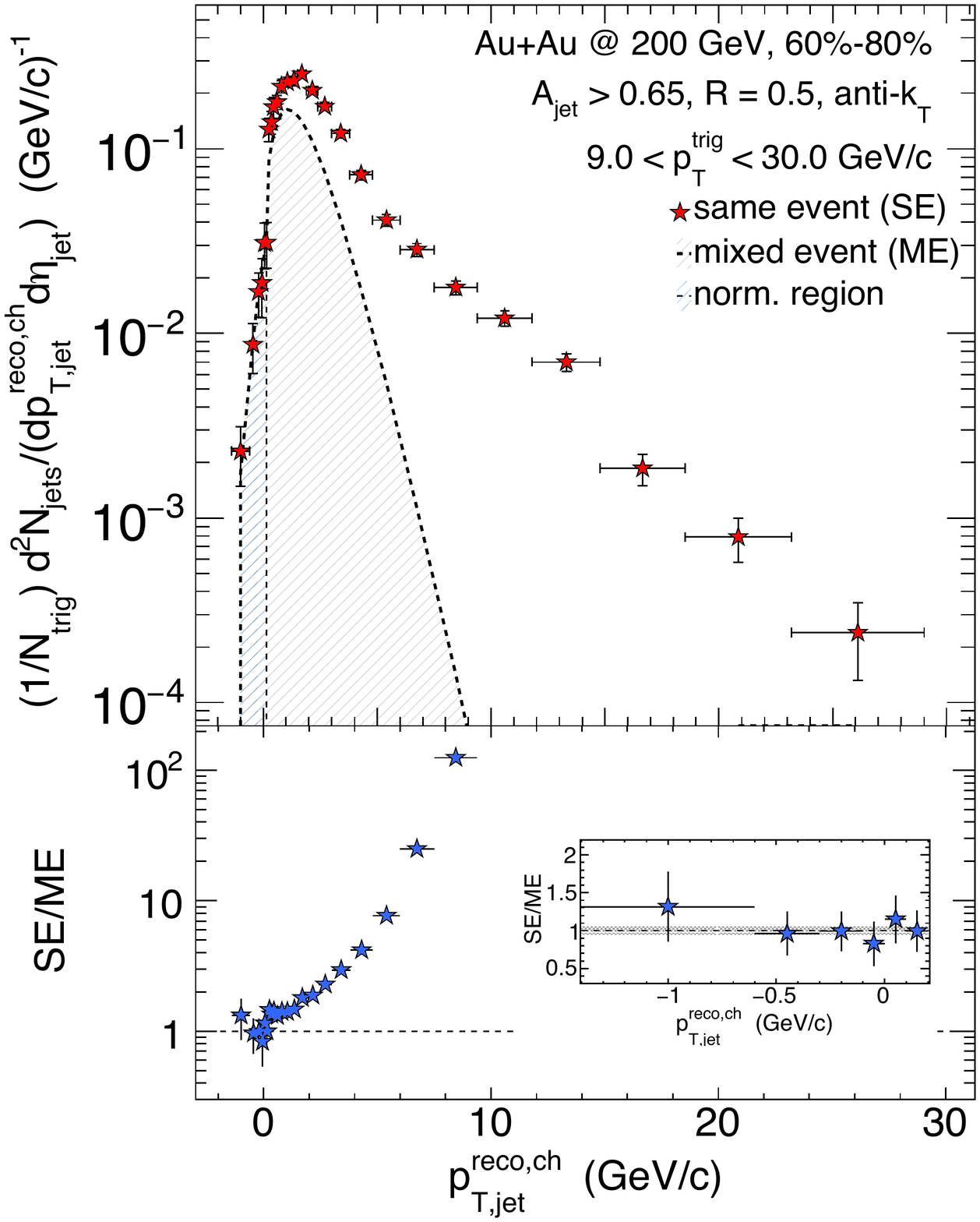}
\caption{(Color online) Same as Fig. \ref{fig:RawData_2_3}, but for \rr\ = 0.4 and 0.5.}
\label{fig:RawData_4_5}
\end{figure*}

Figures \ref{fig:RawData_2_3} and \ref{fig:RawData_4_5} show distributions of 
the uncorrected recoil jet yield in \AuAu\ collisions projected onto \pTreco,
for \rr\ between 0.2 and 0.5. The upper sub-panels show the distributions separately for data (red points) and mixed-event background (shaded histogram). The lower sub-panels 
are discussed below.

The number of jet candidates found in an event is necessarily bounded, 
due to the area subtended by each jet candidate and by the total
experimental acceptance. Table \ref{Tab:Integral_fME} shows the integral over 
\pTreco\ for the SE and ME distributions shown in Figs. \ref{fig:RawData_2_3} 
and \ref{fig:RawData_4_5}. The integral is the average number of observed recoil 
jet 
candidates per trigger hadron, including both 
correlated and uncorrelated. The integrals 
decrease with increasing \rr, as expected since jets with larger \rr\ subtend 
larger area. The integral values are larger for central than for
peripheral \AuAu\ collisions at the same \rr, corresponding to larger jet 
density for central collisions, which is expected since peripheral collisions 
are more 
sparsely populated.
 
The integrals of the SE and ME distributions in central \AuAu\
collisions agree to better than 1\% for each value of \rr. Invariance of such 
integrals for event classes with differing jet-like correlations
has also been observed for high-multiplicity events in model 
studies~\cite{deBarros:2012ws}, and in the analysis of \PbPb\ 
collisions at 2.76 TeV~\cite{Adam:2015doa}. At high multiplicity this integral, 
like the \Ajet\ distribution, is evidently driven predominantly by 
geometric factors, specifically the experimental acceptance, 
characteristic jet size \rr, and the robustness of the shape of \antikT\ jets in 
the presence of background~\cite{FastJetAntikt}, but not by the presence of 
multi-hadron correlations, whose contribution is different in different event 
classes and is absent entirely in the ME population. 

\begin{table*}
\caption{Integral of SE and ME distributions in Figs. \ref{fig:RawData_2_3} and \ref{fig:RawData_4_5}, together with the ME normalization factor \fME. The uncertainty of \fME\ is systematic. 
\label{Tab:Integral_fME}}
\begin{tabular}{ |c|c|c|c|c| }
\hline
\AuAu\ centrality & \rr & \multicolumn{2}{|c|}{Integral} & \fME \\ \hline 
\multicolumn{2}{|c|}{} & SE & ME & \\ \hline
\multirow{4}{*}{peripheral (60\%-80\%)} & 0.2 & 0.446 & 0.397 & $0.72 \pm 0.05$ \\ 
 & 0.3 & 0.269 & 0.252 & $0.67 \pm 0.07$ \\ 
 & 0.4 & 0.184 & 0.175 & $0.61 \pm 0.02$\\
 & 0.5 & 0.094 & 0.089 & $0.49 \pm 0.07$ \\ \hline
\multirow{4}{*}{central (0\%-10\%)} & 0.2 & 1.26 & 1.26 & $0.86 \pm 0.01$ \\ 
 & 0.3 & 0.392 & 0.391 & $0.85\pm 0.03$\\ \
 & 0.4 & 0.228 & 0.227 & $0.80 \pm 0.03$ \\ 
 & 0.5 & 0.119 & 0.119 & $0.80 \pm 0.02$ \\ \hline
\end{tabular}
\end{table*}

In each panel of Figs. 
\ref{fig:RawData_2_3} and \ref{fig:RawData_4_5}, the shape of the ME distribution is very similar to that of the SE distribution in the
region $\pTreco<0$, where the yield is expected to arise predominantly from 
uncorrelated background. The shapes differ significantly at large positive \pTreco, 
where an appreciable contribution from correlated true jets is expected. 
Additionally, the absolutely normalized ME distributions are observed to have larger 
yield than the SE distributions in the region $\pTreco<0$, consistent with the smaller yield in ME 
at large positive \pTreco\ and agreement of the SE and ME integrals within 
better than 1\% for central collisions and within about 10\% for peripheral collisions. 
These features have also been observed for high-multiplicity events in model studies~\cite{deBarros:2012ws} and in analysis of LHC data for \PbPb\ collisions~\cite{Adam:2015doa}. 

In order to utilize the ME distribution to 
determine the contribution of uncorrelated background in the SE 
distribution, the absolutely normalized  ME distribution is therefore scaled 
downwards by a scalar factor \fME, determined by a fit in the blue 
shaded regions in the upper sub-panels. The range in \pTcorr\ for determining 
the central value of \fME\ is chosen as the left-most region of the 
spectrum in 
which the SE/ME yield ratio is uniform within 10\%. The lower sub-panels show 
the SE/ME yield ratio after normalization by \fME, while the inserts show the 
ratio in the fit region, also after normalization. Tab.~\ref{Tab:Integral_fME} 
gives the values of \fME. The systematic uncertainty of \fME\ in
Tab.~\ref{Tab:Integral_fME} is determined by varying the normalization region. 

For jets in central collisions and \rr\ = 0.5, the ratio of normalized ME and
SE distributions is within 10\% of unity in the region $-20<\pTreco<-5$ \gev, 
over which the distributions themselves vary by two orders of magnitude 
(Fig.~\ref{fig:RawData_4_5}, lower left). 
Similarly good agreement of the shapes of the SE and ME distributions over a significant range in 
\pTreco\ is observed for the other values \rr. This good 
agreement indicates that the normalized ME distributions represent the uncorrelated background accurately, and can therefore be used over the
full range of \pTreco\ for correction of uncorrelated background in the SE 
distribution.

For peripheral collisions, the SE distributions fall more rapidly in the region 
$\pTreco<0$ and the ME distributions are overall much narrower than for central 
collisions, as expected since the uncorrelated background level
is much lower. The width of the \fME\ normalization region is 
correspondingly much narrower than for central collisions, with a weaker 
constraint imposed on \fME. However, the precision required for \fME\ is much 
reduced for perpheral collisions, precisely because of the much smaller 
uncorrelated background contribution.

\begin{figure}[htbp] 
 \includegraphics[width=0.5\textwidth]{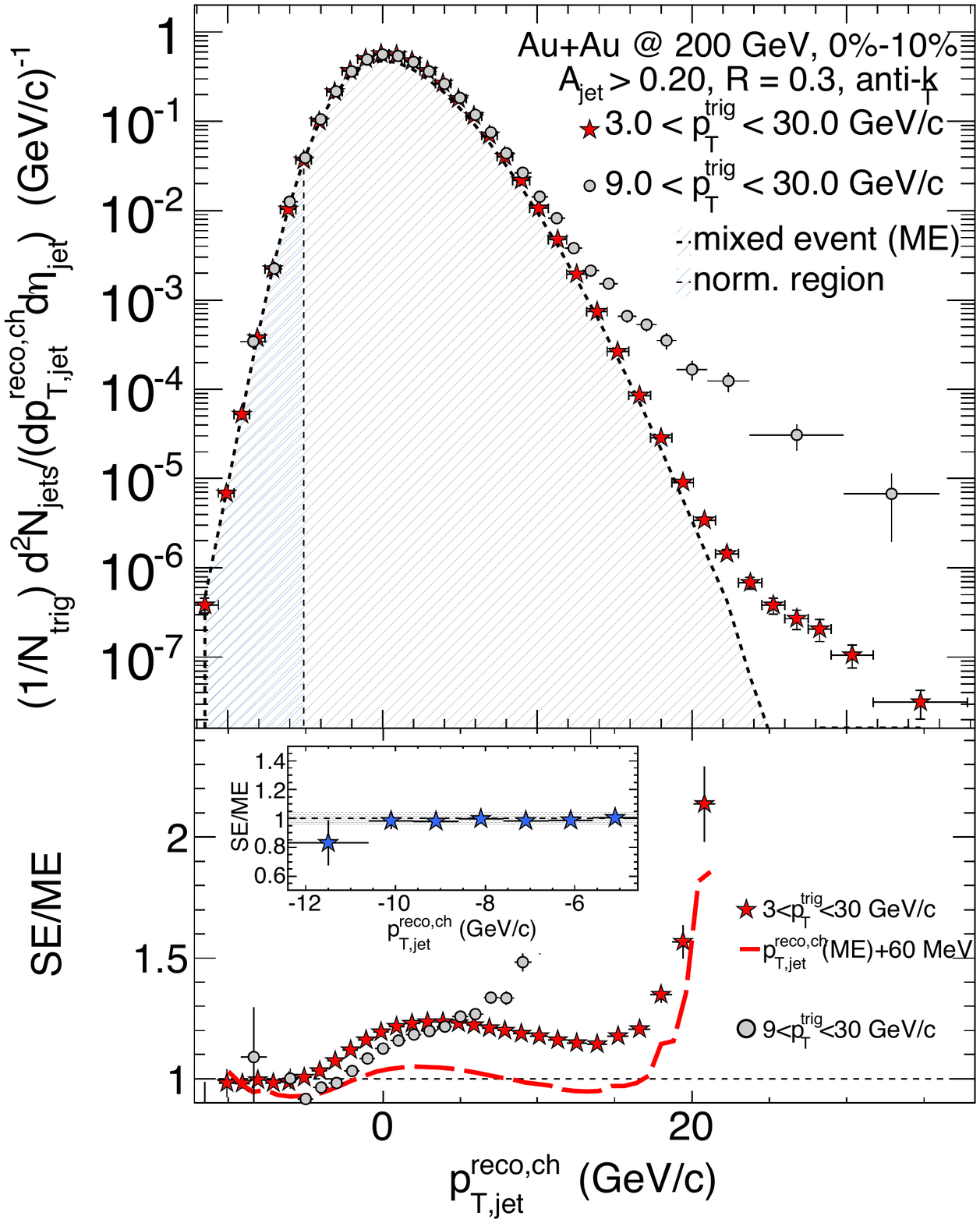}
\caption{(Color online) Same as Fig.~\ref{fig:RawData_2_3}, for central \AuAu\ collisions and \rr\ = 0.3. Upper panel: SE distribution is shown for two different ranges of \pTtrig: $9<\pTtrig<30$ \gev\ (grey points), which is used in the primary analysis, and $3<\pTtrig<30$ \gev\ (red points). The ME distribution is the same as Fig.~\ref{fig:RawData_2_3}, lower left. Lower panel: ratio SE/ME for $9<\pTtrig<30$ \gev\ (grey points), and for $3<\pTtrig<30$ \gev\ with $\rho$ as defined in the primary analysis (red points) and $\rho$ shifted by 60 MeV/$c$ (dashed line). See text for details.
The insert shows the ratio for the $3<\pTtrig<30$ \gev\ SE distribution, in the region of \fME\ normalization.}
\label{fig:RawLowTrigpT}
\end{figure}

Figure \ref{fig:RawLowTrigpT} shows the uncorrected recoil jet distribution for 
central \AuAu\ collisions and \rr\ = 0.3, for two different ranges in 
\pTtrig, $9<\pTtrig<30$ \gev\ and $3<\pTtrig<30$ \gev. The SE distribution for 
the higher-\pTtrig\ interval and the ME distribution are 
the same as in Fig.~\ref{fig:RawData_2_3}, lower 
left. Lower values of \pTtrig\ are expected to select processes with 
smaller $Q^2$ on average, and indeed are observed to generate a lower 
rate of correlated recoil jets in both \pp\ and \PbPb\ collisions at LHC 
energies~\cite{Adam:2015doa}. By measuring the SE distribution for different ranges of \pTtrig, as in Fig. \ref{fig:RawLowTrigpT}, we therefore vary the rate of correlated jet yield in the recoil jet candidate population, while keeping the distribution of uncorrelated jet 
candidates unchanged.

In Fig. \ref{fig:RawLowTrigpT}, upper panel, the ME distribution and the SE distribution with $3<\pTtrig<30$ \gev\ are very similar in the range 
$-10<\pTreco<15$ \gev, over which the distributions themselves vary 
by more than five orders of magnitude. It is only in the region $\pTreco>20$ \gev\ that this SE distribution exceeds the ME distribution by a significant factor, indicative of 
a correlated recoil jet component with relative yield compared to all jet 
candidates of less than $10^{-6}$.

The SE distributions with different lower bound for \pTtrig\ are likewise 
similar in the region $-10<\pTreco<10$ \gev, but differ for larger \pTreco, as 
expected. 
The good agreement of the ME distribution and both SE distributions for
negative and small positive \pTcorr\ confirms that the yield in this 
region is dominated strongly by uncorrelated background. Their 
ordering in magnitude at larger \pTcorr\ also shows that the SE distribution approaches the ME distribution as the lower bound of \pTtrig\ is reduced towards zero.

Figure \ref{fig:RawLowTrigpT}, lower panel, shows ratios of the SE and 
ME distributions for the two different trigger hadron \pT\ ranges. The 
distributions utilize the primary 
analysis approach described in Sect.~\ref{sect:JetReco}, including the choices 
specified there for determining the background density $\rho$ 
(Eq.~\ref{eq:rho}). The ratios 
exhibit a variation of 20\%-30\% in the region $\pTreco<5$ \gev. While the 
distributions themselves vary by several orders of magnitude over this range and 
this variation is small in relative terms, it is nevertheless observable.

Variation in the ratio is related to the ambiguity in defining $\rho$ for 
the SE and ME populations. In Sect.~\ref{sect:ME} we noted that the tails of the 
$\rho$ distribution are slightly narrower for the ME than the SE population, by 
less than 60 MeV/($c$ sr). To assess the influence of this difference, the red 
dashed line in Fig.~\ref{fig:RawLowTrigpT}, lower panel, shows the ratio 
of the SE and ME recoil jet distributions for $3<\pTtrig<30$ \gev, 
but with the value of $\rho$ for each event shifted systematically by 60 MeV/($c$ 
sr)  as in Fig.~\ref{fig:rho}. In this 
case, variation in the SE/ME recoil jet yield ratio is 
reduced to less than 5\% for $\pTreco<15$ \gev. The ratio increases 
rapidly at larger \pTreco, due to significant correlated yield in the SE 
distribution.

The influence of the slightly narrower $\rho$ distribution in the ME population 
on  correction of the recoil jet spectra was assessed by carrying out the full 
analysis (described in the following sections) for representative cases, with 
and without a 60 
MeV/($c$ sr) shift in $\rho$. The resulting change in the fully corrected recoil 
jet yield is significantly smaller than its systematic 
uncertainties due to other sources. An effective shift in $\rho$ can 
also arise from azimuthal anisotropy (\vtwo) of the trigger, which is considered 
below. We therefore do not consider the effect of the narrower $\rho$ 
distribution in the ME population further in the analysis.

The ALICE Collaboration has measured semi-inclusive h+jet distributions  
for \PbPb\ collisions at \sqrtsNN\ = 2.76 TeV with a correction 
procedure for uncorrelated background that utilizes the difference between 
normalized recoil jet distributions for exclusive ranges of 
\pTtrig~\cite{deBarros:2012ws,Adam:2015doa}. 
Compared to the current analysis, the ALICE analysis differs in its use of an SE 
jet distribution recoiling from lower \pTtrig\  to measure 
uncorrelated background, rather than the ME distribution. This approach results 
in a different observable, 
\Drecoil~\cite{Adam:2015doa}, in which the small correlated component of the 
lower threshold SE distribution is also removed by the subtraction. However, the
low-threshold SE and ME distributions in Fig. \ref{fig:RawLowTrigpT} are similar 
in the current analysis, so that the difference between \Drecoil\ calculated 
with this choice of kinematics for the low-threshold SE and \Yjet\ is expected 
to be negligible. Direct comparison 
of these related correction procedures will be explored in future analysis, with 
larger data sets.

We note in addition that these two approaches differ in their treatment of 
multiple partonic  interactions (MPI). Background due to MPI arises when a 
trigger hadron and a jet in the recoil acceptance are generated by two different, 
incoherent high-$Q^2$ processes in the same collision. This background  
is expected to be 
independent of \dphinovar, and to be larger in heavy ion than in \pp\ 
collisions. Since \Drecoil\ is the difference of two SE distributions, which have the same  MPI background by definition~\cite{Adam:2015doa}, the MPI background is removed from \Drecoil\ by construction. In contrast, in the current analysis the event mixing procedure destroys all jet-like correlations, and the ME distribution does not contain an MPI component. However, comparison of the $3<\pTtrig<30$ \gev\ SE and the ME distribution in Fig. \ref{fig:RawLowTrigpT} shows that their difference, which contains the MPI background component, is negligible compared to the correlated yield for the SE $9<\pTtrig<30$ \gev\ distribution. Background due to MPI is therefore negligible in this measurement, and no correction for it is warranted in the analysis.

\begin{figure*}[htbp] 
\includegraphics[width=0.8\textwidth]{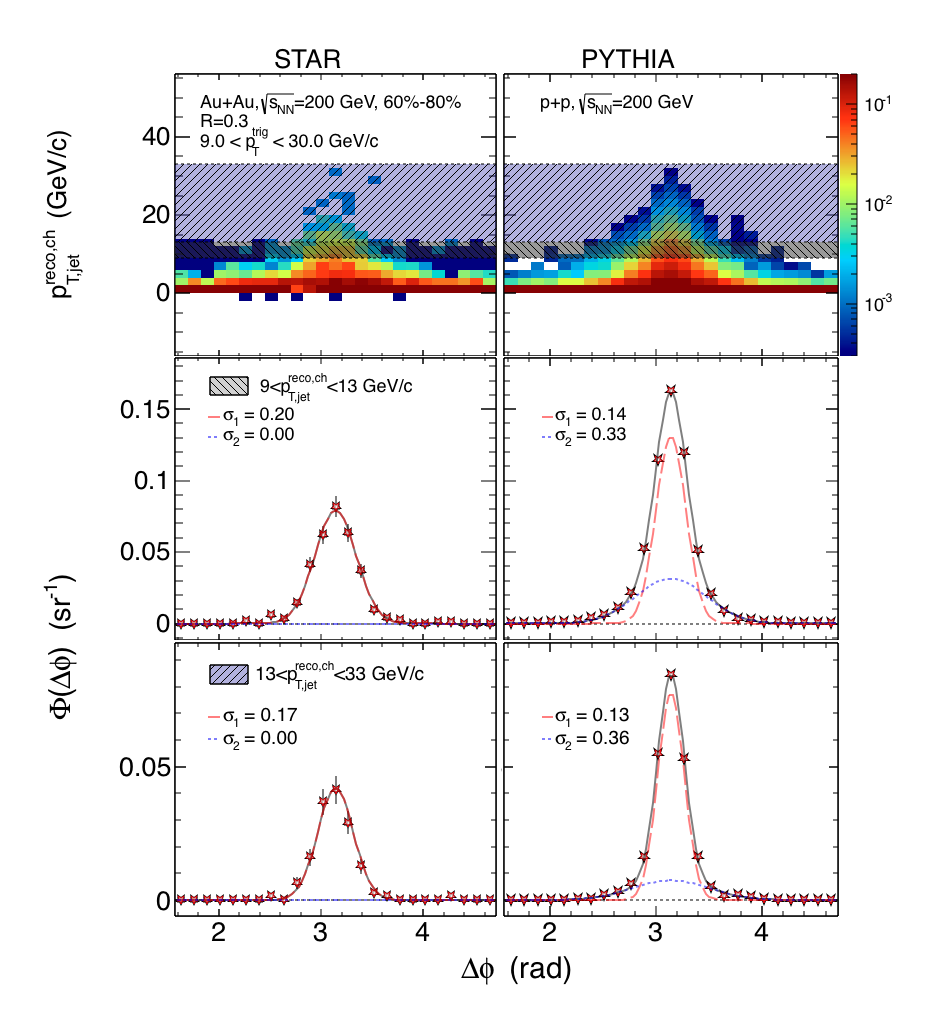} 
\caption{(Color online) Upper panels: recoil jet distributions after mixed event subtraction 
for peripheral 
\AuAu\ STAR events (left)
and for \pp\ collisions generated by PYTHIA detector-level simulations (right). 
Middle and lower panels: projections onto
\dphi\ for two different ranges in \pTreco, indicated by 
the blue and grey shaded areas in the upper plot. The projected distributions are 
fitted with a function that is the sum of two Gaussian distributions, with 
fit widths $\sigma_1$ and $\sigma_2$. The values of $\sigma_1$ and $\sigma_2$ are highly correlated, with negligible statistical error.}
\label{fig:Raw_Delta_phi_per}
\end{figure*}

Figure \ref{fig:Raw_Delta_phi_per}, upper panels, show distributions of the 
background-subtracted recoil jet yield  for \rr\ = 0.3 in peripheral \AuAu\ 
collisions at \sqrtsNN\ = 200 GeV 
from STAR data, and in \pp\ events at \sqrts\ = 200 GeV simulated with 
PYTHIA at the detector level. The middle and lower panels show the projection 
onto \dphinovar\ for selected intervals in \pTreco. Correction of \Phijet\ for 
uncorrelated background by subtraction of the ME distribution is discussed in 
Sect.~\ref{sect:Acoplanarity}. No correction is carried out for the 
effects of underlying event in the PYTHIA-generated \pp\ collision events.

\begin{figure*}[htbp] 
\includegraphics[width=0.8\textwidth]{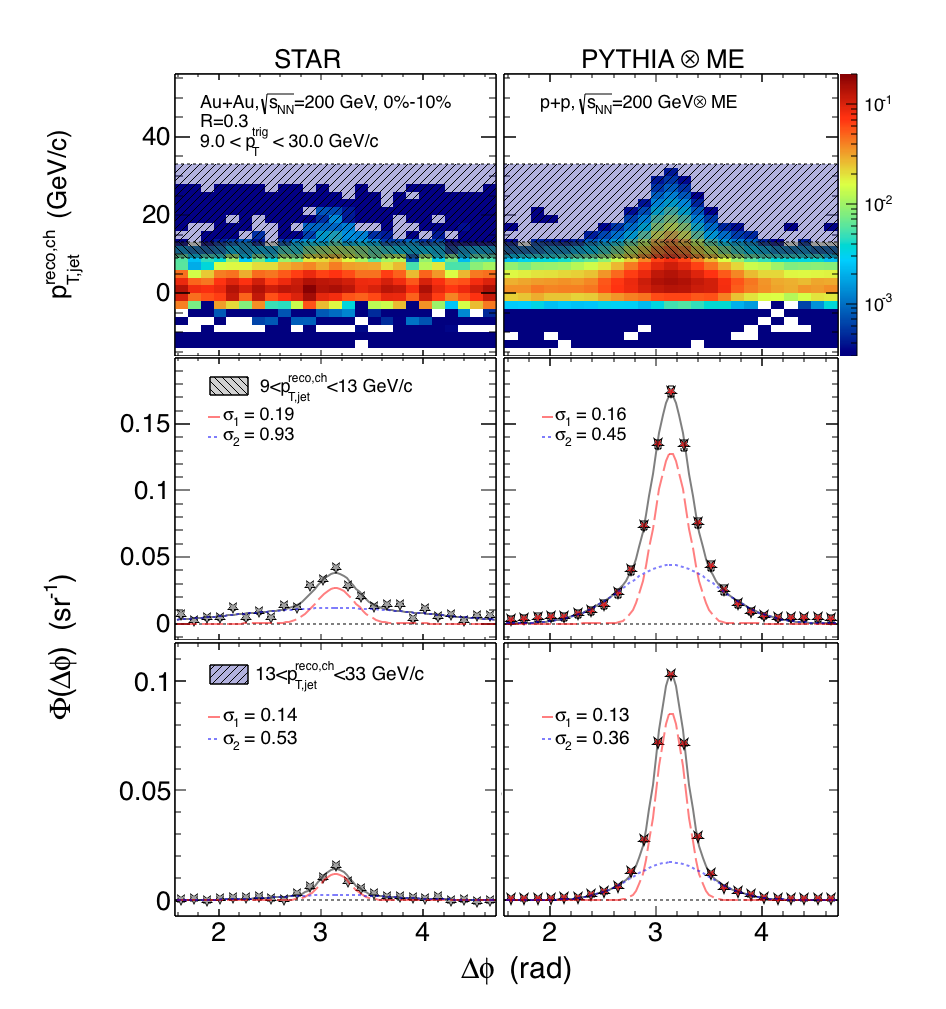} 
\caption{(Color online) Same as Fig. \ref{fig:Raw_Delta_phi_per}, but for central \AuAu\ STAR 
data (left) and detector-level PYTHIA simulations of \pp\ collisions at 
\sqrts\ = 200 GeV embedded into mixed events from central \AuAu\ STAR data at the track level (right).}
\label{fig:Raw_Delta_phi_cen}
\end{figure*}

Figure \ref{fig:Raw_Delta_phi_cen} shows the same distributions as in 
Fig.~\ref{fig:Raw_Delta_phi_per}, but for central \AuAu\ STAR data with 
background subtraction, and for PYTHIA-generated events at the
detector level for \sqrts\ = 200 GeV \pp\ collisions  embedded into central \AuAu\ 
STAR data at the track level. 

The middle and lower panels of Figs.~\ref{fig:Raw_Delta_phi_per} and 
\ref{fig:Raw_Delta_phi_cen} show fits to the \Phijet\ distributions with a 
function that is the sum of two Gaussian distributions, both centered on 
$\dphinovar=\pi$, with fitted widths $\sigma_1$ and $\sigma_2$. The values of 
$\sigma_1$ and $\sigma_2$ are correlated. The fit provides a 
qualitative characterization of the azimuthal distributions. The widths of the
central peaks are seen to be similar in the 
peripheral data and PYTHIA distributions, and in the central data and PYTHIA embedded in 
central events. The recoil yield is 
suppressed for both peripheral and central collisions relative to the yield 
predicted by the PYTHIA calculation, with greater suppression for central 
collisions. Quantitative analyses of these features is presented in Sect. 
\ref{sect:Results}.

%% file: Corrections.tex
\section{Corrections}
\label{sect:Corrections}

\begin{figure}[htbp] 
\includegraphics[width=0.5\textwidth]{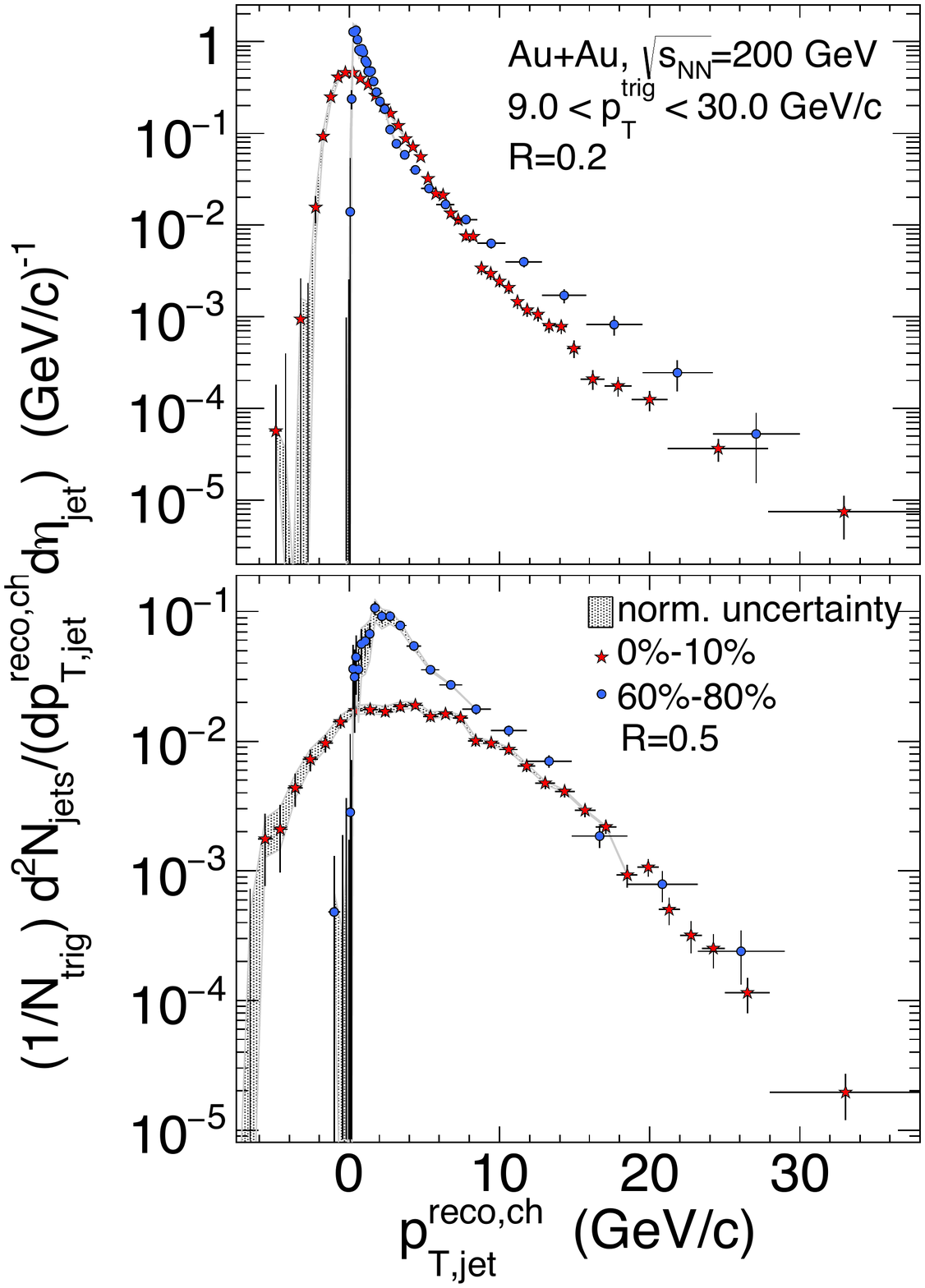} 
\caption{(Color online) Raw correlated jet yield distributions for \rr\ = 0.2 (upper) and \rr\ = 0.5 (lower) in central and peripheral \AuAu\ collisions at \sqrtsNN\ = 200 GeV. The uncorrelated background has been removed by subtraction of the scaled ME distribution from the SE distribution, but no other corrections have been applied. The gray shaded band shows the mixed event normalization uncertainty. 
\label{fig:Raw_sub_spectra}}
\end{figure}

Figure~\ref{fig:Raw_sub_spectra} shows the raw correlated recoil jet yield 
distributions for \rr\ = 0.2 and \rr\ = 0.5 in central and peripheral \AuAu\ 
collisions, determined by subtracting the \fME-normalized ME distribution from the 
SE distribution. The SE-ME distributions for \rr\ = 0.3 and 
\rr\ = 0.4 (not shown) are similar, with features that interpolate 
between the distributions in the figure. 

In the region where the SE and ME distributions have similar magnitude, their 
difference can be negative due to statistical fluctuations. However, the 
vertical axis of Fig.~\ref{fig:Raw_sub_spectra} is logarithmic, and negative 
entries are not displayed. Negative values only occur in the region $\pTreco<0$ 
\gev\ for peripheral \AuAu\ collisions, and in $\pTreco<-10$ to $-20$ \gev\ 
(\rr-dependent) in central \AuAu\ collisions. The negative values after 
subtraction are consistent with zero within statistical uncertainty in all 
cases, and carry negligible weight in the correction and unfolding procedures 
discussed below. All negative entries are therefore set to zero, to simplify the 
unfolding procedure.

These distributions must still be corrected for the effects of local 
fluctuations in background energy density and for 
instrumental response. The corrections are carried out using regularized 
unfolding methods~\cite{Cowan:2002in,Hocker:1995kb}. In this approach, the measured jet 
distribution $M$ and true jet distribution $T$ are related by a response matrix,

\begin{widetext}
\begin{equation}
M(\pTreco) = \Big[ \Rbkg(\pTreco,\pTdetch) \times \Rdet(\pTdetch,\pTpartch) \Big] \times T(\pTpartch),
\label{eq:foldequation}
\end{equation}
\end{widetext}

\noindent
where the square brackets express the cumulative response matrix as the product of matrices separately encoding background and instrumental response effects; \pTpartch\ is the particle-level charged jet \pT; \pTdetch\ is
the detector-level charged jet \pT; and \pTreco\ the reconstructed jet \pT\ at the detector level, including \pT-smearing due to uncorrelated background. Factorization of the response into two separate matrices was studied in simulations and found to have negligible influence on the corrected distributions.

The corrected spectrum, which is a measurement of $T$, is determined by inverting Eq. \ref{eq:foldequation}. However, exact inversion of Eq. \ref{eq:foldequation} can result in a solution which has large fluctuations in central values and large variance, due to statistical noise in $M(\pTrec)$~\cite{Cowan:2002in}. A physically interpretable solution can be obtained by regularized unfolding, which imposes an additional smoothness constraint on the solution.

\subsection{Uncorrelated background response matrix \Rbkg}
\label{sect:dpT}

Central \AuAu\ collisions have large uncorrelated background energy 
density, with significant local fluctuations. While the scalar quantity $\rho$ accounts 
approximately for the event-to-event variation of  
uncorrelated background energy, it
does not account for local background fluctuations that smear \pTreco. 
Full background correction requires unfolding of these fluctuations. 

The response matrix for fluctuations in uncorrelated energy density is calculated by embedding detector-level simulated jets into real events at the track level, 
reconstructing the hybrid events, and matching each embedded jet with 
a reconstructed jet. The matching is carried out in the same way as for \Rdet, described below. The response matrix elements are the probability distribution of \dpT, the \pT-shift from the embedding procedure: 

\begin{equation}
\dpT=\pTreco-\pTembed.
\label{eq:dpT}
\end{equation}

High-\pT\ hadrons can be correlated in azimuth with the EP
orientation. The strength of this correlation is characterized by \vtwo, the 
second-order coefficient  of the Fourier expansion of 
the azimuthal distribution between the hadron and the EP~\cite{Adare:2014bga}. If \vtwo\ is non-zero for $\pT>9$ \gev, selection of a trigger hadron will bias 
the EP orientation in the accepted event population, thereby biasing the level of 
uncorrelated background in the recoil acceptance opposite to the trigger. This 
bias is taken into account in the calculation of the \dpT\ probability distribution by weighting the relative 
orientation of the trigger axis and EP 
orientation according to $1+\vtwo\cdot\cos\left(2\dphi\right)$. 

Observables based on reconstructed jets measure energy flow associated 
with a high-$Q^2$ process, independent of the specific distribution of hadrons 
arising from jet fragmentation. For accurate correction of local background 
fluctuations, the background response matrix should likewise depend only on the 
energy of the embedded object, and be independent of its specific distribution 
of hadrons. To explore this variation we use two different jet models for 
embedding: charged jets generated by PYTHIA, and single tracks carrying the 
entire jet energy \pTpart. Models with 
softer fragmentation than PYTHIA have likewise been explored in simulations, 
giving similar results~\cite{deBarros:2011ph}.

\begin{figure}[htbp] 
\includegraphics[width=0.4\textwidth]{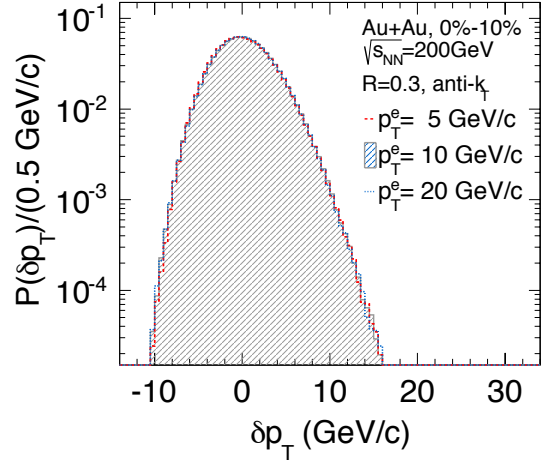}
\includegraphics[width=0.4\textwidth]{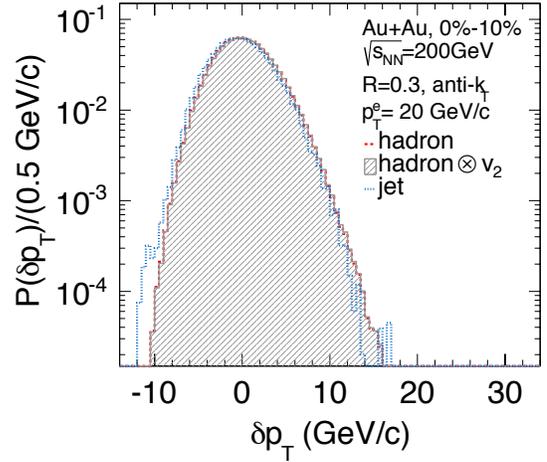}
\caption{(Color online) Probability distributions for \dpT\ in central \AuAu\ collisions. 
Upper: single-track embedding with different values of \pTembed ($p_{T}^{e}$).
Lower: \pTembed\ = 20 \gev\ 
with three different embedded-jet models: PYTHIA-generated detector-level jets, 
single tracks, and single tracks with \vtwo\ modulation of average background 
density. See text for details. 
}
\label{fig:dpT}
\end{figure}

Figure~\ref{fig:dpT}, upper panel shows the \dpT\ probability distribution for 
different values \pTembed\ of the embedded track, in central \AuAu\ collisions. 
Negligible 
dependence on \pTembed\ is observed.
The lower panel shows the \dpT\ probability distribution 
for \pTembed\ = 20 \gev\ with three different 
models for the embedded jet: PYTHIA-generated with no EP-bias; single particles 
with no EP-bias; and single particles with EP-bias corresponding to 
\vtwo\ = 0.04 for the trigger hadron, which is the largest \vtwo\ value for 
hadrons with $\pT>9$ \gev\ that is compatible with the uncertainty band measured 
in~\cite{Adare:2014bga}. The three distributions are similar, supporting this 
approach to correction for background fluctuations. Unfolding is carried out 
using all three distributions, with the variation between them contributing to 
the systematic uncertainty. Measurements of \vthree\ and higher 
harmonics for high-\pT\ hadrons are not presently available at RHIC energies. 
However, non-zero \vthree\ for the trigger hadron would only offset the 
influence in the recoil direction of trigger hadron \vtwo. 

\begin{figure}[htbp]
\includegraphics[width=0.5\textwidth]{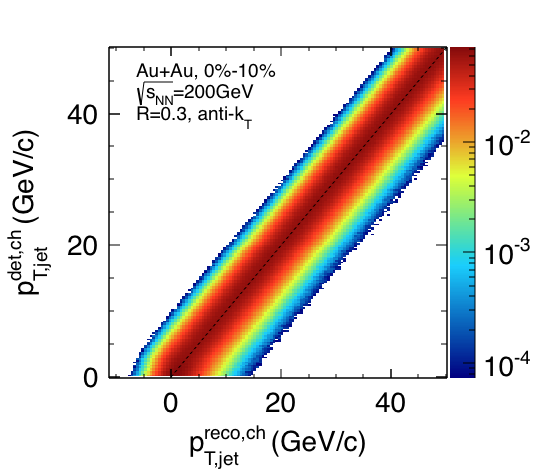}
\includegraphics[width=0.5\textwidth]{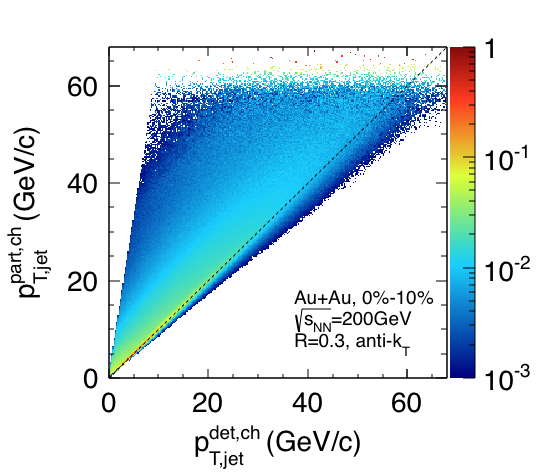}
\caption{(Color online) Response matrices for \rr\ = 0.3 jets in central \AuAu\ 
collisions. Upper: uncorrelated background response matrix \Rbkg. Lower: instrumental response matrix \Rdet.}
\label{fig:pt_embed_vs_pt_rec}
\end{figure}

Figure \ref{fig:pt_embed_vs_pt_rec}, upper panel, shows the full background response matrix \Rbkg, calculated by embedding single tracks.

\subsection{Instrumental response matrix \Rdet}
\label{sect:Rdet}

The largest contribution to the instrumental response matrix \Rdet\ is from
tracking efficiency, which shifts the spectrum lower in \pTreco. There is a smaller 
contribution from track momentum resolution, which smears \pTreco. 

The matrix \Rdet\ is determined using PYTHIA-generated events for \pp\ 
collisions at \sqrts\ = 200 GeV. Jet reconstruction is carried out at the particle 
level 
with the \antikT\ algorithm. Detector-level jets are generated by fast 
simulation,
applying the effects of tracking efficiency and track \pT\ resolution on the 
constituents of each 
particle-level jet. Jet reconstruction is then carried out on the detector-level 
event. Jets from this procedure are rejected if they lie outside the experimental acceptance, for both the particle-level and detector-level populations.

Tracks in particle-level jets are matched to detector-level 
tracks. For each particle-level jet, the detector-level jet with the largest 
fraction of the 
particle-level jet energy is matched to it, with the additional requirement that 
the fraction be greater than 15\%. 
The elements of \Rdet\ are the probability 
for a particle-level jet with \pTjetpart\ to have matched detector-level partner 
with \pTjetdet. Elements of \Rdet\ are normalized such that, for each bin in 
\pTjetpart, the sum over all bins in \pTjetdet\ is unity. The inefficiency 
arising from particle-level jets without a detector-level match is corrected on 
a statistical basis (Sect.~\ref{sect:JetMatching}), in a separate 
correction step.

As discussed in Sect.~\ref{sect:Interp}, the approach of this analysis results in corrected distributions for $\pTjetch>0$, while interpretation of such distributions in terms of parton showers and their modification in-medium is restricted to $\pTjetch>10$ \gev. In order to avoid the introduction of arbitrary cuts, \Rdet\ is constructed as described above for $\pTpart>0$, though jet-like objects with $\pTpart<10$ \gev\ should be interpreted with caution in terms of the fragmentation of quarks and gluons.

The contribution of secondary decays was determined using PYTHIA. The effect of 
feed-down from weak decays is 
negligible compared to other
systematic uncertainties, and no correction for this effect is applied.

Figure \ref{fig:pt_embed_vs_pt_rec}, lower panel, shows the matrix \Rdet\ for 
central \AuAu\ collisions. Matrix elements with $\pTjetdet<\pTjetpart$ arise largely due 
to tracking efficiency, which causes tracks to be lost from the jet. Matrix 
elements with $\pTjetdet>\pTjetpart$, which is less probable, arise from the 
effect of momentum resolution, for cases in which \pT-loss due to tracking 
efficiency is small.

\subsection{Unfolding}
\label{sect:Unfolding}

Unfolding is carried out using two different methods: an iterative method 
based on Bayes's Theorem~\cite{D'Agostini:1994zf}, and a method based on Singular 
Value Decomposition (SVD)~\cite{Hocker:1995kb}. For iterative Bayesian 
unfolding, regularization is imposed by
limiting the number of iterations, while for 
SVD unfolding, regularization is imposed by truncating the expansion to $k$ terms.

The unfolding procedure requires specification of a prior distribution. In order 
to assess the dependence of the unfolded solution on the choice of prior, 
several different prior distributions were used for both the Bayesian and SVD methods (see 
Sect.~\ref{sect:SysUncertUnfold}).

\subsection{Jet reconstruction efficiency}
\label{sect:JetMatching}

The matching procedure between particle-level and detector-level jets in 
Sect.~\ref{sect:Rdet} does not generate a match for every particle-level jet. 
The corresponding detector-level jet can be lost due to fiducial cuts and 
instrumental response, most notably tracking efficiency: especially for low-\pT\ jets containing few tracks, there is a non-zero probability that none of the tracks will be detected due to tracking efficiency less than unity. In addition, the jet area cut generates a small 
inefficiency for $\pTjetpart<4$ \gev, with negligible inefficiency at larger 
\pTjetpart\ (Sect. \ref{sect:ME}).

\begin{figure}[htbp]
\includegraphics[width=0.49\textwidth]{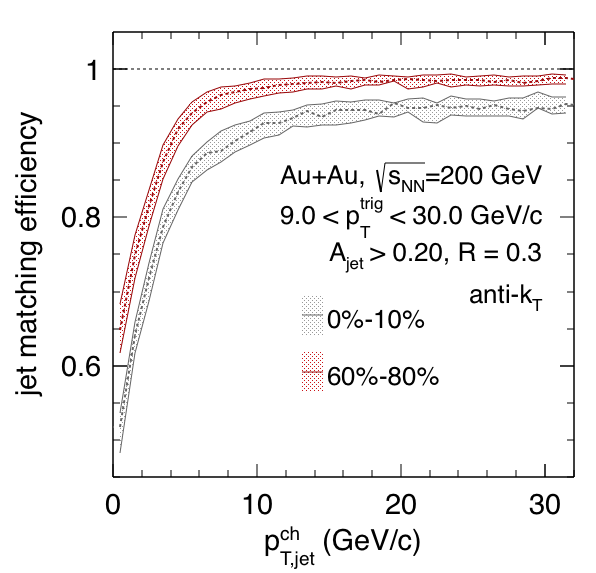}
\caption{(Color online) Jet reconstruction efficiency for peripheral 
and central \AuAu\ collisions, as a function of particle-level \pTjetpart. See 
text for details.}
\label{fig:Jet_match_eff}
\end{figure}

Figure~\ref{fig:Jet_match_eff} shows the jet reconstruction efficiency 
for central and peripheral \AuAu\ collisions, defined as the matching efficiency 
between particle-level and detector-level jets. The efficiency is calculated for particle-level jets whose centroid is within the experimental acceptance, $|\eta_\mathrm{jet}|<1-\rr$. The systematic uncertainty in 
efficiency, indicated by the bands, is due predominantly to uncertainty in the 
tracking efficiency. The correction for inefficiency is applied bin-by-bin to 
ensemble-averaged distributions, after the unfolding step.

\subsection{Estimated magnitude of corrections}
\label{sect:EstCorrections}

\begin{figure}[htbp] 
\includegraphics[width=0.5\textwidth]{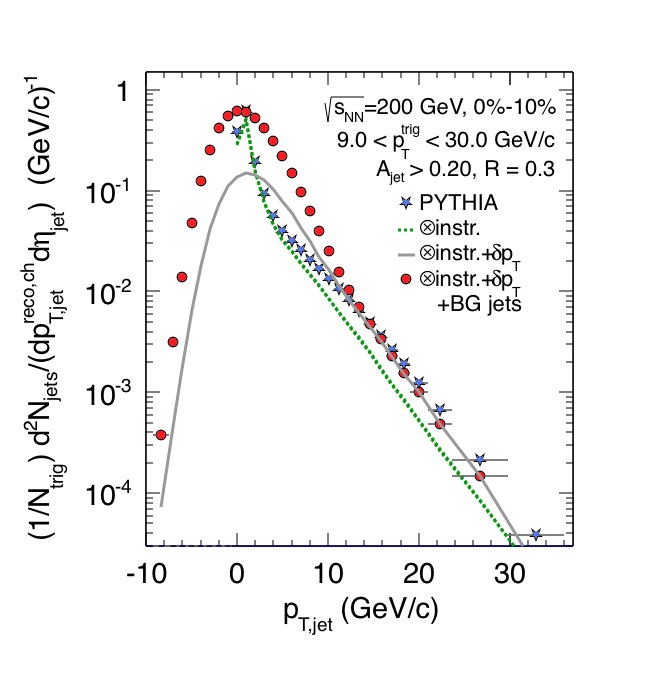} 
\caption{(Color online) Estimation of the magnitude of corrections for jets with 
\rr\ = 0.3, in central \AuAu\ collisions.}
\label{fig:RawDistributions}
\end{figure}

We conclude this section by estimating the magnitude of corrections. 
The estimate, shown in 
Fig, \ref{fig:RawDistributions}, is based on the recoil 
jet distribution (\rr\ = 0.3) for \pp\ collisions at \sqrts\ = 200 GeV calculated by 
PYTHIA at the particle level (blue stars), which is then modified by the inverse 
of the corrections discussed above. The effects correspond to a measurement in 
central \AuAu\ collisions. Instrumental effects, which are dominated by tracking 
efficiency, shift the distribution to lower \pTjet\ (blue stars $\rightarrow$ 
green dashed). Fluctuations due to uncorrelated background, as characterized by the
\dpT\ distribution, smear \pTjet\ but do not change the integrated yield of the distribution (green dashed $\rightarrow$ grey solid). Finally, the large population of uncorrelated 
background jet candidates in central \AuAu\ collisions modifies the spectrum 
significantly for $\pTjet<10$ \gev\ (grey solid $\rightarrow$ red circles). The 
cumulative correction for instrumental response and uncorrelated 
background therefore corresponds to the transformation from red circles to
blue stars. If considered on a bin-by-bin basis, the cumulative correction modifies the magnitude of the distribution by a factor less than two for $\pTreco>10$ \gev.

%% file: SystematicUncertainties.tex
\section{Systematic Uncertainties}
\label{sect:SysUncert}

Systematic uncertainties arise from the corrections for instrumental response and 
uncorrelated background, and from the different algorithmic choices in the unfolding 
procedure. This section discusses the significant systematic uncertainties, with representative 
values given in Tab.~\ref{Tab:SysUncert}.

\subsection{Instrumental response}
\label{sect:SysUncertInstr}

The systematic uncertainty due to track reconstruction efficiency is determined 
by varying the efficiency by $\pm$5\% relative to its
central value (Sect. \ref{sect:Rdet}). This variation generates a shift 
in \pTjetch, corresponding to variation in yield at fixed \pTjetch\ of less than 10\% for 
all \pTjetch, in both central and peripheral \AuAu\ collisions. Variation of other instrumental 
response corrections, including track \pT\ 
resolution and the contribution of secondary decays, generate smaller systematic 
uncertainties. The systematic uncertainty due to instrumental effects is labeled ``Instr" in 
Tab.~\ref{Tab:SysUncert}.

\subsection{Mixed events}
\label{sect:SysUncertME}

Correction for uncorrelated background by subtraction of the ME from the SE 
distribution requires normalization of the ME distribution by the factor \fME\ (Tab.~\ref{Tab:Integral_fME}). 
Variation of the normalization region for determining \fME\ 
results in a systematic uncertainty in corrected recoil jet yield of less than 10\% 
(``ME norm" in Tab.~\ref{Tab:SysUncert}).

The track population used to generate the ME data set includes
high-\pT\ tracks that arise predominantly from the fragmentation of jets, and 
their inclusion  means that not all jet-specific structure has been removed from 
the ME distributions. In
order to assess the importance of this contribution, the ME events
were modified to remove all tracks with $\pT>3$
\gev\ and the analysis was repeated. No significant change in the
distribution of reconstructed jets was observed from this modification.

\subsection{\dpT}
\label{sect:dpT}

The probability distribution of \dpT, which represents the fluctuations in uncorrelated background energy, was varied  
by using different models for embedded jets: single hadrons with the full jet 
energy, distributed either uniformly in azimuth or with anisotropic azimuthal 
distribution relative to the EP corresponding to \vtwo\ of the trigger 
hadron~\cite{Adare:2014bga}, or PYTHIA-simulated jets at the particle level with 
uniform azimuthal distribution. This variation of the \dpT\ distribution 
generates a systematic uncertainty in corrected jet yield of up to 19\% for 
central \AuAu\ collisions (``\dpT" in Tab.~\ref{Tab:SysUncert}).

\subsection{Unfolding}
\label{sect:SysUncertUnfold}

Systematic uncertainty due to the unfolding procedure was determined by 
varying the choice of unfolding algorithm, choice of prior, and regularization cutoff. 
Two different unfolding algorithms were used: iterative Bayesian and SVD. 
Two different functional forms of the prior were used: the recoil jet distribution for
 \pp\ collisions at \sqrts\ = 200 GeV, calculated by PYTHIA, and a parameterized Levy distribution,
 
 \begin{equation}
 f(p_{T},T,n) = \frac{p_{T}B}{[1+(\sqrt{p_{T}^{2}+m_{\pi}^{2}}-m_{\pi})/(nT)]^{n}}
 \label{func:Levy}
 \end{equation}
 
\noindent
The parameters $T$ and $n$, which determine the spectrum shape at low and high \pT\ respectively, were varied independently but constrained to $0.6<T<1.5$ GeV and $6<n<7$. These parameter ranges generate priors whose shapes bracket the resulting unfolded solutions, indicating convergence of the unfolding procedure.

For iterative Bayesian unfolding, the regularization limit on the number of iterations is varied between 1 to 5. For SVD 
unfolding, regularization is imposed by truncating the number of terms in the series expansion between 2 to 5.

The systematic uncertainty in corrected recoil jet yield resulting from these variations in unfolding procedure is \pTjetch-dependent, and is labeled ``Unfold" in Tab.~\ref{Tab:SysUncert}.

\subsection{Cumulative uncertainties}
\label{sect:SysUncertCumulative}

There is a complex interplay between the various components of the correction 
procedure. 
To determine the cumulative systematic uncertainty, each of the components was 
varied independently, thereby sampling the parameter space of corrections. The 
unfolding process was carried out multiple times, varying the
choices for tracking 
efficiency, ME normalization, \dpT\ algorithm, unfolding 
algorithm, prior, and regularization cutoff. 

For each specific set of choices, convergence 
of the unfolded distribution was evaluated by convoluting it  
with the same set of corrections (``backfolding") and comparing the result to 
the initial raw 
distribution using a $\chi^{2}$ test. The errors used to calculate $\chi^{2}$ 
are the diagonal elements of the covariance matrix from the 
unfolding procedure. The off-diagonal covariance elements, representing the 
correlation between bins, were not considered in this test. 
A set of choices was accepted if the comparison had $\chi^{2}/\mathrm{nDOF}$ 
less than a threshold which varied between 1.8 and 6.5, depending upon jet 
radius 
and collision centrality. For SVD unfolding, if an unfolded spectrum 
with regularization parameter $k$ was accepted, variations with the same prior but larger value of $k$ were rejected.

\begin{center}
\begin{table*}
\caption{Representative values for components of the cumulative systematic 
uncertainty in corrected recoil jet yield for \rr\ = 0.2 and 0.5 in central and 
peripheral \AuAu\ collisions, for various ranges in \pTjetch.  
See text for details. \label{Tab:SysUncert}}
\begin{tabular}{ |c|c|c|c|c|c|c||c| }
\hline
\multicolumn{3}{|c|}{} & \multicolumn{5}{c|}{Systematic uncertainty (\%)} \\ 
\hline 
 \rr & \pTjetch\ range [\gev] & centrality & Instr & ME norm & \dpT & Unfold & Cumulative\\ 
\hline
\multirow{6}{*}{0.2} & \multirow{2}{*}{[5,10]} & peripheral (60\%-80\%) & 4 & 2 & 1 & 6 & 10 \\ 
 & & central (0\%-10\%) & 7 & 10 & 19 & 41 & 47 \\ \cline{2-8}
 & \multirow{2}{*}{[10,20]} & peripheral (60\%-80\%) & 6 & 2 & 2 & 12 & 18 \\ 
 & & central (0\%-10\%) & 7 & 5 & 10 & 31 & 36 \\ \cline{2-8}
 & \multirow{2}{*}{[20,25]} & peripheral (60\%-80\%) & 11 & 8 & 6 & 25 & 33 \\ 
 & & central (0\%-10\%) & 10 & 7 & 16 & 47 & 49 \\ \hline 
 \multirow{6}{*}{0.5} & \multirow{2}{*}{[5,10]} & peripheral (60\%-80\%) & 4 & 3 & 4 & 22 & 23 \\ 
 & & central (0\%-10\%) & 6 & 5 & 3 & 21 & 27 \\ \cline{2-8}
 & \multirow{2}{*}{[10,20]} & peripheral (60\%-80\%) & 7 & 1 & 4 & 31 & 35 \\ 
 & & central (0\%-10\%) & 4 & 2 & 7 & 28 & 34 \\ \cline{2-8}
 & \multirow{2}{*}{[20,25]} & peripheral (60\%-80\%) & 9 & 3 & 5 & 29 & 35 \\ 
 & & central (0\%-10\%) & 8 & 1 & 10 & 30 & 39 \\ \hline 
\end{tabular}
\end{table*}
\end{center}
Due to the interplay between various components of the correction procedure, the 
contribution of each component to the cumulative systematic uncertainty of the 
recoil jet yield cannot be uniquely specified. Nevertheless, it is instructive to 
identify the principal factors that drive the cumulative systematic uncertainty. 
Table~\ref{Tab:SysUncert} shows representative values of each uncertainty 
component, for \rr\ = 0.2 and 0.5 in central and peripheral \AuAu\ 
collisions. These values are calculated by varying only the specified 
component, and keeping all other components in the correction procedure fixed. 
The uncertainties are averaged over three different 
ranges of \pTjetch, weighted by the spectrum shape.
It is seen that the unfolding procedure generates the largest systematic 
uncertainty in the recoil jet yield.

The rightmost column of Tab.~\ref{Tab:SysUncert} shows the cumulative 
systematic uncertainty in recoil jet yield. However, the unfolding process generates 
significant off-diagonal covariance, especially for large \rr, arising predominantly from correction of fluctuations in uncorrelated background. In order to indicate the significant correlation between different values of \pTjetch, in the following sections we represent the 
unfolded distributions graphically as bands rather than as binned histograms, 
with the width of the band representing the outer envelope of all distributions 
that were accepted by the above procedure. 

\section{Closure test}
\label{sect:SysUncertClosure}

\begin{figure}[htbp]
\includegraphics[width=0.5\textwidth]{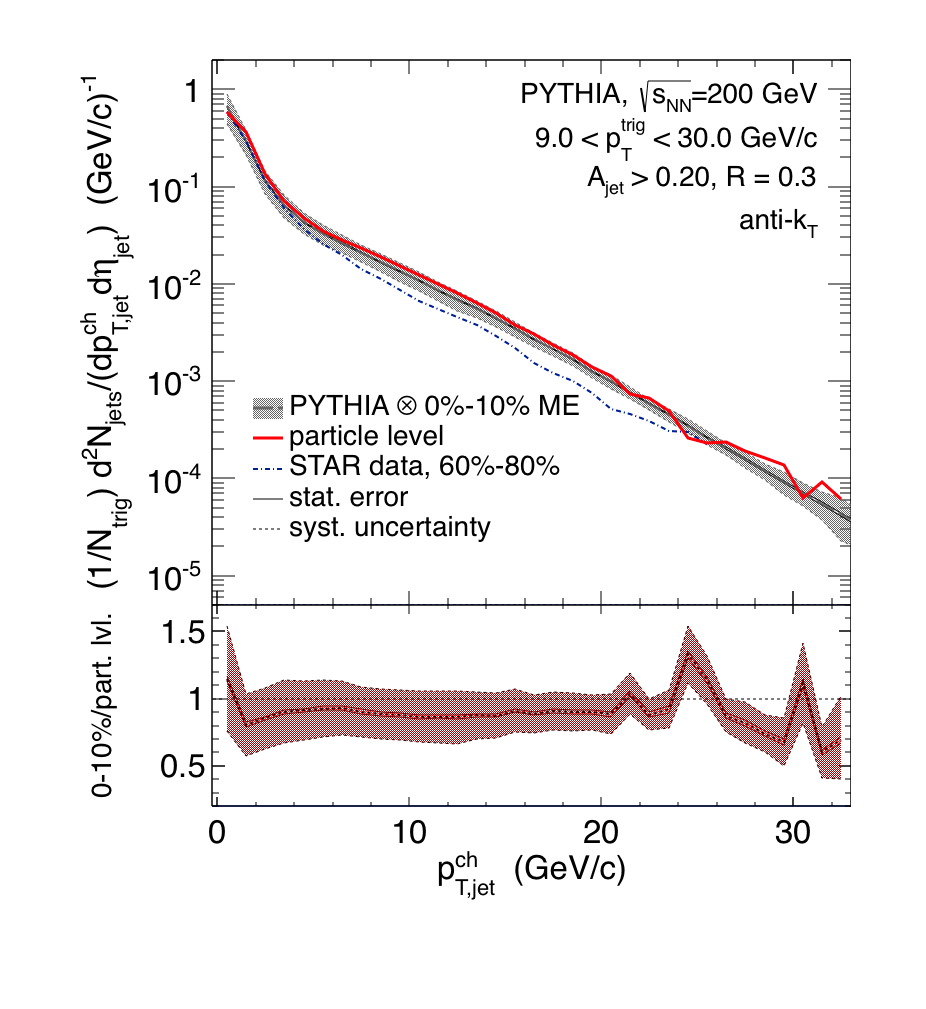}
\caption{(Color online) Closure test for central \AuAu\ collisions. Upper panel: 
particle-level input distribution from PYTHIA (red line), unfolded spectrum for 
\AuAu\ detector effects and background (grey band), and central value for fully 
corrected peripheral STAR data (blue dashed, systematic uncertainty not shown 
for clarity). Lower panel: ratio of unfolded 
over input distribution from upper panel. See text for details.}
\label{fig:Closure_test}
\end{figure}

Convergence of the full correction procedure was validated by a closure test on 
simulated data, utilizing events for \pp\ collisions at \sqrts\ = 200 GeV 
generated by PYTHIA. Figure \ref{fig:Closure_test}, upper panel, shows the 
particle-level distribution of these events for jets with \rr\ = 0.3, which is 
similar in shape to the fully corrected distribution from data for peripheral 
\AuAu\ collisions. 

Detector-level events were generated with
tracking efficiency and \pT-resolution\ corresponding to those of central \AuAu\ 
collisions. Each detector-level simulated event containing an 
accepted trigger hadron was embedded into a mixed event from the central 
\AuAu\ data set. The hybrid dataset has the same number of trigger 
hadrons as the real dataset, so that effects arising from finite event 
statistics are modeled accurately. The complete analysis chain, including 
generation of \dpT\ and the full set of corrections via unfolding, was then run 
on the hybrid events to generate the fully corrected recoil jet spectrum, as 
shown in the upper panel.

Figure \ref{fig:Closure_test}, lower panel, shows the 
ratio of the fully-corrected recoil jet distribution to the particle-level 
input distribution. The band shows the systematic uncertainty of the corrected 
distribution. For $\pTjetch>20$ \gev, fluctuations in the central 
value arise from the finite number of 
events in 
the input spectrum of the simulation, since the 
corrected distribution in the numerator is smoothed by regularized unfolding. 
For $\pTjetch<20$ \gev, the 
ratio is consistent with unity within the uncertainty of about 20\%, with no 
indication of a \pT-dependent bias in central value.

%% file: NLO.tex
\section{Perturbative QCD calculation}
\label{sect:pQCD}

The semi-inclusive recoil jet distribution is the ratio of 
cross sections for h+jet and inclusive hadron production 
(Eq.~\ref{eq:hJetDefinition}). The spin-dependent cross section for h+jet 
production in \pp\ collisions at \sqrts\ = 200 GeV has been calculated 
perturbatively at NLO~\cite{deFlorian:2009fw}. We utilize this NLO approach to 
calculate the spin-averaged h+jet and inclusive hadron cross sections, and 
their ratio. 

This measurement reports charged-particle jets. Although charged-particle jets 
are not infrared-safe in perturbation theory, non-perturbative track functions 
have been defined that represent the energy fraction of a parton 
carried by charged tracks and that account for infrared divergences, enabling 
calculation of infrared-safe charged-jet observables~\cite{Chang:2013rca}. 
PYTHIA-based calculations have been compared to such 
track functions and have similar evolution~\cite{Chang:2013rca}. 
For comparison of these measurements to NLO pQCD calculations, we therefore 
utilize PYTHIA to transform perturbatively calculated distributions from the 
parton to the charged-particle level.

\begin{figure}[htbp] 
\includegraphics[width=0.45\textwidth]{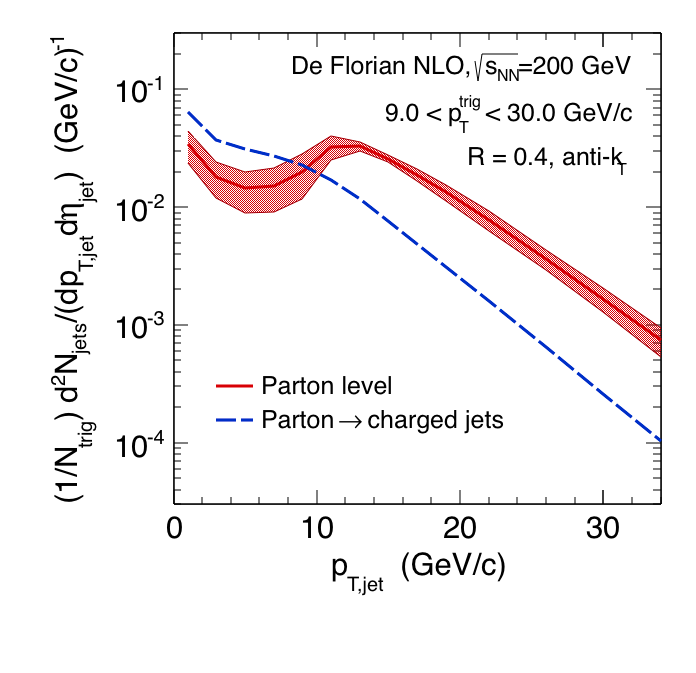}
\caption{(Color online) Calculation of the semi-inclusive recoil jet distribution in \pp\ collisions at \sqrts\ = 200 GeV, for jets 
with \rr\ = 0.4. The parton-level distribution 
is calculated perturbatively at NLO~\cite{deFlorian:2009fw}. The band shows the theoretical uncertainty due to scale variations. The 
charged-jet distribution is the transformation of the parton-level 
jet distribution using PYTHIA.}
\label{fig:NLO}
\end{figure}

Figure~\ref{fig:NLO} shows the distribution of \Yjet (Eq.~\ref{eq:hJetYield}) for jets 
with \rr\ = 0.4 in \pp\ collisions at \sqrts\ = 200 GeV (Eq.~\ref{eq:hJetDefinition}, 
RHS). The NLO pQCD formalism in~\cite{deFlorian:2009fw} is used for both the h+jet and inclusive hadron cross 
sections, with CTEQ6M parton distribution functions~\cite{Pumplin:2002vw} and DSS 
fragmentation functions~\cite{deFlorian:2007ekg}.
Variation of a factor 2 in the renormalization and factorization scales gives a 
variation in the ratio of 30\%-40\%, which represents the theoretical 
uncertainty. The figure also shows \Yjet\ at the 
charged-particle level, obtained by transforming the NLO distribution of recoil 
jets to charged-particle jets using PYTHIA, in this case version 
6.4.26, tune Perugia-0~\cite{Skands:2010ak}.

At LO, the trigger hadron threshold of 9 \gev\ sets a lower bound for \pTjet\ of 
the recoil jet. The parton-level recoil jet distribution at NLO indeed exhibits 
a peak around 
\pTjet\ = 9 \gev, reflecting this kinematic constraint. However, yield at lower 
\pTjet\ is also observed, indicating a contribution from higher-order processes. The peak is significantly reduced by the transformation from 
parton-level to charged-particle level, which both reduces and smears 
\pTjet. We note that, in this calculation, each parton-level 
jet is transformed into only one particle-level jet. The transformation from parton-level to 
particle-level distributions based on PYTHIA therefore does not account for jet splitting, 
which may contribute at low \pTjet\ and for small \rr. 

Comparison of these distributions to measurements is made in the 
following section.

%% file: Results.tex
\section{Results}
\label{sect:Results}

\subsection{Jet yield suppression}
\label{sect:YieldSuppress}

\begin{figure*}[htbp] %
\includegraphics[width=0.45\textwidth]{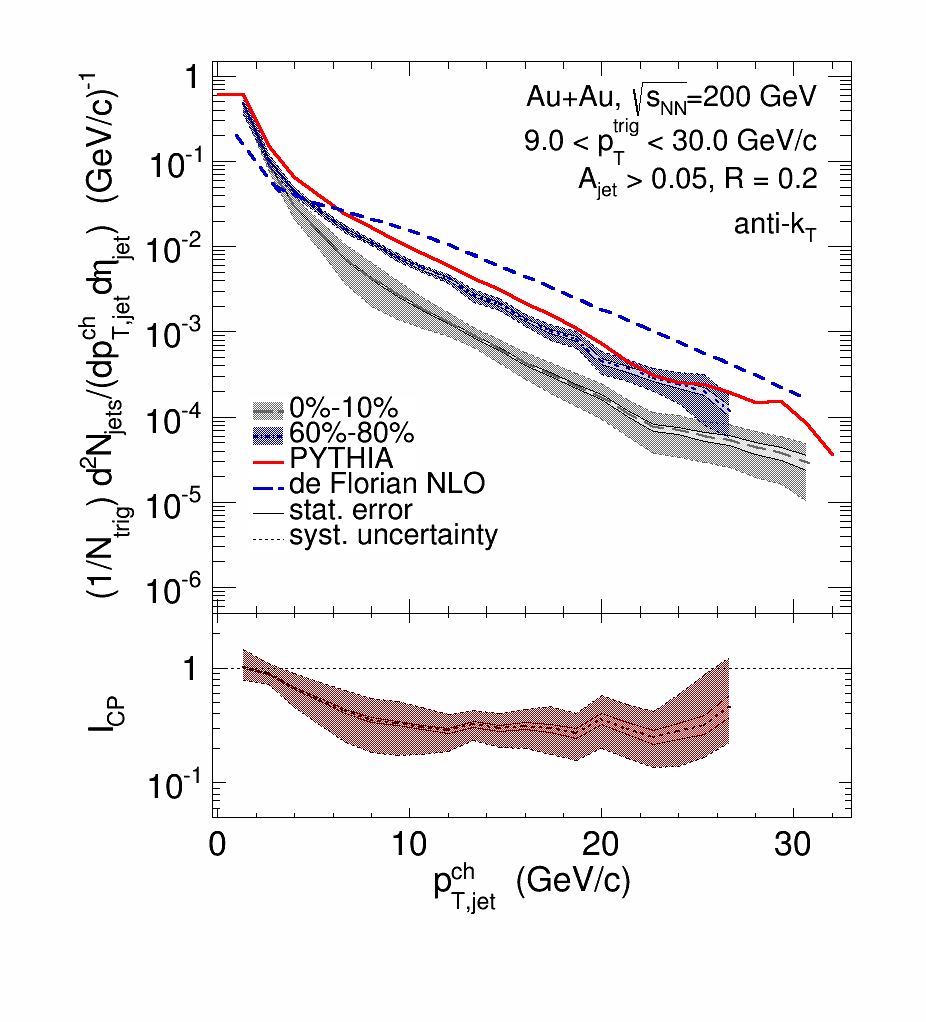}
\includegraphics[width=0.45\textwidth]{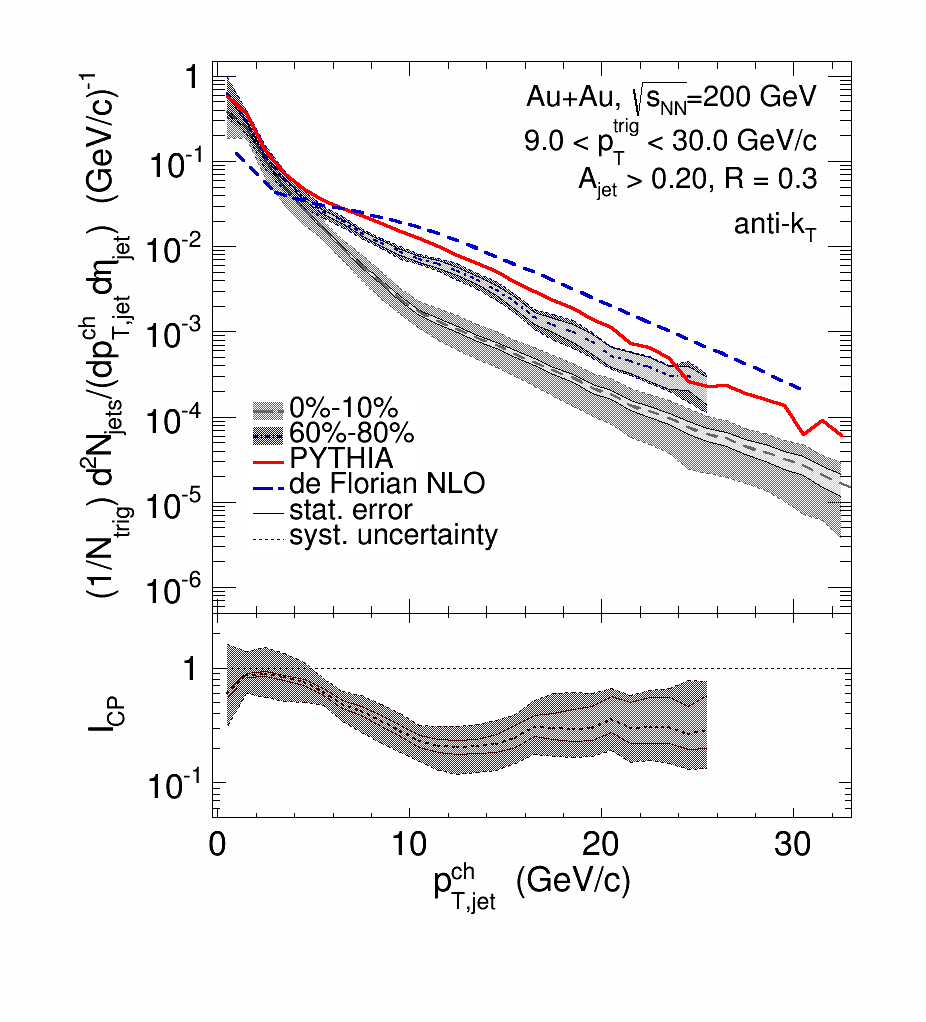}

\includegraphics[width=0.45\textwidth]{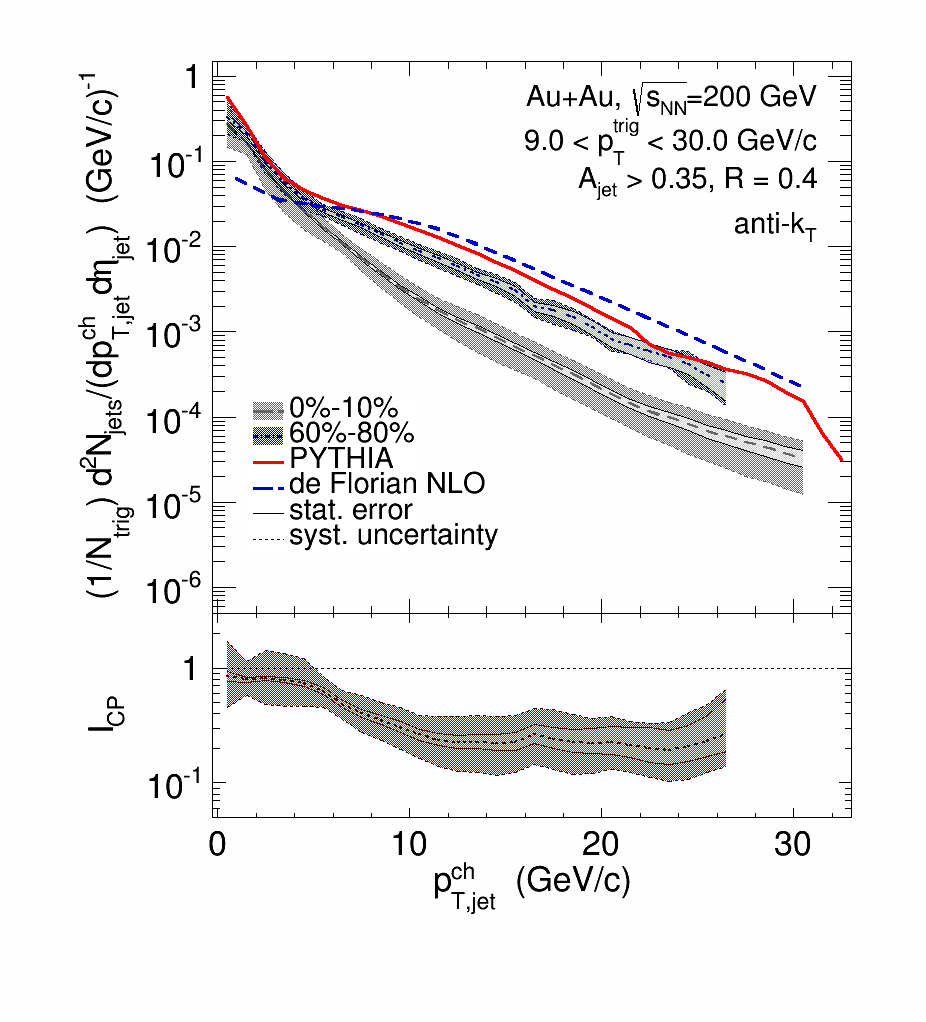}
\includegraphics[width=0.45\textwidth]{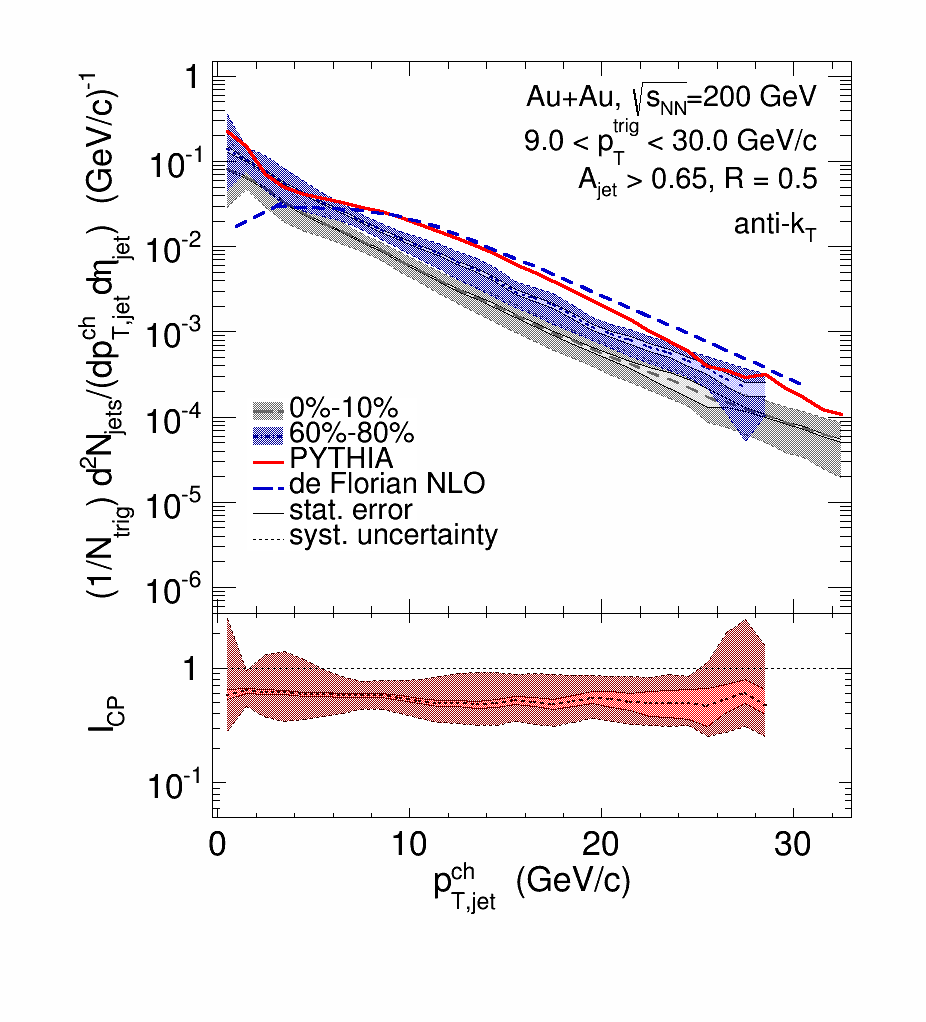}
\caption{(Color online) Fully corrected distributions of \Yjet (upper panels) 
and its ratio \ICP\ (lower panels) for central and peripheral \AuAu\ collisions at 
\sqrtsNN\ = 200 GeV, for \antikT\ jets with \rr\ = 0.2, 0.3, 0.4 and 0.5.  The upper 
panels also show \Yjet\ for \pp\ collisions at 
\sqrts\ = 200 GeV, calculated using PYTHIA at the charged-particle level 
and NLO pQCD transformed to the charged-particle level (Sect.~\ref{sect:pQCD}). The uncertainty of the NLO calculation is not shown.}
\label{fig:ICP}
\end{figure*}

Figure \ref{fig:ICP}, upper panels, show fully corrected distributions of \Yjet\ 
for \rr\ = 0.2, 0.3, 0.4 and 0.5, in peripheral 
and central \AuAu\ collisions at \sqrtsNN\ = 200 GeV. The lower panels show \ICP, 
the ratio of \Yjet\ in central to peripheral 
distributions. The systematic 
uncertainty of \ICP\ takes into account the correlated uncertainties of 
numerator and denominator. 
The recoil jet yield in central collisions is strongly suppressed in the region $\pTjetch>10$ 
\gev\ for \rr\ between 0.2 and 0.5, with less suppression for 
\rr\ = 0.5 than for \rr=0.2. 

\begin{center}
\begin{table*}
\caption{Shift of \Yjet\ in \pTjetch\  from peripheral to 
central collisions in
Fig.~\ref{fig:ICP}. Statistical and systematic uncertainties are shown. 
The systematic uncertainty takes 
into account correlated uncertainties between the peripheral and central 
distributions, in particular the tracking efficiency. Also shown is the equivalent 
shift between \pp\ and central \PbPb\ collisions at \sqrtsNN\ = 2.76 TeV~\cite{Adam:2015doa}. \label{Tab:pTjetShift}}
\begin{tabular}{ |c|c|c|c| }
\hline
\multicolumn{2}{|c|}{System} & \AuAu\ \sqrtsNN\ = 200 GeV & \PbPb\ \sqrtsNN\ = 2.76 
TeV \\ 
\hline 
\multicolumn{2}{|c|}{\pTjetch\ range (\gev)} & [10,20] & [60,100] \\ \hline 
\hline
\multicolumn{2}{|c|}{} & \multicolumn{2}{c|}{\pT-shift of \Yjet\ (\gev)} \\ 
\cline{3-4}
\multicolumn{2}{|c|}{} & peripheral$\rightarrow$central & \pp$\rightarrow$central \\ 
\hline
\multirow{4}{*}{\rr} & 0.2 & $-4.4\pm0.2\pm1.2$ & \\ \cline{2-4}
 & 0.3 & $-5.0\pm0.5\pm1.2$ & \\ \cline{2-4}
 & 0.4 & $-5.1\pm0.5\pm1.2$ & \\ \cline{2-4}
 & 0.5 & $-2.8\pm0.2\pm1.5$ & $-8\pm2$ \\ \hline
\end{tabular}
\end{table*}
\end{center}

The upper panels also show \Yjet\ distributions for \pp\ 
collisions at \sqrts\ = 200 GeV, calculated by PYTHIA and by pQCD at NLO 
transformed to charged jets (Sect.~\ref{sect:pQCD}). The uncertainty of the NLO calculation (Fig.~\ref{fig:NLO}) is not shown, for visual clarity.
The central value of the PYTHIA-generated
distribution lies about 20\% above the peripheral \AuAu\ 
distribution for all values of \rr. The NLO-generated distribution lies yet 
higher for \rr\ = 0.2, but agrees better with PYTHIA  for \rr\ = 0.5.  A similar 
comparison was carried out for \pp\ collisions at \sqrts\ = 7 
TeV, with PYTHIA found to agree better than NLO with data~\cite{Adam:2015doa}. 

\begin{figure*}[htbp] 
\includegraphics[width=0.48\textwidth]{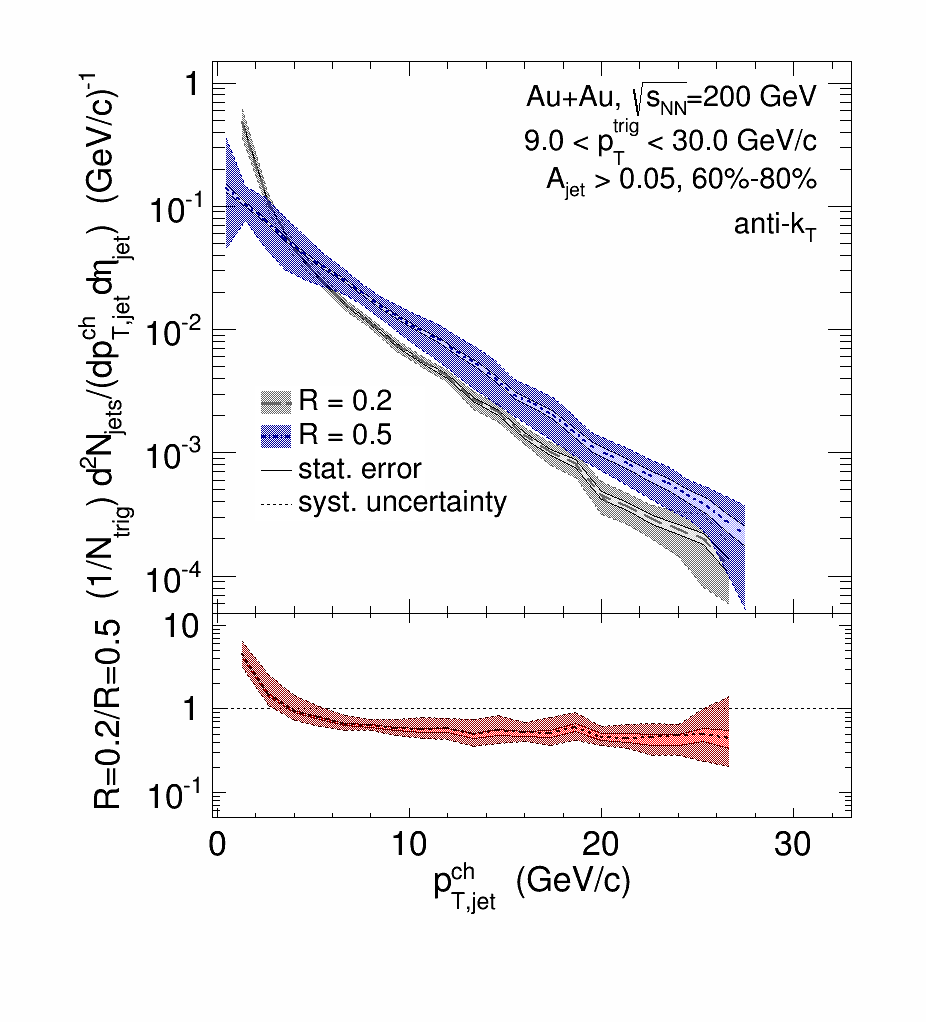}
\includegraphics[width=0.48\textwidth]{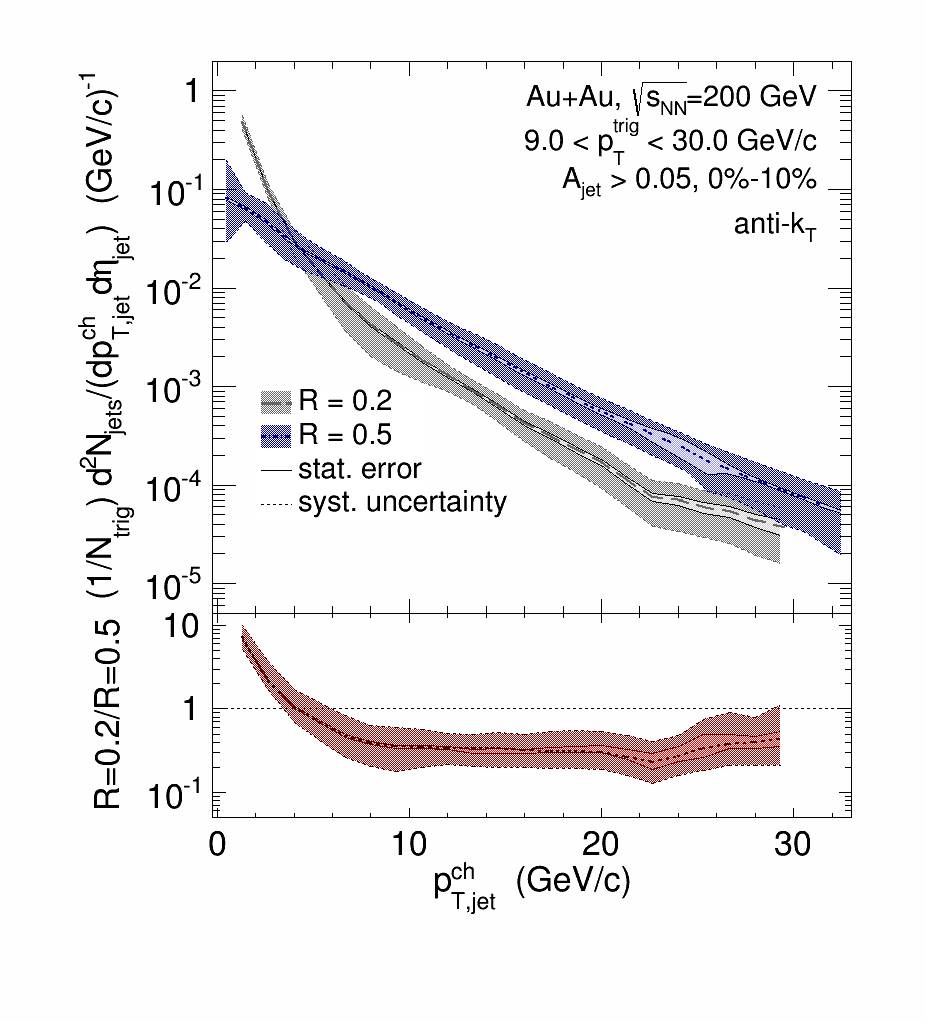}
\caption{(Color online) Distributions of \Yjet\ for \rr\ = 0.2 and 0.5 (upper 
panels) 
and their ratios (lower panels) in peripheral (left) and central (right) \AuAu\ 
collisions at \sqrtsNN\ = 200 GeV.}
\label{fig:Ratio_2_5}
\end{figure*}

Since the shape of the \Yjet\ distributions is approximately exponential, for a range of 
\pTjetch\ in which \ICP\ is constant, suppression of \ICP\ can be expressed 
equivalently as a shift of \Yjet\ in \pTjetch\ between the peripheral and 
central distributions. Tab.~\ref{Tab:pTjetShift} gives values of the shift 
for the distributions in Fig.~\ref{fig:ICP}, together with the 
shift measured for \rr\ = 0.5 between \pp\ and central \PbPb\ 
collisions at \sqrtsNN\ = 2.76 TeV. The peripheral-central shifts are 
consistent within uncertainties for the various \rr\ in \AuAu\ collisions at 
\sqrtsNN\ = 200 GeV, and are systematically smaller than the \pp\ to central \PbPb\ 
shift measured 
at \sqrtsNN\ = 2.76 TeV.

\begin{figure}[htbp] 
\includegraphics[width=7.5cm]{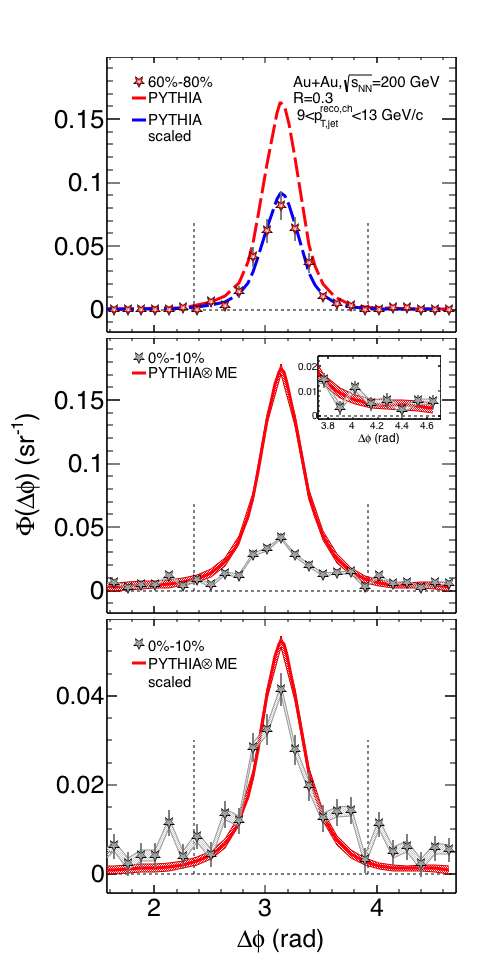} 
\caption{(Color online) Distribution of \Phijet\ at \sqrts\ = 
200 GeV, for \AuAu\ collisions measured by STAR and \pp\ collisions generated by PYTHIA (detector level). Vertical dashed lines show 
limits of integration for \Yjet. Top panel: peripheral \AuAu\ 
compared to \pp. Blue dashed curve shows PYTHIA 
distribution scaled to have the same integral as data between the vertical dashed lines.
Middle panel: central \AuAu\ compared 
to \pp\ detector-level events embedded
into central \AuAu\ 
mixed events. Shaded bands show systematic uncertainty due to 
mixed-event normalization. Bottom panel: same as middle panel, but with PYTHIA 
distribution scaled to have the same integral as data between the vertical dashed lines.}
\label{fig:dphi}
\end{figure}

In light of the low infrared cutoff of jet constituents 
in this analysis (track $\pT>0.2$ \gev), we interpret the shift as the 
charged-particle energy 
transported to angles larger than \rr\ by interaction of the jet with the 
medium, averaged over the recoil jet population. 
In this interpretation, the  spectrum shift represents the average out-of-cone 
partonic energy loss for central relative to peripheral 
collisions. Table~\ref{Tab:pTjetShift} presents the first quantitative 
comparison of the quenching of reconstructed jets at RHIC and the LHC, indicating 
reduced medium-induced energy transport to large angles at RHIC,
though the different ranges in \pTjetch\ and the different reference spectra (\pp\ vs. peripheral) should be noted.

\subsection{Modification of jet shape}
\label{sect:IntraJet}

The ratio of inclusive jet cross sections with small \rr\ relative to large \rr\ 
has been measured to be less than unity in \pp\ collisions at \sqrts\ = 2.76 and 
7 
TeV~\cite{Abelev:2013fn,Chatrchyan:2014gia}, 
reflecting the distribution of jet energy transverse to the jet axis.
These measurements are well-described by pQCD 
calculations at NLO and NNLO~\cite{Soyez:2011np,Dasgupta:2016bnd}. 
Inclusive measurements of small-radius jets are also well-described by an approach based on soft collinear effective theory \cite{Kang:2017frl}.
The ratio of semi-inclusive recoil jet yields  with small relative to large \rr\ 
is likewise less than unity in
\pp\ collisions at \sqrts\ = 7 TeV~\cite{Adam:2015doa}, exhibiting sensitivity 
to 
the transverse distribution of jet energy in the recoil jet population.
PYTHIA provides a better description than NLO of this 
ratio~\cite{Adam:2015doa,deFlorian:2009fw}. 
A jet quenching calculation using a hybrid weak/strong-coupling 
approach indicates that the ratio of (semi-)inclusive yields with different 
values of \rr\ has smaller theoretical uncertainties than other jet shape 
observables~\cite{Casalderrey-Solana:2016jvj}.
The \rr-dependent ratios of inclusive jet cross sections and semi-inclusive jet 
yields therefore
provide discriminating jet shape observables that can be calculated 
theoretically 
for \pp\ collisions, and that provide sensitive probes of medium-induced 
broadening of the jet shower. We note that this approach to measuring jet shapes is different than the differential jet shape observable employed by CMS to measure medium-induced modification of jet shapes in \PbPb\ collisions at \sqrtsNN=2.76 TeV~\cite{Chatrchyan:2013kwa}.

Figure \ref{fig:Ratio_2_5} shows distributions of 
\Yjet\ for \rr\ = 0.2 and 0.5, for peripheral and central 
\AuAu\ collisions at \sqrtsNN\ = 200 GeV. Their ratio, shown in the lower panels, 
is less than 
unity, also reflecting the intra-jet distribution of 
energy transverse to the jet axis. Comparison of the distributions for peripheral and central collisions measures 
medium-induced broadening of the jet shower in an angular range between 0.2 and 0.5 rad of the recoil 
jet axis. For quantitative comparison, we again express the change in \Yjet\ 
between \rr\ = 0.2 and 0.5 as  a horizontal shift of the spectra. 
In the range $10<\pTjetch<20$ \gev, the \pT-shift in \Yjet\  
from \rr\ = 0.2 to \rr\ = 0.5 is $2.9\pm{0.4\mathrm{(stat)}}\pm{1.9\mathrm{(sys)}}$ \gev\ in 
peripheral collisions and $5.0\pm{0.5\mathrm{(stat)}}\pm{2.3\mathrm{(sys)}}$ \gev\ in central 
collisions, which are consistent within uncertainties. From this measurement 
we find no evidence of broadening of the jet shower due to jet quenching. A 
similar picture was obtained for \PbPb\ collisions at the LHC~\cite{Adam:2015doa}.

\subsection{Medium-induced acoplanarity}
\label{sect:Acoplanarity}

In this section we discuss the measurements of \Phijet\ (Eq.~\ref{eq:hJetPhi}), the 
azimuthal distribution of the recoil jet centroid relative to the 
axis of the trigger hadron. In \pp\ collisions, the azimuthal distribution of back-to-back di-jet pairs is peaked at
$\dphinovar\sim\pi$, with initial-state and final-state radiative processes 
generating acoplanarity that 
broadens the \dphinovar\ distribution. In nuclear collisions, additional acoplanarity may 
be induced by jet interactions in hot QCD matter~\cite{D'Eramo:2012jh,Wang:2013cia,Mueller:2016gko,Chen:2016vem,Casalderrey-Solana:2016jvj}, with magnitude
related to the jet transport parameter 
\qhat~\cite{Mueller:2016gko,Chen:2016vem,Casalderrey-Solana:2016jvj}. Acoplanarity from vacuum 
radiation grows with both jet energy and \sqrts, so that low energy jets may
have greatest sensitivity to \qhat~\cite{Chen:2016vem,Casalderrey-Solana:2016jvj}. The \rr\ dependence of 
acoplanarity may probe the distribution of both vacuum and medium-induced gluon radiation within the jet shower~\cite{Chen:2016vem}, and may also probe different quenching effects for initially narrow or wide jets~\cite{Casalderrey-Solana:2016jvj}.

Scattering of a jet off quasi-particles in the hot QCD medium is conjectured 
to dominate
the azimuthal distribution at large angles from the trigger axis
(QCD Moli{\`e}re scattering), with radiative processes and 
soft multiple scattering making smaller
contributions in that region~\cite{D'Eramo:2012jh}. 
Measurement of jet acoplanarity at large angles can potentially discriminate 
between a 
medium with distinct quasi-particles and one that is effectively continuous at 
the length scale being probed by the scattering~\cite{D'Eramo:2012jh}. It is 
important 
to perform such large-angle scattering measurements over a large 
range of jet energy, which varies the length scale of the probe. Such 
measurements can only be carried out using reconstructed jets recoiling from a 
trigger object; observables based on the distribution of single recoil 
hadrons convolute the effects of intra-jet broadening and scattering of the 
parent, and cannot 
discriminate the two processes.

We note that the trigger hadron, with $\pTtrig>9$ \gev, most likely 
arises from fragmentation of a jet, but that the direction of such a trigger hadron 
and its parent jet centroid are not necessarily coincident. In order to quantify the 
difference, the correlation between the axis defined by jet centroid and the 
direction of the leading hadron in the jet was studied using PYTHIA-generated events for \pp\ collisions at \sqrts\ = 200 GeV. The distribution of the angular difference between jet 
centroid and leading hadron has RMS = 10 mrad for hadrons with  $\pT>9$ \gev\ and jets with \rr\ = 0.3. 
Since high-\pT\ hadrons in \AuAu\ collisions are expected to bias towards jets that have lost 
relatively little energy due to quenching~\cite{Baier:2002tc}, we expect a similar correlation in 
central \AuAu\ collisions. The trigger hadron direction in this analysis 
therefore corresponds closely to the axis of the jet that generates it. 

In order to measure the distribution of \Phijet, the contribution of 
uncorrelated background must be removed 
from the raw \dphinovar\ distribution. As in the \Yjet\ analysis, this correction is 
carried out by subtracting the scaled ME distribution from the SE distribution. 
However, to correct \Phijet\ we utilize an ME scaling factor that is determined 
separately for each 
bin in \dphinovar, rather than applying \fME\ (Tab.~\ref{Tab:Integral_fME}), which 
is the scale factor averaged over the 
\dphinovar\ range of the recoil acceptance for \Yjet. This modified 
procedure is used because the ME scale factor depends upon the interplay between 
conservation of total jet number and the enhanced yield at large positive 
\pTreco\ for the SE distribution relative to ME. At large angles to the 
trigger axis the SE enhancement is small, and the ME scale factor approaches 
unity in that region. By utilizing a \dphinovar-dependent scaling of the ME 
distribution we track this effect 
accurately, resulting in an accurate ME normalization for correction of 
uncorrelated background yield.

Figure \ref{fig:dphi} shows \Phijet\ distributions  
for \rr\ = 0.3 and $9<\pTreco<13$ \gev\ measured in peripheral and central 
\AuAu\ collisions at \sqrtsNN\ = 200 GeV, compared to \Phijet\ 
distributions for \pp\ collisions at \sqrts\ = 200 GeV generated by PYTHIA. The 
data are the same as those in Figs. 
\ref{fig:Raw_Delta_phi_per} and \ref{fig:Raw_Delta_phi_cen}. 
The data are corrected for uncorrelated background yield using ME subtraction,
but no correction is applied for instrumental response or uncorrelated 
background fluctuations. Rather, for comparison to data, the PYTHIA \pp\ 
distribution is used at the detector level, which incorporates the effects of 
instrumental response. In addition, for comparison to the central \AuAu\ data, 
the effects of  
uncorrelated background fluctuations are imposed 
by embedding the \pp\ events generated by PYTHIA at the detector level 
into \AuAu\ mixed events. These reference events based on PYTHIA are analysed in the same way as real data; in particular, the effect of correlated recoil jets on the calculation of $\rho$ is the same as that in real data analysis.

The top and middle panels of Figure \ref{fig:dphi} compare absolutely 
normalized  \Phijet\ distributions for \AuAu\ and \pp. The yield for the 
PYTHIA-generated \pp\ distribution in this 
region is significantly larger than that of the \AuAu\ data for both peripheral 
and central collisions, with larger difference for central collisions. This is 
in 
qualitative agreement 
with Fig.~\ref{fig:ICP}, though quantitative comparison is not possible because 
these data are not fully corrected.

For detailed comparison of the shape of the central peaks of the 
\Phijet\ distributions, we scale the PYTHIA-generated \pp\ distributions to have 
the same integrated yield as the data in the range $|\pi-\dphinovar|<\pi/4$. 
The top panel of Figure \ref{fig:dphi} shows scaled \pp\ compared to peripheral 
\AuAu, which agree well. The bottom panel shows the 
scaled embedded \pp\ and central \AuAu\ distributions, indicating a slightly 
broader central peak in data. A recent calculation suggests that such 
comparisons may be used to constrain $\langle\qhat\cdot{L}\rangle$, where \qhat\ 
is the jet transport parameter and 
$L$ is the in-medium path length~\cite{Chen:2016vem}. However, quantitative 
comparison of such measurements and calculations requires correction of the data for 
instrumental and background fluctuation effects, which requires higher 
statistical precision than the data presented here and is beyond the scope of 
the current analysis.

Finally, we turn to the search for large-angle Moli{\`e}re 
scattering in the hot QCD medium~\cite{D'Eramo:2012jh}. Absolutely 
normalized \Phijet\ 
distributions are required for this measurement. We focus on the \Phijet\ 
distribution at large angles 
relative to the trigger axis, in the range $|\pi-\dphinovar|>0.56$. 
Fig.~\ref{fig:dphi}, upper panel, shows no 
significant yield in this range for both peripheral 
\AuAu\ events and  PYTHIA-generated \pp\ events. 
The insert in the middle panel shows the \Phijet\ 
distribution in this range for central 
\AuAu\ collisions and PYTHIA-generated \pp\ events embedded into central \AuAu\ 
mixed events. Both distributions have 
non-zero yield and are consistent with each other within the uncertainty band. 
We therefore do not observe significant evidence for 
large-angle Moli{\`e}re scattering in central \AuAu\ collisions. A similar measurement by the ALICE Collaboration for 
\PbPb\ collisions 
at \sqrtsNN\ = 2.76 TeV likewise found no evidence for large-angle Moli{\`e}re 
scattering in nuclear collisions at the LHC~\cite{Adam:2015doa}.

The comparison of central \AuAu\ and embedded \pp\ distributions can however be 
used to establish a limit on the magnitude of large-angle scattering, under two 
assumptions. The first assumption 
is that PYTHIA provides an accurate reference distribution. The second 
assumption, which we make for simplicity, is that the distribution of excess 
yield from large angle scattering is a constant fraction of the \pp\ reference 
yield, independent of \dphinovar\ for $|\pi-\dphinovar|>0.56$. We then form the 
ratio 
of the central \AuAu\ yield over that for PYTHIA-generated and 
embedded \pp\ collisions. No scaling of the \pp\ distribution is applied, since 
this measurement requires absolutely normalized distributions. This ratio is 
indeed 
independent of \dphinovar\ within uncertainties, consistent with the second 
assumption. Averaged over the eight data points shown in the inset of Fig.~\ref{fig:dphi}, the ratio is measured to be 
$1.2\pm0.2\mathrm{(stat)}\pm0.3\mathrm{(sys)}$. In order to express 
this measurement as a limit, we 
consider only the statistical error to be Gaussian-distributed, and cite the 
systematic uncertainty separately. At 90\% statistical confidence 
level (one-sided), the excess yield due to medium-induced large angle scattering 
is less than  
$50\pm30\mathrm{(sys)\%}$ of the large-angle yield for \pp\ collisions 
predicted by PYTHIA.

Future measurements, based on larger \AuAu\ data sets, will reduce the 
statistical error and systematic uncertainty of this measurement. The two 
assumptions used in the analysis can be relaxed by measurement of the reference 
distribution in \pp\ collisions, and by theoretical calculations of the expected 
distribution.

%% file: Summary.tex
\section{Summary}
\label{sect:Summary}

We have reported the measurement of jet quenching in peripheral and central 
\AuAu\ collisions at \sqrtsNN\ = 200 GeV, based on the semi-inclusive distribution 
of reconstructed charged jets 
recoiling from a high-\pT\ trigger hadron. Jets were reconstructed with low 
infrared cutoff of constituents, $\pT>0.2$ \gev. Uncorrelated background was 
corrected at the level of ensemble-averaged distributions using a new 
event-mixing method. Comparison is made to similar distributions for \pp\ 
collisions 
at \sqrts\ = 200 GeV, calculated using PYTHIA and NLO pQCD, and to similar 
measurements for \PbPb\ collisions at \sqrtsNN\ = 2.76 TeV.

The recoil jet yield is suppressed in central \AuAu\ collisions for 
jet radii \rr\ between 0.2 and 0.5. Taking into account the low IR-cutoff for 
jet 
constituents, the suppression corresponds to medium-induced energy transport to 
large angles relative to the jet axis of $\sim3-5$ \gev, smaller than that 
measured 
for central \PbPb\ collisions at the LHC. Comparison of recoil jet yields for 
different \rr\ 
exhibits no evidence of significant intra-jet broadening within an angle of 0.5 
relative to the jet axis. 

Yield excess in the tail of the recoil jet azimuthal distribution would 
indicate large-angle jet scattering in 
the medium, which could probe its quasi-particle nature. However, no evidence 
for such a process is seen within the current experimental precision.
The 90\% statistical confidence upper limit from this measurement for the excess jet 
yield at large deflection angles is $50\pm30\mathrm{(sys)\%}$ of the large-angle yield in 
PYTHIA-generated \pp\ events. This is the first quantitative limit on 
large-angle  Moli{\`e}re scattering of jets in heavy ion collisions at RHIC.

Future measurements, based on data sets with high integrated luminosity and 
incorporating the 
STAR electromagnetic calorimeter, will explore these observables with greater 
statistical and systematic precision and with greater kinematic reach, providing 
further quantification of 
jet quenching effects and clarification of their underlying mechanisms.

%% file: Acknowledgements.tex
\section{Acknowledgments}
\label{sect:Acknowledgments}

We thank the RHIC Operations Group and RCF at BNL, the NERSC Center at LBNL, and the Open Science Grid consortium for providing resources and support. This work was supported in part by the Office of Nuclear Physics within the U.S. DOE Office of Science, the U.S. National Science Foundation, the Ministry of Education and Science of the Russian Federation, National Natural Science Foundation of China, Chinese Academy of Science, the Ministry of Science and Technology of China and the Chinese Ministry of Education, the National Research Foundation of Korea, GA and MSMT of the Czech Republic, Department of Atomic Energy and Department of Science and Technology of the Government of India; the National Science Centre of Poland, National Research Foundation, the Ministry of Science, Education and Sports of the Republic of Croatia, RosAtom of Russia and German Bundesministerium fur Bildung, Wissenschaft, Forschung and Technologie (BMBF) and the Helmholtz Association.

%% file: STAR_hJet_PRC_Master.bbl
\begin{thebibliography}{10}

\bibitem{Majumder:2010qh}
A.~Majumder and M.~Van~Leeuwen,
\newblock Prog.Part.Nucl.Phys. {\bf 66}, 41 (2011), arXiv:1002.2206.

\bibitem{Adare:2010ry}
A.~Adare {\em et~al.}, PHENIX,
\newblock Phys.Rev.Lett. {\bf 104}, 252301 (2010), arXiv:1002.1077.

\bibitem{Adler:2002xw}
C.~Adler {\em et~al.}, STAR,
\newblock Phys.Rev.Lett. {\bf 89}, 202301 (2002), arXiv:nucl-ex/0206011.

\bibitem{Adler:2002tq}
C.~Adler {\em et~al.}, STAR,
\newblock Phys.Rev.Lett. {\bf 90}, 082302 (2003), arXiv:nucl-ex/0210033.

\bibitem{Adams:2003kv}
J.~Adams {\em et~al.}, STAR,
\newblock Phys.Rev.Lett. {\bf 91}, 172302 (2003), arXiv:nucl-ex/0305015.

\bibitem{Adams:2006yt}
J.~Adams {\em et~al.}, STAR,
\newblock Phys. Rev. Lett. {\bf 97}, 162301 (2006), arXiv:nucl-ex/0604018.

\bibitem{Adamczyk:2013jei}
L.~Adamczyk {\em et~al.}, STAR,
\newblock Phys.Rev.Lett. {\bf 112}, 122301 (2014), arXiv:1302.6184.

\bibitem{Adcox:2001jp}
K.~Adcox {\em et~al.}, PHENIX,
\newblock Phys.Rev.Lett. {\bf 88}, 022301 (2002), arXiv:nucl-ex/0109003.

\bibitem{Adare:2012wg}
A.~Adare {\em et~al.}, PHENIX,
\newblock Phys.Rev. {\bf C87}, 034911 (2013), arXiv:1208.2254.

\bibitem{Adare:2012qi}
A.~Adare {\em et~al.}, PHENIX,
\newblock Phys.Rev.Lett. {\bf 111}, 032301 (2013), arXiv:1212.3323.

\bibitem{Abelev:2012hxa}
B.~Abelev {\em et~al.}, ALICE,
\newblock Phys.Lett. {\bf B720}, 52 (2013), arXiv:1208.2711.

\bibitem{CMS:2012aa}
S.~Chatrchyan {\em et~al.}, CMS,
\newblock Eur.Phys.J. {\bf C72}, 1945 (2012), arXiv:1202.2554.

\bibitem{Aamodt:2011vg}
K.~Aamodt {\em et~al.}, ALICE,
\newblock Phys.Rev.Lett. {\bf 108}, 092301 (2012), arXiv:1110.0121.

\bibitem{Chatrchyan:2012wg}
S.~Chatrchyan {\em et~al.}, CMS,
\newblock Eur.Phys.J. {\bf C72}, 2012 (2012), arXiv:1201.3158.

\bibitem{Burke:2013yra}
K.~M. Burke {\em et~al.},
\newblock Phys.Rev. {\bf C90}, 014909 (2014), arXiv:1312.5003.

\bibitem{Baier:2002tc}
R.~Baier,
\newblock Nucl.Phys. {\bf A715}, 209 (2003), arXiv:hep-ph/0209038.

\bibitem{Abelev:2013kqa}
B.~Abelev {\em et~al.}, ALICE,
\newblock JHEP {\bf 03}, 013 (2014), arXiv:1311.0633.

\bibitem{Aad:2014bxa}
G.~Aad {\em et~al.}, ATLAS,
\newblock Phys. Rev. Lett. {\bf 114}, 072302 (2015), arXiv:1411.2357.

\bibitem{Adam:2015ewa}
J.~Adam {\em et~al.}, ALICE,
\newblock Phys.Lett {\bf B746}, 1 (2015), arXiv:1502.01689.

\bibitem{Khachatryan:2016jfl}
V.~Khachatryan {\em et~al.}, CMS,
\newblock Submitted to: Phys. Rev. C  (2016), arXiv:1609.05383.

\bibitem{Aad:2010bu}
G.~Aad {\em et~al.}, ATLAS,
\newblock Phys.Rev.Lett. {\bf 105}, 252303 (2010), arXiv:1011.6182.

\bibitem{Chatrchyan:2012nia}
S.~Chatrchyan {\em et~al.}, CMS,
\newblock Phys.Lett. {\bf B712}, 176 (2012), arXiv:1202.5022.

\bibitem{Chatrchyan:2012gt}
S.~Chatrchyan {\em et~al.}, CMS,
\newblock Phys.Lett. {\bf B718}, 773 (2013), arXiv:1205.0206.

\bibitem{Adam:2015doa}
J.~Adam {\em et~al.}, ALICE,
\newblock JHEP {\bf 09}, 170 (2015), arXiv:1506.03984.

\bibitem{Wang:2013cia}
X.-N. Wang and Y.~Zhu,
\newblock Phys.Rev.Lett. {\bf 111}, 062301 (2013), arXiv:1302.5874.

\bibitem{Kurkela:2014tla}
A.~Kurkela and U.~A. Wiedemann,
\newblock Phys. Lett. {\bf B740}, 172 (2015), arXiv:1407.0293.

\bibitem{Casalderrey-Solana:2016jvj}
J.~Casalderrey-Solana, D.~Gulhan, G.~Milhano, D.~Pablos, and K.~Rajagopal,
\newblock (2016), arXiv:1609.05842.

\bibitem{D'Eramo:2012jh}
F.~D'Eramo, M.~Lekaveckas, H.~Liu, and K.~Rajagopal,
\newblock JHEP {\bf 05}, 031 (2013), arXiv:1211.1922.

\bibitem{Chen:2016vem}
L.~Chen, G.-Y. Qin, S.-Y. Wei, B.-W. Xiao, and H.-Z. Zhang,
\newblock (2016), arXiv:1607.01932.

\bibitem{Sjostrand:2006za}
T.~Sj{\"o}strand, S.~Mrenna, and P.~Z. Skands,
\newblock JHEP {\bf 05}, 026 (2006), arXiv:hep-ph/0603175.

\bibitem{deFlorian:2009fw}
D.~de~Florian,
\newblock Phys.Rev. {\bf D79}, 114014 (2009), arXiv:0904.4402.

\bibitem{Ackermann:2002ad}
K.~H. Ackermann {\em et~al.}, STAR,
\newblock Nucl. Instrum. Meth. {\bf A499}, 624 (2003).

\bibitem{Llope:2003ti}
W.~J. Llope {\em et~al.},
\newblock Nucl. Instrum. Meth. {\bf A522}, 252 (2004), arXiv:nucl-ex/0308022.

\bibitem{Anderson:2003ur}
M.~Anderson {\em et~al.},
\newblock Nucl. Instrum. Meth. {\bf A499}, 659 (2003), arXiv:nucl-ex/0301015.

\bibitem{Aggarwal:2010vc}
M.~M. Aggarwal {\em et~al.}, STAR,
\newblock Phys. Rev. Lett. {\bf 106}, 062002 (2011), arXiv:1009.0326.

\bibitem{Miller:2007ri}
M.~L. Miller, K.~Reygers, S.~J. Sanders, and P.~Steinberg,
\newblock Ann.Rev.Nucl.Part.Sci. {\bf 57}, 205 (2007), arXiv:nucl-ex/0701025.

\bibitem{GEANT3}
{R. Brun, F. Bruyant, M. Maire, A.C. McPherson, and P. Zanarini},
\newblock CERN Data Handling Division DD/EE/84-1  (1985).

\bibitem{Cacciari:2011ma}
M.~Cacciari, G.~P. Salam, and G.~Soyez,
\newblock Eur.Phys.J. {\bf C72}, 1896 (2012), arXiv:1111.6097.

\bibitem{FastJetAntikt}
M.~Cacciari, G.~P. Salam, and G.~Soyez,
\newblock JHEP {\bf 04}, 063 (2008), arXiv:0802.1189.

\bibitem{FastJetArea}
M.~Cacciari, G.~P. Salam, and G.~Soyez,
\newblock JHEP {\bf 04}, 005 (2008), arXiv:0802.1188.

\bibitem{FastJetPileup}
M.~Cacciari and G.~P. Salam,
\newblock Phys. Lett. {\bf B659}, 119 (2008), arXiv:0707.1378.

\bibitem{Cowan:2002in}
G.~Cowan,
\newblock Conf.Proc. {\bf C0203181}, 248 (2002).

\bibitem{Hocker:1995kb}
A.~H{\"o}cker and V.~Kartvelishvili,
\newblock Nucl.Instrum.Meth. {\bf A372}, 469 (1996), arXiv:hep-ph/9509307.

\bibitem{deBarros:2012ws}
G.~de~Barros, B.~Fenton-Olsen, P.~Jacobs, and M.~Ploskon,
\newblock Nucl.Phys. {\bf A910-911}, 314 (2013), arXiv:1208.1518.

\bibitem{deFlorian:2005yj}
D.~de~Florian and W.~Vogelsang,
\newblock Phys.Rev. {\bf D71}, 114004 (2005), arXiv:hep-ph/0501258.

\bibitem{d'Enterria:2013vba}
D.~d'Enterria, K.~J. Eskola, I.~Helenius, and H.~Paukkunen,
\newblock Nucl.Phys. {\bf B883}, 615 (2014), arXiv:1311.1415.

\bibitem{Renk:2011gj}
T.~Renk, H.~Holopainen, R.~Paatelainen, and K.~J. Eskola,
\newblock Phys.Rev. {\bf C84}, 014906 (2011), arXiv:1103.5308.

\bibitem{Armesto:2007dt}
N.~Armesto, L.~Cunqueiro, C.~A. Salgado, and W.-C. Xiang,
\newblock JHEP {\bf 02}, 048 (2008), arXiv:0710.3073.

\bibitem{Chang:2014fba}
N.-B. Chang, W.-T. Deng, and X.-N. Wang,
\newblock Phys.Rev. {\bf C89}, 034911 (2014), arXiv:1401.5109.

\bibitem{Adamczyk:2013gw}
L.~Adamczyk {\em et~al.}, STAR,
\newblock Phys. Rev. {\bf C88}, 014902 (2013), arXiv:1301.2348.

\bibitem{Adare:2014bga}
A.~Adare {\em et~al.}, PHENIX,
\newblock Phys. Rev. {\bf C92}, 034913 (2015), arXiv:1412.1043.

\bibitem{deBarros:2011ph}
G.~de~Barros,
\newblock AIP Conf.Proc. {\bf 1441}, 825 (2012), arXiv:1109.4386.

\bibitem{D'Agostini:1994zf}
G.~D'Agostini,
\newblock Nucl.Instrum.Meth. {\bf A362}, 487 (1995).

\bibitem{Chang:2013rca}
H.-M. Chang, M.~Procura, J.~Thaler, and W.~J. Waalewijn,
\newblock Phys.Rev.Lett. {\bf 111}, 102002 (2013), arXiv:1303.6637.

\bibitem{Pumplin:2002vw}
J.~Pumplin {\em et~al.},
\newblock JHEP {\bf 07}, 012 (2002), arXiv:hep-ph/0201195.

\bibitem{deFlorian:2007ekg}
D.~de~Florian, R.~Sassot, and M.~Stratmann,
\newblock Phys. Rev. {\bf D76}, 074033 (2007), arXiv:0707.1506.

\bibitem{Skands:2010ak}
P.~Z. Skands,
\newblock Phys.Rev. {\bf D82}, 074018 (2010), arXiv:1005.3457.

\bibitem{Abelev:2013fn}
B.~Abelev {\em et~al.}, ALICE,
\newblock Phys.Lett. {\bf B722}, 262 (2013), arXiv:1301.3475.

\bibitem{Chatrchyan:2014gia}
S.~Chatrchyan {\em et~al.}, CMS,
\newblock Phys. Rev. {\bf D90}, 072006 (2014), arXiv:1406.0324.

\bibitem{Soyez:2011np}
G.~Soyez,
\newblock Phys.Lett. {\bf B698}, 59 (2011), arXiv:1101.2665.

\bibitem{Dasgupta:2016bnd}
M.~Dasgupta, F.~A. Dreyer, G.~P. Salam, and G.~Soyez,
\newblock JHEP {\bf 06}, 057 (2016), arXiv:1602.01110.

\bibitem{Kang:2017frl}
Z.-B. Kang, F.~Ringer, and I.~Vitev,
\newblock (2017), arXiv:1701.05839.

\bibitem{Chatrchyan:2013kwa}
S.~Chatrchyan {\em et~al.}, CMS,
\newblock Phys.Lett. {\bf B730}, 243 (2014), arXiv:1310.0878.

\bibitem{Mueller:2016gko}
A.~H. Mueller, B.~Wu, B.-W. Xiao, and F.~Yuan,
\newblock Phys. Lett. {\bf B763}, 208 (2016), arXiv:1604.04250.

\end{thebibliography}
